\documentclass[reprint,
preprintnumbers,
nofootinbib,
amsmath,amssymb,
aps,
prd,
superscriptaddress,
]{revtex4-2}

\usepackage{graphicx}
\usepackage{dcolumn}
\usepackage{bm}
\usepackage[hidelinks]{hyperref}
\usepackage[all]{hypcap}
\usepackage{xcolor}
\usepackage{enumitem}
\usepackage{multirow}
\usepackage{orcidlink}

\hypersetup{
	colorlinks,
	linkcolor={red!50!black},
	citecolor={blue!50!black},
	urlcolor={blue!80!black}
}

\newcommand{\order}[1]{\mathcal{O}(#1)}

\DeclareMathOperator{\Tr}{Tr}

\begin{document}
	
\preprint{ADP-25-1/T1263}

\title{Center vortices and the \texorpdfstring{$\mathrm{SU}(3)$}{SU(3)} conformal window}

\author{Jackson A. Mickley\;\orcidlink{0000-0001-5294-2823}}
\affiliation{Centre for the Subatomic Structure of Matter, Department of Physics, The University of Adelaide, South Australia 5005, Australia}

\author{Derek B. Leinweber\;\orcidlink{0000-0002-4745-6027}}
\affiliation{Centre for the Subatomic Structure of Matter, Department of Physics, The University of Adelaide, South Australia 5005, Australia}

\author{Daniel Nogradi\;\orcidlink{0000-0002-3107-1958}}
\affiliation{E{\"o}tv{\"o}s Lor{\'a}nd University, Institute of Physics and Astronomy, Department of Theoretical
Physics, Budapest, Hungary}

\begin{abstract}
A novel approach for estimating the lower end of the $\mathrm{SU}(3)$ conformal window is presented through the study of center vortex geometry and its dependence on the number of fermion flavors $N_f$. Values ranging from $N_f = 2$--$8$ are utilized to infer an upper limit for vortex behavior in the low $N_f$ phase, which may inform the transition to the conformal window. The simulations are performed at a single lattice spacing and pion mass, both fixed for all $N_f$. Visualizations of the center vortex structure in three-dimensional slices of the lattice reveal a growing roughness in the vortex matter as a function of $N_f$, embodied by an increase in the density of vortex matter in the percolating cluster and a simultaneous reduction in secondary clusters disconnected from the percolating cluster in 3D slices. This is quantified by various bulk properties, including the vortex and branching point densities. A correlation of the vortex structure reveals a turning point near $N_f \simeq 5$ past which a randomness in the vortex field becomes the dominant aspect of its evolution with $N_f$. As a byproduct, extrapolations to the vortex content of a uniform-random gauge field provide a critical point at which there must be a drastic shift in vacuum field structure. A precise estimate for the critical value is extracted as $N_f^* = 11.43(16)(17)$, close to various other estimates.
\end{abstract}

\maketitle

\section{Introduction} \label{sec:intro}
A generic gauge theory coupled to $N_f$ flavors of massless fermions typically exhibits three distinct phases. For low $N_f$ the system is confining and chiral symmetry is spontaneously broken, while for very high $N_f$ asymptotic freedom is lost and the model only makes sense as a free theory. For intermediate values of $N_f$ between these extremes, chiral symmetry is intact and the theory is conformal with nontrivial anomalous dimensions \cite{Banks:1981nn}. The intermediate $N_f$ range is called the conformal window and its lower end, $N_f^*$, in addition to being an interesting quantity in its own right, is important as an input for various beyond the Standard Model scenarios based on strong dynamics. For a review, see Refs.~\cite{Nogradi:2016qek, Drach:2020qpj} and references therein.

The upper end of the conformal window at which asymptotic freedom is lost, $N_f = 16.5$, is determined by the one-loop beta function. This implies that it arises purely due to perturbative phenomena. On the other hand, the more difficult question of $N_f^*$ is expected to be determined by nonperturbative phenomena. Estimates for $N_f^*$ from different approaches vary \cite{Appelquist:1988yc, Cohen:1988sq, Sannino:2004qp, Dietrich:2006cm, Armoni:2009jn, Braun:2009ns, Frandsen:2010ej, Ryttov:2016ner, Kim:2020yvr, Lee:2020ihn}, though mostly fall within $8 \leq N_f^* \leq 13$ without a clear consensus. The specific case of $N_f = 12$ has been a particular focus in lattice simulations, with extensive effort to determine whether the system is infrared conformal \cite{Appelquist:2007hu, Appelquist:2009ty, Fodor:2009wk, Fodor:2011tu, Hasenfratz:2011xn, DeGrand:2011cu, Lin:2012iw, Aoki:2012eq, LatKMI:2013bhp, Fodor:2016zil, Hasenfratz:2016dou, Fodor:2017gtj, Hasenfratz:2017qyr, Fodor:2017nlp, Hasenfratz:2024fad}. The majority of these studies suggest $N_f = 12$ does lie inside the conformal window, though a minority conclude that it exists in the confining and chirally broken phase or are inconclusive, making this a point of contention.

There is a wealth of literature that indicates the relevant nonperturbative degrees of freedom in the low-$N_f$ phase are center vortices \cite{tHooft:1977nqb, tHooft:1979rtg, Nielsen:1979xu, Greensite:2003bk}. Extracting the center vortex degrees of freedom from configurations generated by lattice simulations has been successful in reproducing many emergent phenomena of QCD both qualitatively and quantitatively, including confinement and chiral symmetry breaking \cite{DelDebbio:1996lih, DelDebbio:1997ep, Langfeld:1997jx, DelDebbio:1998luz, Faber:1997rp, Faber:1998qn, Kovacs:1998xm, Langfeld:1998cz, Engelhardt:1998wu, Bertle:1999tw, Engelhardt:1999fd, Engelhardt:1999wr, Faber:1999sq, deForcrand:1999our, deForcrand:2000pg, Kovacs:2000sy, Langfeld:2001cz, Langfeld:2003ev, Engelhardt:2003wm, Gattnar:2004gx, Bornyakov:2007fz, Bowman:2008qd, Bowman:2010zr, OMalley:2011aa, Hollwieser:2013xja, Hollwieser:2014soz, Trewartha:2015nna, Greensite:2016pfc, Trewartha:2017ive, Biddle:2022zgw, Biddle:2022acd, Mickley:2024zyg}. In this work, we study center vortices for $2 \leq N_f \leq 8$ in the framework of lattice QCD to ascertain how the vortex structure evolves as the conformal window is approached. This will allow us to infer whether center vortices carry any new insight on the nature of the conformal window. We observe a smooth $N_f$ dependence for $N_f \gtrsim 5$ that allows an extrapolation beyond our highest point at $N_f = 8$. By proposing certain bounds on our vortex observables, our results indicate that a drastic shift in behavior must occur at $N_f \simeq 11$--$12$, suggesting an estimate for $N_f^*$.

This paper is structured as follows. In Sec.~\ref{sec:simdetails}, we summarize our lattice simulation details and the line of constant physics. The center vortex model along with their identification on the lattice is reviewed in Sec.~\ref{sec:centervortices}. Thereafter, visualizations of the center vortex structure at increasing $N_f$ are presented in Sec.~\ref{sec:visualizations}. Section~\ref{sec:analysis} contains our primary quantitative analysis on the evolution of intrinsic vortex statistics as a function of $N_f$. An estimate for $N_f^*$ is extracted by performing an extrapolation based on the observed trends. Finally, we conclude our main findings in Sec.~\ref{sec:conclusion}.

\section{Simulation details} \label{sec:simdetails}
The calculations are carried out at fixed lattice spacing and pion mass. Hence, the line of constant physics is implemented by fixing both $a f_\pi$ and $a m_\pi$ for all $N_f$. Staggered fermions are used with two steps of stout-smeared links \cite{Morningstar:2003gk, BMW:2010skj} at smearing parameter $\rho = 0.12$, together with the tree-level Symanzik-improved gauge action. The setup is identical to \cite{Nogradi:2019iek, Nogradi:2019auv, Kotov:2021mgp}. An important consideration with the chosen action is the existence of a bulk phase transition separating the physically relevant $\beta\to\infty$ region and a lattice-artefact phase at finite $\beta$ and fermion mass $m$ \cite{Cheng:2011ic}. In Ref.~\cite{Kotov:2021mgp} the phase diagram in the bare parameter space $(\beta,m)$ was mapped out for all $N_f$ under consideration, which was employed to ensure our present simulations all lie in the physically relevant phase. The question of whether the center vortex degrees of freedom are additionally sufficient to detect this bulk phase transition is an interesting question for future work.

The correct number of continuum flavors is achieved by utilizing both the HMC \cite{Duane:1987de} and RHMC \cite{Clark:2006fx} algorithms or a combination of both, as
needed. Multiple integration time scales \cite{Sexton:1992nu} and the Omelyan integrator \cite{Omelyan:2002qkh, Takaishi:2005tz} are used along the molecular dynamics trajectories. For each $N_f$ a total of 10000 trajectories are generated and every $10$th trajectory is extracted for measurements.

The bare parameters for our simulations are tuned to the line of constant physics given by $a f_\pi = 0.0568$ and $a m_\pi = 0.248$. Tuning of $a m_\pi$ is performed to better than 0.5\% and tuning of $a f_\pi$ to better than 1\%. The precise details are provided in Table~\ref{tab:ensembles}. The lattice volume is uniformly $24^3 \times 48$ and the analysis of finite volume effects as a function of $N_f$ in \cite{Nogradi:2019iek, Nogradi:2019auv, Kotov:2021mgp} indicates that finite volume effects are at most 1\%. As a result, we assign an additional 1\% error to any observable we measure originating from 
both possible mistuning and finite volume effects. The convention for the decay constant corresponds to the choice $f_\pi \simeq 92\,$MeV in QCD.
\begin{table}
	\centering
	\caption{\label{tab:ensembles} The bare parameters (beta value and bare fermion mass) and measured pion masses and decay constants (in lattice units) for each $N_f$ ensemble. These latter two values are tuned to our target line of constant physics $a m_\pi = 0.248$, $a f_\pi = 0.0568$ to within 0.5\% for $a m_\pi$ and 1\% for $a f_\pi$.}
	\begin{ruledtabular}
		\begin{tabular}{ccD{.}{.}{1.5}cD{.}{.}{2.8}}
			$N_f$ & $\beta$ & \multicolumn{1}{c}{$m$} & $a m_\pi$ & \multicolumn{1}{c}{$a f_\pi$} \\
			\colrule \\[-0.9em]
			2 & 4.0638 & 0.0248  & 0.2494(2) & 0.0567(1)  \\
			3 & 3.8825 & 0.0229  & 0.2485(3) & 0.0574(1)  \\
			4 & 3.7091 & 0.0215  & 0.2492(3) & 0.0568(1)  \\
			5 & 3.5013 & 0.0192  & 0.2474(3) & 0.05687(8) \\
			6 & 3.2451 & 0.01692 & 0.2487(2) & 0.05737(8) \\
			7 & 2.9354 & 0.01481 & 0.2478(2) & 0.05622(7) \\
			8 & 2.5014 & 0.01236 & 0.2475(2) & 0.0566(1)  \\[-0.2em]
		\end{tabular}
	\end{ruledtabular}
\end{table}

\section{Center vortices} \label{sec:centervortices}
Center vortices~\cite{tHooft:1977nqb, tHooft:1979rtg, Nielsen:1979xu, Greensite:2003bk} are regions of the gauge field that carry magnetic flux quantized according to the center of SU(3),
\begin{equation}
	\mathbb{Z}_3 = \left\{ \exp\left(\frac{2\pi i}{3}\, n \right) \mathbb{I} \;\middle|\; n = -1,0,1 \right\} \,.
\end{equation}
Physical vortices in the QCD ground-state fields have a finite thickness. Any Wilson loop that encircles a vortex picks up a factor of an element of $\mathbb{Z}_3$,
\begin{equation}
	W(C) \longrightarrow z\,W(C) \,.
\end{equation}

In contrast, on the lattice ``thin" center vortices are extracted through a well-known gauge-fixing procedure that seeks to bring each link variable $U_\mu(x)$ as close as possible to an element of $\mathbb{Z}_3$, known as maximal center gauge (MCG). These thin vortices form closed surfaces in four-dimensional Euclidean spacetime, and thus one-dimensional structures in a three-dimensional slice of the four-dimensional spacetime.

Fixing to MCG is typically performed by finding the gauge transformation $\Omega(x)$ to maximize the functional \cite{Montero:1999by}
\begin{equation}
	R = \sum_{x,\,\mu} \,\left| \Tr U_\mu^{\Omega}(x) \right|^2 \,.
\end{equation}
The links are subsequently projected onto the center,
\begin{equation}
	U_\mu^{\Omega}(x) \longrightarrow Z_\mu(x) = \exp\left(\frac{2\pi i}{3} \, n_\mu(x) \right) \mathbb{I} \in \mathbb{Z}_3 \,,
\end{equation}
with $n_\mu(x) \in \{-1,0,1\}$ identified as the center phase nearest to $\arg \Tr U_\mu(x)$ for each link. Finally, the locations of vortices are identified by nontrivial plaquettes in the center-projected field,
\begin{equation} \label{eq:centerprojplaq}
	P_{\mu\nu}(x) = \prod_\square Z_\mu(x) = \exp\left(\frac{2\pi i}{3} \, m_{\mu\nu}(x) \right)\mathbb{I}
\end{equation}
with $m_{\mu\nu}(x) = \pm 1$. The value of $m_{\mu\nu}(x)$ is referred to as the \textit{center charge} of the vortex, and we say the plaquette is pierced by a vortex.

Due to a Bianchi identity satisfied by the vortex fields~\cite{Engelhardt:1999wr, Spengler:2018dxt}, the center charge is conserved such that the vortex topology constitutes closed sheets in four dimensions, or closed lines in three-dimensional slices of the lattice. Although gauge dependent, numerical evidence strongly suggests that the projected vortex locations correspond to the physical vortices of the original fields~\cite{DelDebbio:1998luz, Langfeld:2003ev, Montero:1999by, Faber:1999gu}. This allows one to investigate the significance of center vortices through the vortex-only field $Z_\mu(x)$.

Center vortices are understood to underpin both confinement and dynamical chiral symmetry breaking. Percolating center vortices, in which a single connected cluster fills the volume, naturally generate an area-law falloff for large Wilson loops~\cite{Bertle:1999tw, Engelhardt:1999fd, Engelhardt:1999wr, Engelhardt:1998wu}. This is often taken as an indicator of confinement for static heavy quarks \cite{Wilson:1974sk}, allowing the extraction of a linear quark-antiquark potential $V(r)$ through asymptotic space-time Wilson loops,
\begin{align} \label{eq:staticquarkpotential}
	\langle W(r,t) \rangle \sim \exp\left(-V(r) \, t\right) \,, && t \text{ large} \,.
\end{align}
Such behavior is realized in numerical simulations both with and without dynamical fermions \cite{DelDebbio:1996lih, DelDebbio:1998luz, Bertle:1999tw, Engelhardt:1999fd, Langfeld:2003ev, Biddle:2019gke, Biddle:2022zgw, Biddle:2023lod}, though the quantity of vortex matter and in particular the density of vortices in the percolating cluster is substantially greater in the presence of dynamical fermions \cite{Biddle:2023lod}. This raises the interesting question of how the vortex geometry will continue to evolve as the number of fermion flavors is varied.

In addition, the removal of center vortices through $R_\mu(x) = U_\mu(x) Z_\mu^\dagger(x)$ results in a vanishing string tension \cite{Langfeld:2003ev, Biddle:2022zgw}. It is here also that the connection to chiral symmetry breaking is made apparent. Calculations of the quark mass function on the vortex-removed field show a substantial infrared suppression, indicating chiral symmetry has been recovered \cite{Bowman:2008qd, Bowman:2010zr, OMalley:2011aa, Trewartha:2015nna}. Furthermore, computations of the low-lying hadron spectrum on the vortex-only and vortex-removed fields demonstrate that center vortices alone are able to reproduce the salient features of the hadron spectrum, while the vortex-removed configurations feature hadronic degeneracies that reflect restoration of chiral symmetry \cite{Trewartha:2017ive}.

These findings strongly suggest that center vortices will be sensitive to the conformal window, and motivate investigating how vortex structure evolves as $N_f$ increases towards its lower end. By placing justifiable limits on vortex properties based on the observed evolution, we are able to elicit an estimate for the point at which these extremes are reached with consistency across several different aspects of vortex geometry. This provides a novel approach to constraining $N_f^*$.

\section{Visualizations} \label{sec:visualizations}
We start by visualizing the center vortex structure for various choices of $N_f$, revealing the qualitative features of vortex evolution as $N_f$ increases. It will also be interesting to draw a comparison to the prior work in $2+1$ dynamical lattice simulations \cite{Biddle:2023lod}. The visualizations are constructed utilizing techniques previously established in Ref.~\cite{Biddle:2019gke}. To produce a three-dimensional visualization, we take a cross section of the full four-dimensional lattice by holding one of the four coordinates fixed. For each plaquette in the three-dimensional slice, this leaves one orthogonal direction that is used to identify the plaquette. A vortex is then rendered as an arrow existing on the dual lattice and piercing the associated nontrivial plaquette. One can interpret these arrows as the vortex ``axis of rotation".

\begin{figure}
	\centering
	\includegraphics[width=\linewidth]{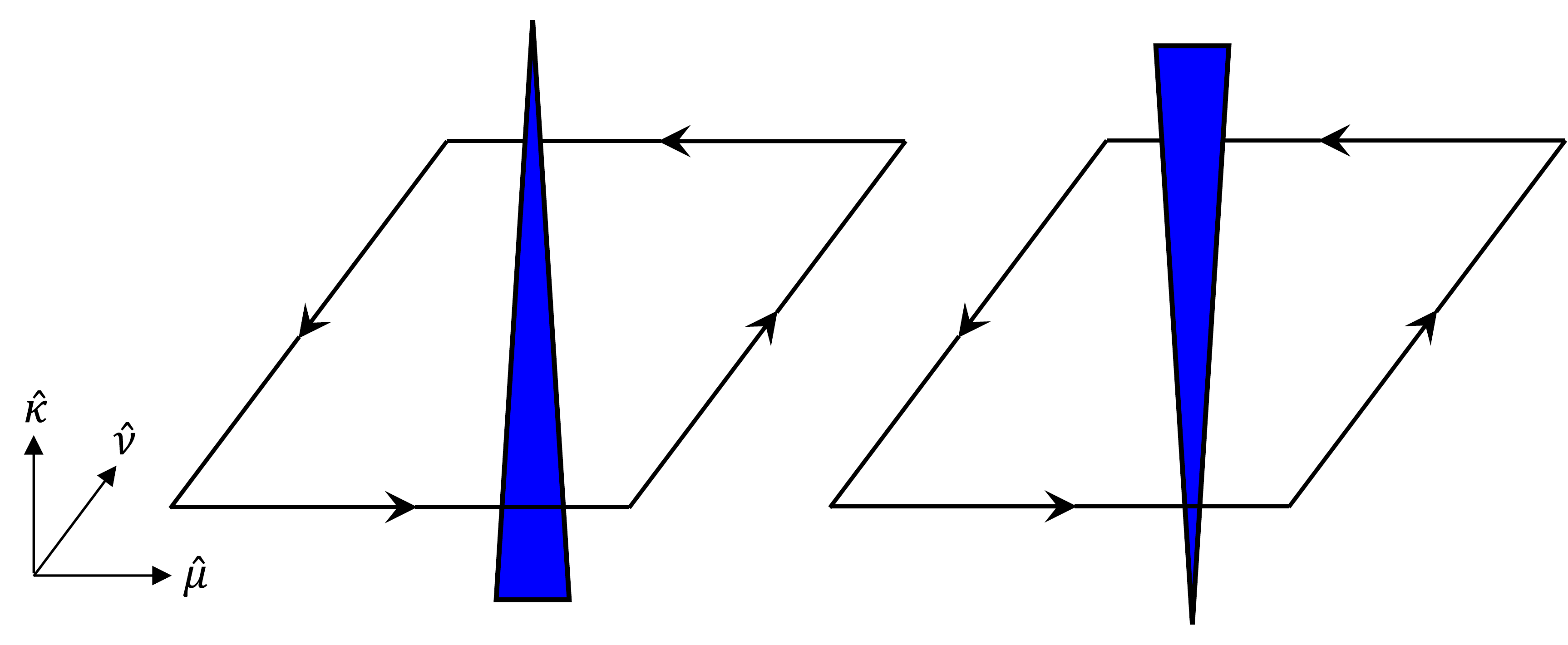}
	\caption{\label{fig:visconvention} The visualization convention for center vortices. An $m=+1$ vortex (\textbf{left}) is represented by a jet in the available orthogonal dimension, with the direction given by the right-hand rule. An $m=-1$ vortex (\textbf{right}) is rendered by a jet in the opposite direction, such that we always show the flow of $m=+1$ center charge.}
\end{figure}

The orientation of an $m = +1$ vortex is determined by applying the right-hand rule. Since the flow of $m = -1$ center charge is indistinguishable from an opposite flow of $m = +1$ center charge, we display an $m = -1$ vortex as a jet pointing in the opposite direction from the right-hand rule. In other words, the visualizations exclusively show the flow of $m = +1$ center charge. This convention is demonstrated in Fig.~\ref{fig:visconvention}.

\begin{figure*}
	\centering
	\includegraphics[width=0.48\linewidth]{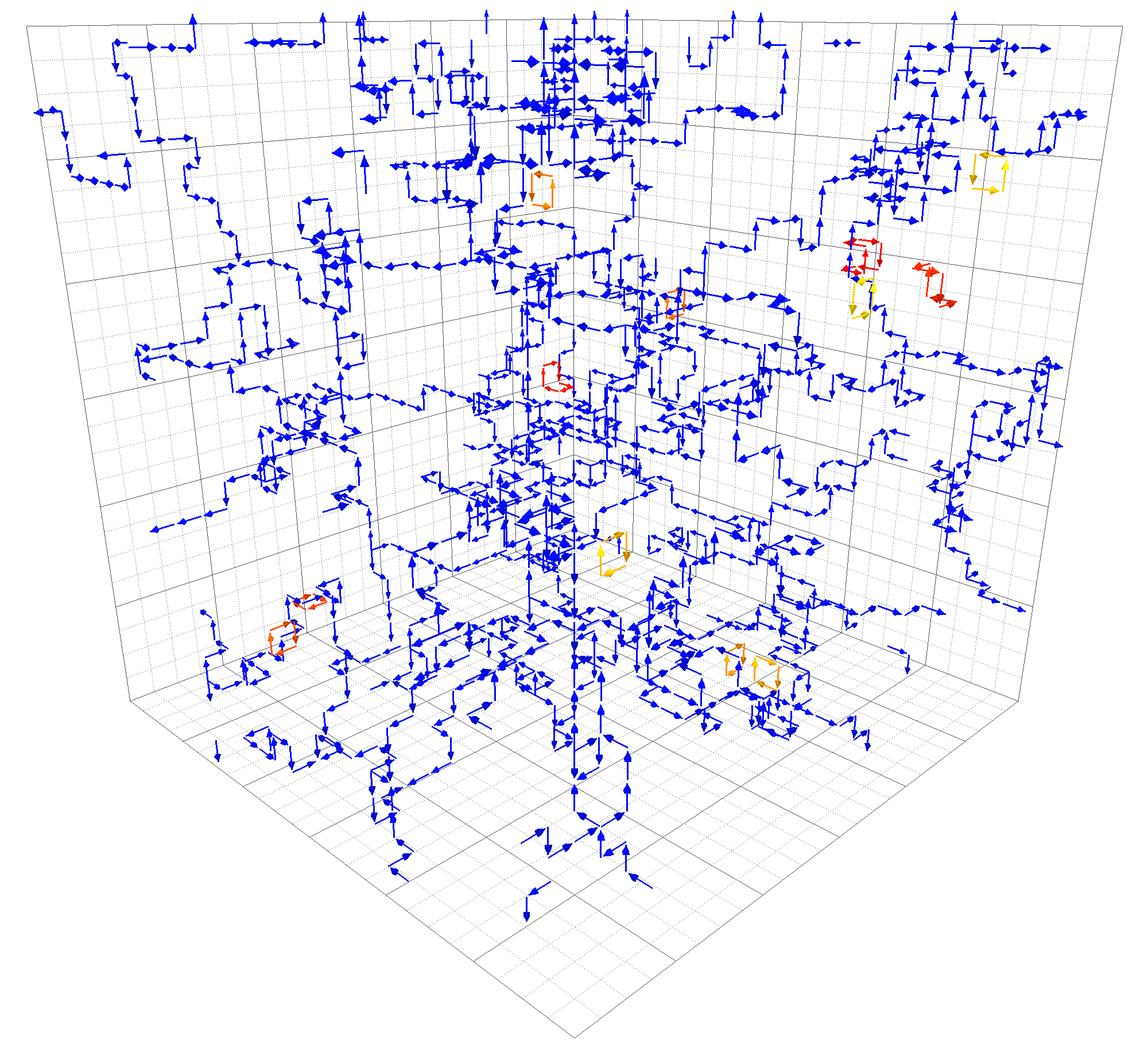}
	\includegraphics[width=0.48\linewidth]{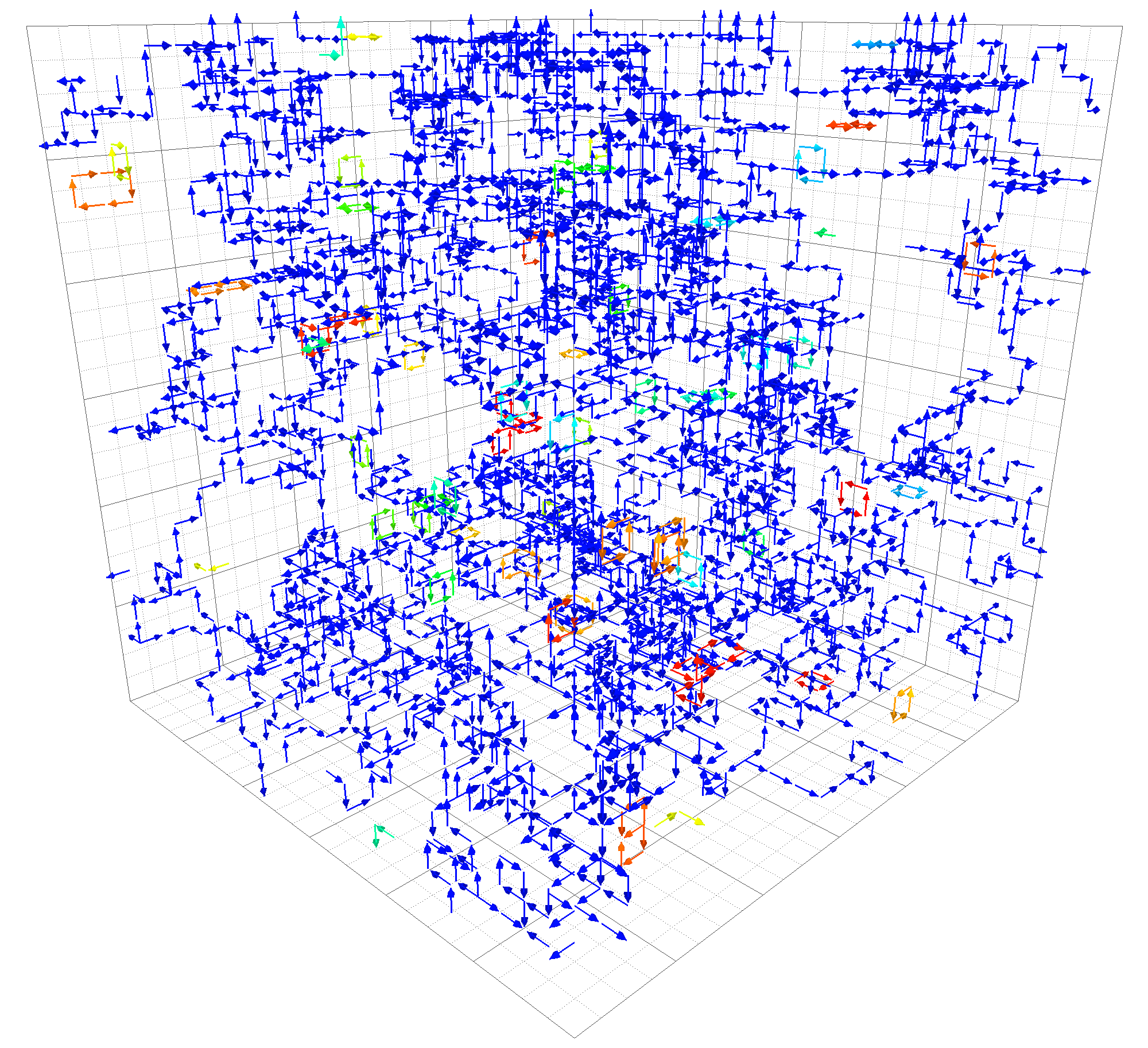}
	
	\vspace{1em}
	
	\includegraphics[width=0.48\linewidth]{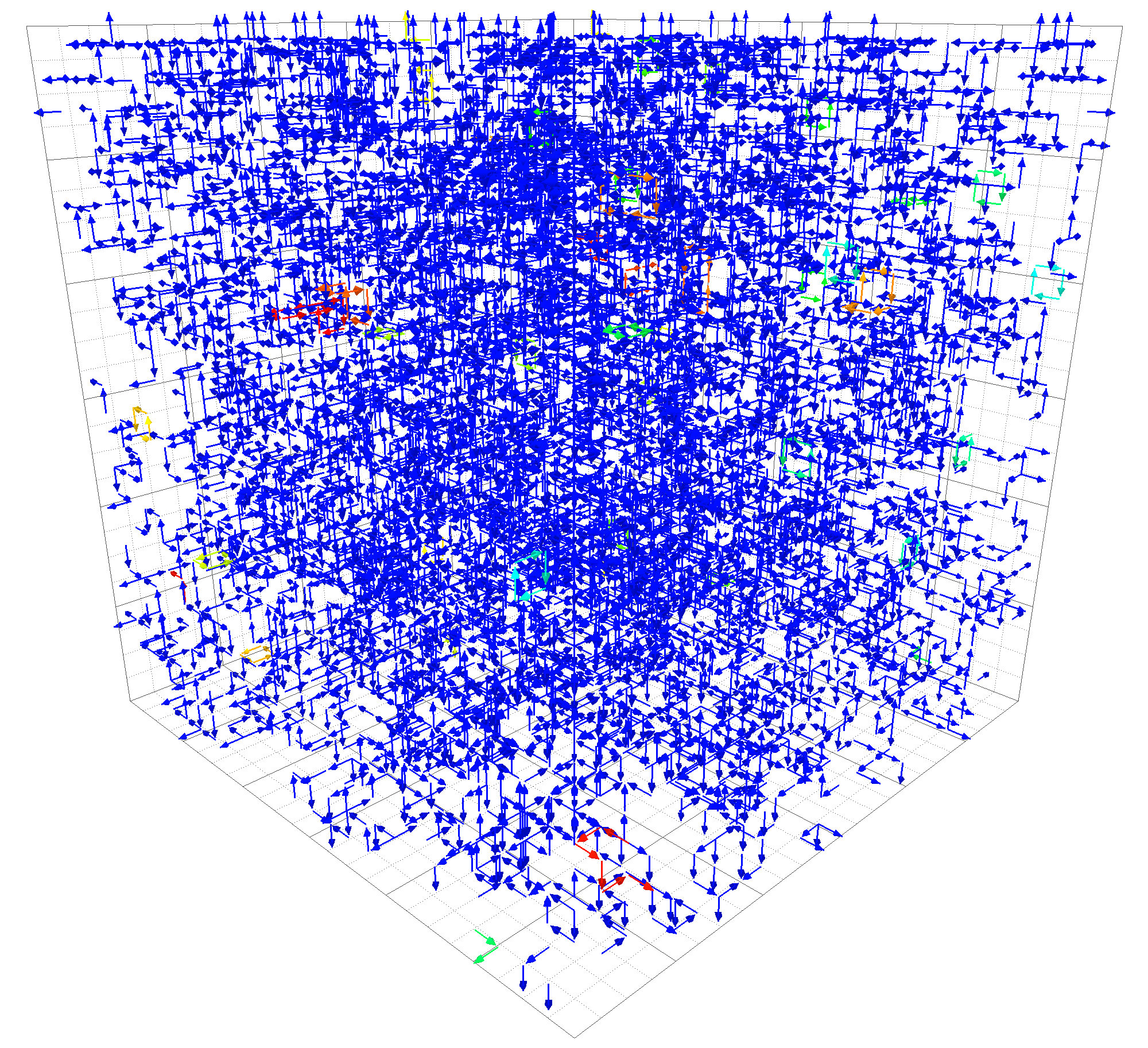}
	\includegraphics[width=0.48\linewidth]{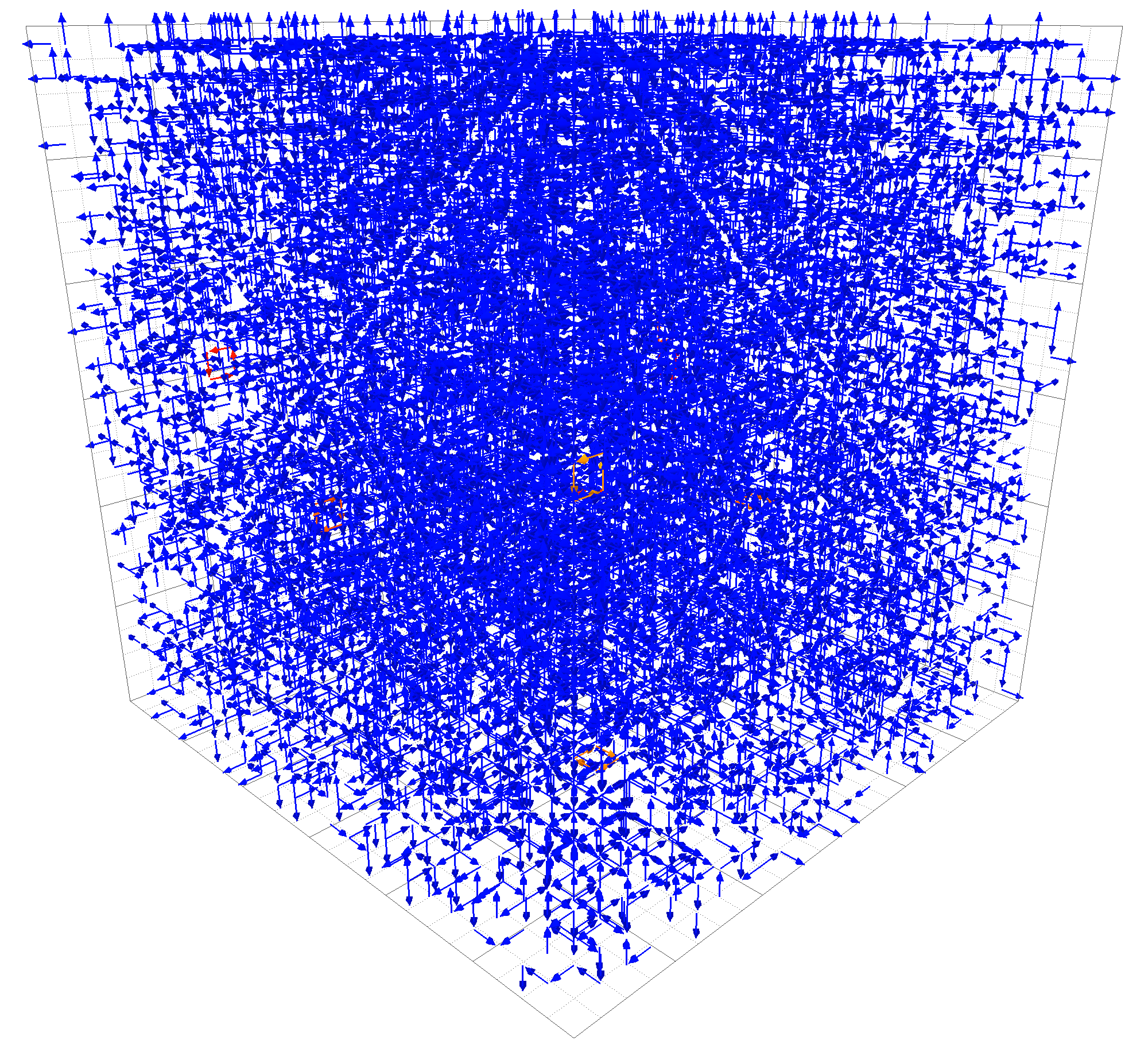}
	
	\caption{\label{fig:nfvis} Center vortex structure in temporal slices of the lattice for increasing number of fermion flavors: $N_f = 2$ (\textbf{top left}), $4$ (\textbf{top right}), $6$ (\textbf{bottom left}) and $8$ (\textbf{bottom right}). The amount of vortex matter rapidly grows as a function of $N_f$, especially the density of the percolating cluster which is colored in dark blue. Initially, the number of secondary clusters also increases; however, as the percolating cluster continues to fill the volume, these secondary clusters diminish as less space is available for them to pervade.}
\end{figure*}

At low temperatures the vortex sheet percolates all four dimensions \cite{Langfeld:1998cz, Bertle:1999tw, Engelhardt:1999fd, Biddle:2019gke, Mickley:2024zyg}, and as such the observed three-dimensional structure is insensitive to the choice of dimension sliced over. We show visualizations obtained from slicing through the temporal dimension, holding the Euclidean time coordinate fixed. The resulting three-dimensional volume comprises the three spatial dimensions ($x$-$y$-$z$). Typical vortex structures are presented for the four even numbers of fermion flavors under consideration in Fig.~\ref{fig:nfvis}. This is sufficient to establish the bulk changes in vortex geometry that arise from increasing $N_f$, before proceeding to a detailed quantitative analysis of the full evolution in Sec.~\ref{sec:analysis}. 

For small $N_f$, the structures are comparable to those previously found with $2+1$ flavors of dynamical fermions \cite{Biddle:2023lod, Mickley:2024vkm}. This includes a percolating cluster colored in dark blue that dominates the structure, with small ``secondary clusters" scattered throughout the lattice. In relation to the pure gauge sector, which can be treated as a theory with $N_f = 0$, there is a substantial increase in vortex matter and the prevalence of secondary clusters. With that understanding, it is perhaps unsurprising that the density of vortex matter continues to rise as the number of flavors increases. This is particularly reflected in the percolating cluster, which rapidly grows in density. By $N_f = 8$, it is so pervasive that one can barely see into the volume.

The behavior of secondary clusters is also fascinating. It is clear, especially from comparing $N_f = 2$ and $4$, that the number of secondary clusters initially experiences an increase as a function of $N_f$. However, at the upper end of our $N_f$ range the proliferation of secondary clusters starts to decline. This is a natural consequence of the growing percolating cluster. As it saturates the volume, beyond a certain point there will be insufficient space for additional secondary clusters. Therefore, these two observations are inherently intertwined, with the decrease in secondary clusters necessitated by the continued amplification of the percolating cluster.

These findings can be intuitively summarized with the observation that center vortex matter is becoming ``rougher" as $N_f$ increases. We propose that this is a necessity in maintaining a line of constant physics to draw a fair comparison as the number of fermion flavors is varied. Indeed, as seen in Table~\ref{tab:ensembles}, the $\beta$ value must decrease as $N_f$ is increased in order to hold $a m_\pi$ and $a f_\pi$ fixed. That is, we are actively working against the ``perturbative limit" realized by $\beta \to \infty$.

However, we stress that the particular manner in which this ``roughness" manifests is nonetheless nontrivial. For instance, we understand that even if clusters appear as disconnected in a three-dimensional slice, it is possible that they lie in the same connected surface in four dimensions due to the surface's curvature \cite{Mickley:2024vkm}. It would be reasonable to propose that as the vortex matter becomes rougher, there are greater instances of the surface curving to produce disconnected clusters in the three-dimensional cross sections, effectively breaking the percolating cluster in three dimensions apart. The fact that we observe the percolating cluster increase in density with a notable absence of disconnected clusters is remarkable, and is indicative of center vortex geometry in the full four dimensions.
	
\section{Quantitative analysis} \label{sec:analysis}
Having identified the qualitative changes in center vortex structure as the number of flavors increases, we now move to a quantitative investigation of intrinsic vortex statistics. We start by introducing the measures used to describe vortex behavior in Sec.~\ref{subsec:statistics}. Based on their observed $N_f$ dependence, in Sec.~\ref{subsec:haar} we propose bounds to these quantities that cannot be exceeded. Then, in Sec.~\ref{subsec:fits}, we will perform extrapolations utilizing these limits to provide an estimate for $N_f^*$.

Throughout this section, all statistics are obtained using 100 bootstrap ensembles, with statistical errors calculated through the standard deviation of the bootstrap estimates. As mentioned in Sec.~\ref{sec:simdetails}, we include 1\% systematic errors on our results to account for mistuning and possible finite volume artefacts. These are combined with the statistical errors in quadrature, and form the dominant contribution.
	
\subsection{Vortex statistics} \label{subsec:statistics}
\begin{table}
	\centering
	\caption{\label{tab:statistics} The vortex density $\rho_\mathrm{vortex}$, branching point density $\rho_\mathrm{branch}$, and correlation on consecutive slices $C(\Delta = 1)$, for each $N_f$ ensemble. Quoted errors represent a 1\% systematic error combined with statistical uncertainties in quadrature. The numerical values from the Haar-random ensemble described in Sec.~\ref{subsec:haar} are also provided. Here, the errors are purely statistical.}
	\begin{ruledtabular}
		\begin{tabular}{cD{.}{.}{2.8}D{.}{.}{2.8}D{.}{.}{2.8}}
			$N_f$ & \multicolumn{1}{c}{$\rho_\mathrm{vortex}$} & \multicolumn{1}{c}{$\rho_\mathrm{branch}$} & \multicolumn{1}{c}{$C(\Delta = 1)$} \\
			\colrule \\[-0.9em]
			2 & 0.0324(3)  & 0.0073(1)  & 0.1833(19) \\
			3 & 0.0492(5)  & 0.0134(2)  & 0.1666(17) \\
			4 & 0.0755(8)  & 0.0245(3)  & 0.1559(16) \\
			5 & 0.1269(13) & 0.0504(5)  & 0.1545(15) \\
			6 & 0.2057(21) & 0.0983(10) & 0.1646(16) \\
			7 & 0.2908(29) & 0.1594(16) & 0.1819(18) \\
			8 & 0.3779(38) & 0.2297(23) & 0.2057(21) \\
			\colrule \\[-0.9em]
			Haar & 0.66669(3) & 0.50208(4) & 0.33337(4) \\[-0.2em]
		\end{tabular}
	\end{ruledtabular}
\end{table}
The values of the quantities introduced herein evaluated on each of our $N_f$ ensembles are provided in Table~\ref{tab:statistics}.

\subsubsection{Vortex density}
\begin{figure}
	\centering
	\includegraphics[width=\linewidth]{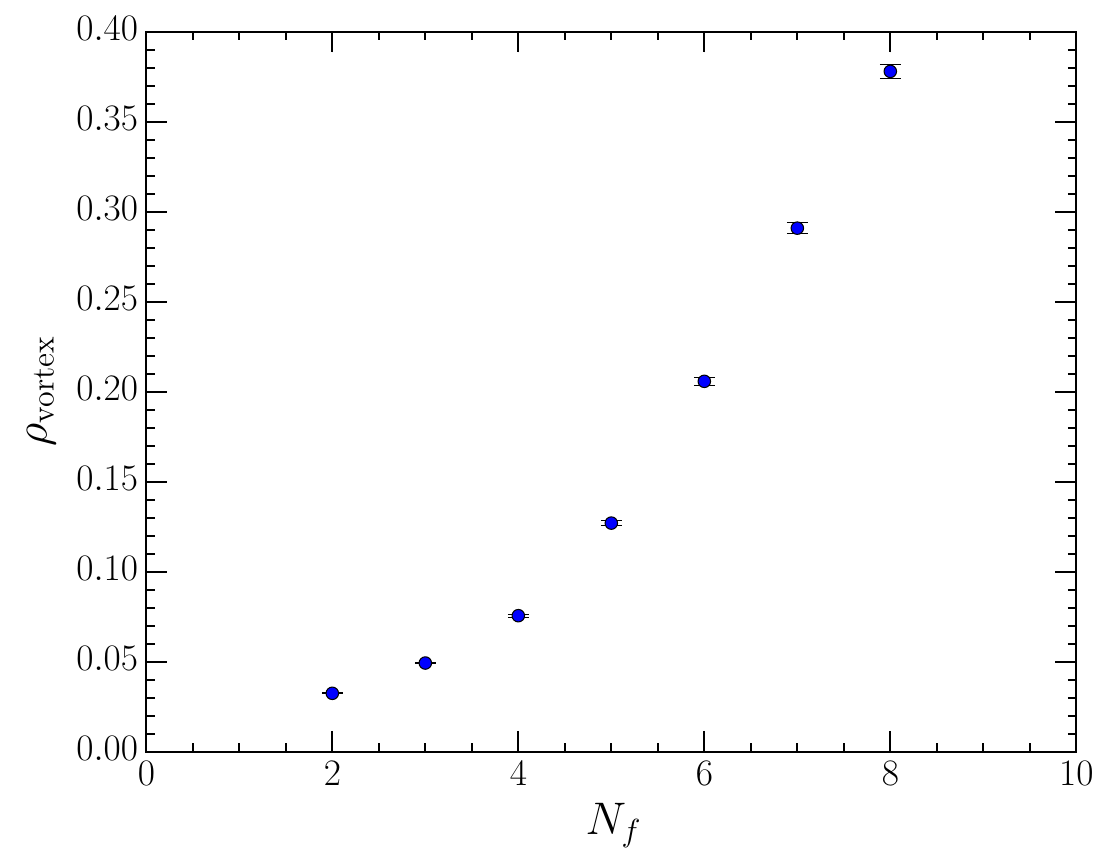}
	
	\vspace{-0.5em}
	
	\caption{\label{fig:vortexdensity} The dimensionless vortex density defined in Eq.~(\ref{eq:vortexdensity}) as a function of $N_f$. It rises quickly as the number of flavors grows, and has appeared to settle into a stable linear trend for $N_f \gtrsim 5$.}
\end{figure}

The first quantity we consider is the vortex density, defined simply as the proportion of plaquettes pierced by a vortex,
\begin{equation} \label{eq:vortexdensity}
	\rho_\mathrm{vortex} = \frac{\text{Number of nontrivial plaquettes}}{6\,N_\mathrm{sites}} \,,
\end{equation}
where $N_\mathrm{sites}$ is the dimensionless lattice volume, and $\binom{4}{2} = 6$ counts the number of distinct plaquette orientations. $\rho_\mathrm{vortex}$ can be converted to a physical quantity through dividing by the area of a single plaquette, $a^2$, though it is the dimensionless version defined in Eq.~(\ref{eq:vortexdensity}) that will be needed later. Its evolution with increasing $N_f$ is presented in Fig.~\ref{fig:vortexdensity}.

As gleaned from the visualizations, the vortex density quickly rises as $N_f$ increases. By the largest $N_f$ considered, over $1/3$ of all plaquettes are pierced by a vortex. It is interesting to note that for $N_f \gtrsim 5$, the vortex density appears to increase approximately linearly with $N_f$. At this stage, it is unclear whether this is specific to the vortex density or if it will be a persistent feature across multiple measures.

Before proceeding, we mention that $\rho_\mathrm{vortex}$ can also be defined by making contact with the visualizations. For a given slice dimension $\mu$, one can define the average vortex density in the respective three-dimensional cross section as
\begin{equation}
	\rho_\mathrm{vortex}(\mu) = \frac{1}{N_\mu} \sum_\mathrm{slices} \frac{\text{Number of vortices in slice}}{3 \, N_\mathrm{slice}} \,,
\end{equation}
where $N_\mu$ is the number of lattice sites in dimension $\mu$, $N_\mathrm{slice}$ is the three-dimensional volume of the slice, and here $\binom{3}{2} = 3$ counts the number of plaquette orientations in the three-dimensional cross section. This can subsequently be averaged over all four slice dimensions to obtain another method of calculating $\rho_\mathrm{vortex}$,
\begin{equation}
	\rho_\mathrm{vortex} = \frac{1}{4} \sum_\mu \rho_\mathrm{vortex}(\mu) \,.
\end{equation}
Note that this is completely equivalent to the direct calculation of Eq.~(\ref{eq:vortexdensity}), as every plaquette has been accounted for exactly once in both cases. However, this procedure of decomposing the density into three-dimensional slices and averaging will be beneficial for our next quantity, which is inherently three-dimensional.

\subsubsection{Branching point density}
\begin{figure}
	\centering
	\includegraphics[width=0.49\linewidth]{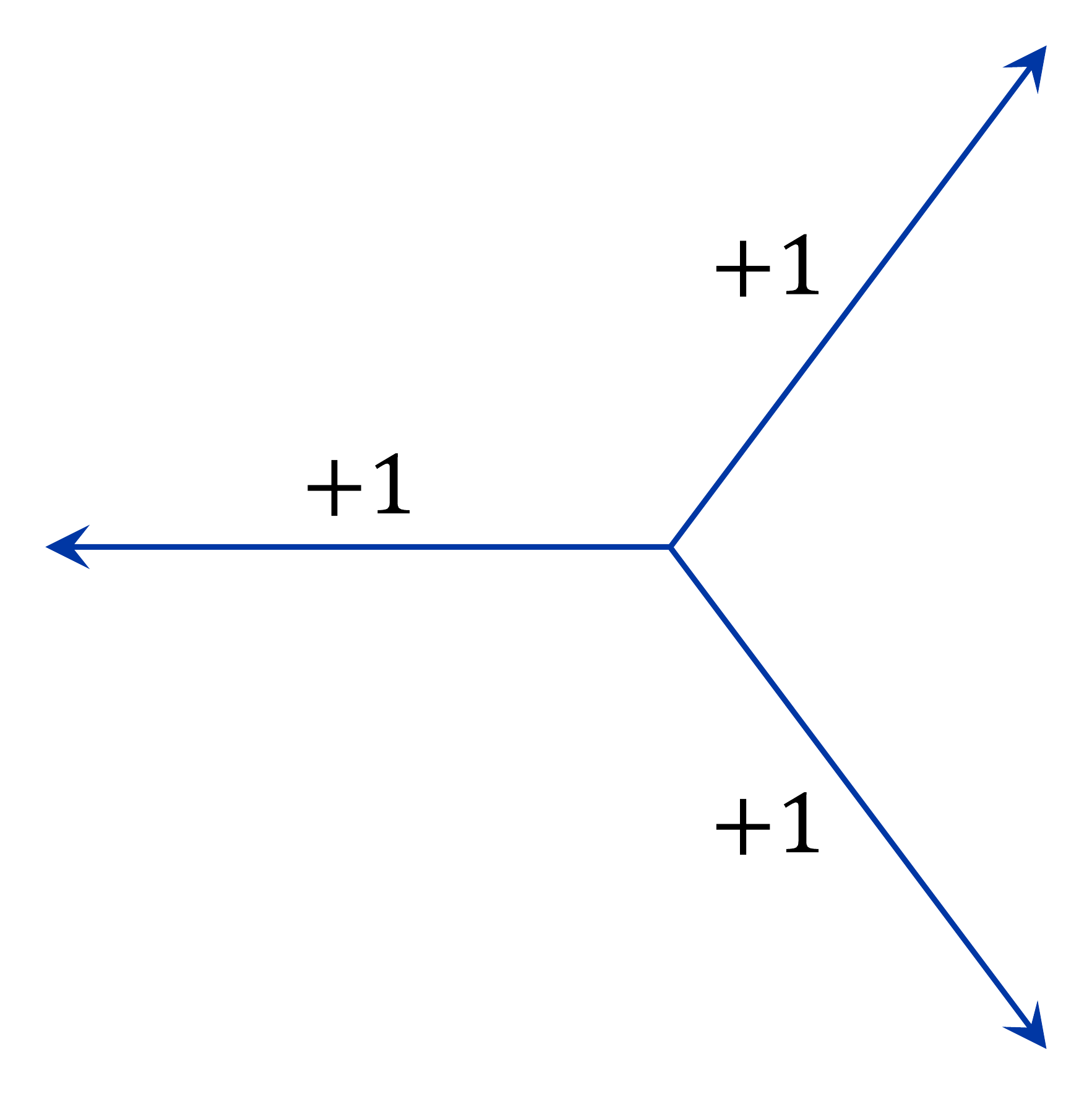}
	\hfill
	\includegraphics[width=0.49\linewidth]{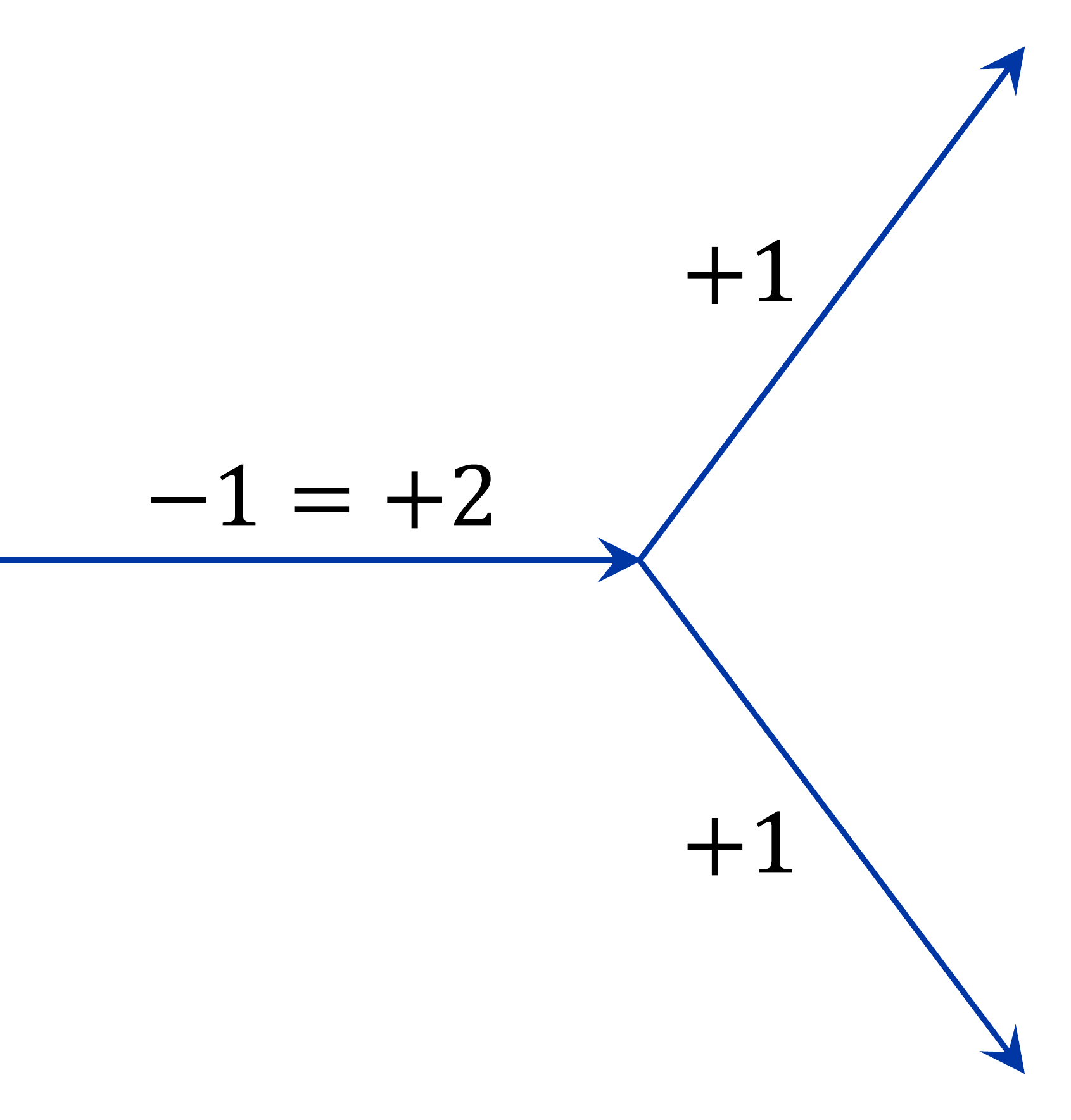}
	\caption{\label{fig:branching} Schematic of a monopole vertex (\textbf{left}) versus a branching point (\textbf{right}). The monopole follows our convention to show the directed flow of $m = +1$ center charge. Reversal of the left-hand arrow indicates the flow of $m = -1$ charge, as seen on the right. Due to periodicity in the center charge, $m = -1$ is equivalent to $m = +2$. Thus, the right-hand diagram depicts the branching of center charge.}
	
	\vspace{4em}
	
	\includegraphics[width=\linewidth]{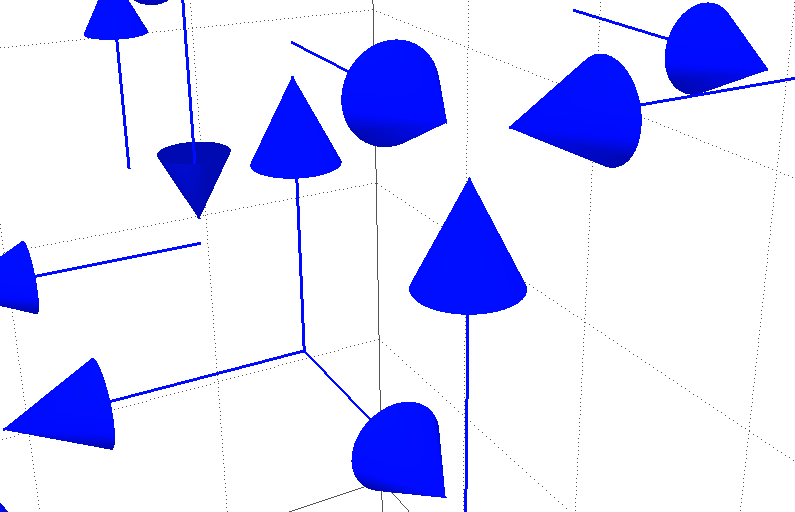}
	\caption{\label{fig:branching_example} An example of two branching points located near each other in a three-dimensional visualization. One cube contains three vortices converging to a point (monopole), while the other cube contains three vortices emerging from a point (anti-monopole).}
\end{figure}

Due to the existence of two distinct nontrivial center phases, $\mathrm{SU}(3)$ vortices are allowed to branch, in which an $m=\pm 1$ vortex splits into two $m=\mp 1$ vortices. This is allowed due to the conservation of center charge modulo $N$ in $\mathrm{SU}(N)$ gauge theory. Since reversing the orientation of a jet indicates the flow of the opposite center charge, branching points can equivalently be interpreted as monopoles in which three vortices of the same center charge emerge from, or converge to, a single point. This is how they manifest in the visualizations of Fig.~\ref{fig:nfvis}. The equivalence between branching and monopole points is illustrated in Fig.~\ref{fig:branching}. An example of branching points as they appear in the visualizations is further provided in Fig.~\ref{fig:branching_example}.

It is natural then to define a branching point density as the proportion of elementary cubes in a given three-dimensional slice that contain a branching point. This can thereafter be averaged, as described with the vortex density,
\begin{equation} \label{eq:branchingpointmu}
	\rho_\mathrm{branch}(\mu) = \frac{1}{N_\mu} \sum_\mathrm{slices} \frac{\text{Number of branchings in slice}}{N_\mathrm{slice}} \,,
\end{equation}
\begin{equation} \label{eq:branchingpoint}
	\rho_\mathrm{branch} = \frac{1}{4} \sum_\mu \rho_\mathrm{branch}(\mu) \,.
\end{equation}
Similarly to $\rho_\mathrm{vortex}$, $\rho_\mathrm{branch}$ can also be converted to a physical quantity, this time dividing by $a^3$ (as a volume density), though we leave it dimensionless.

\begin{figure}
	\centering
	\includegraphics[height=0.45\linewidth]{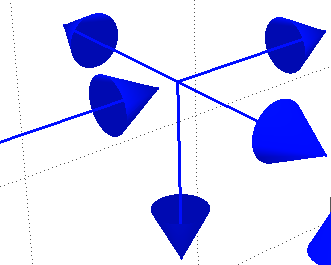}
	\hfill
	\includegraphics[height=0.45\linewidth]{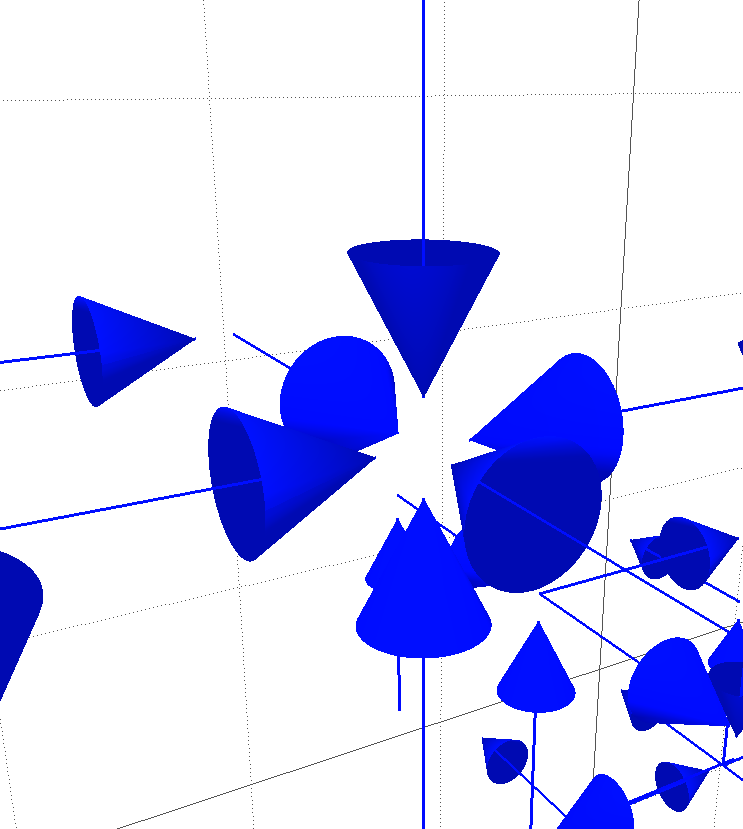}
	\caption{\label{fig:branching_examples} Examples of five- and six-way branching points. The former comprises a monopole and continuous vortex line that touches the same point, and the latter is a ``double monopole" in which six vortices converge to a single point.}
\end{figure}

We note that intersections of five and six vortices are also counted in the definition of the branching point density. Examples of how these manifest in the visualizations are provided in Fig.~\ref{fig:branching_examples}. These are comparatively rare relative to the standard three-way branching point. The case with five faces of a cube pierced can be unambiguously resolved as a branching point and a continuous vortex line that touches this same point. The instance of all six faces pierced shown in Fig.~\ref{fig:branching_examples} is a ``double monopole", with the six vortices converging to a point. Another possibility for a six-way branching point is a monopole plus anti-monopole, in which three of the vortices converge while the other three diverge. This last possibility possesses some ambiguity, as it could equivalently be interpreted as three continuous vortex lines that intersect at a common point. Nevertheless, we still choose to count this as a branching point for the purposes of defining $\rho_\mathrm{branch}$.

\begin{figure}
	\centering
	\includegraphics[width=\linewidth]{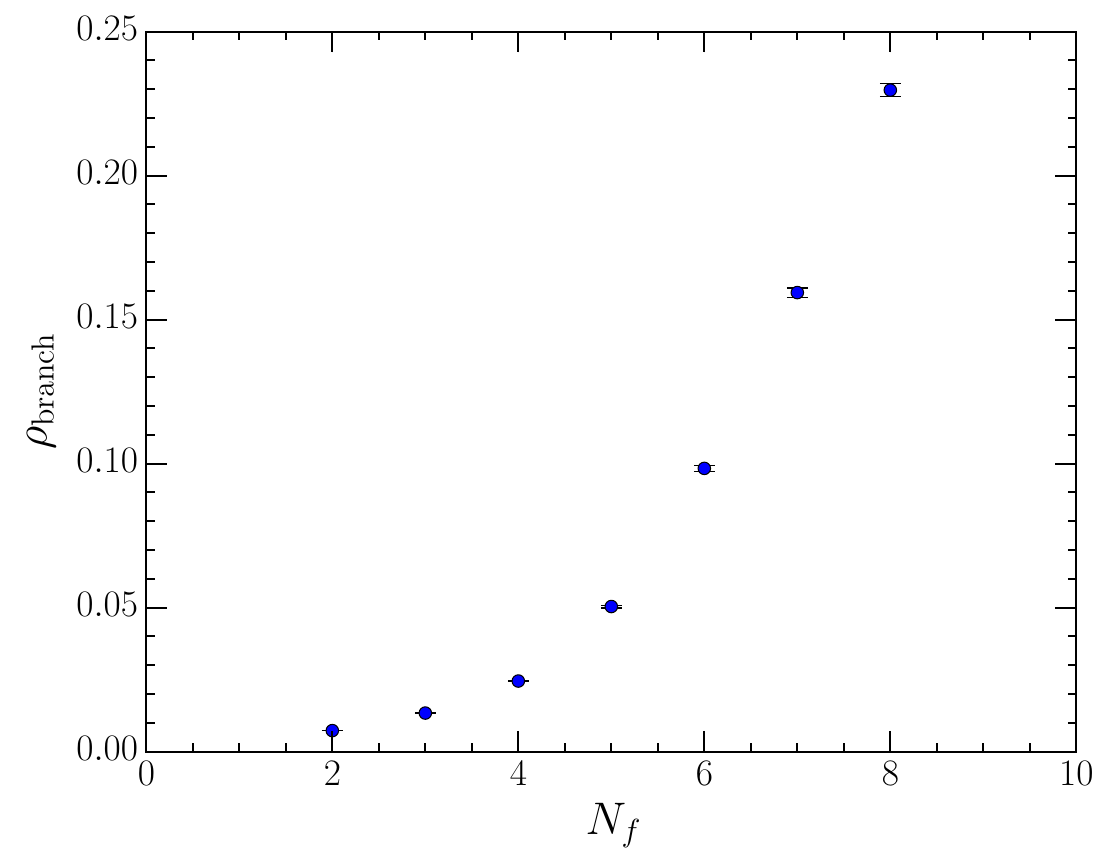}
	
	\vspace{-0.5em}
	
	\caption{\label{fig:branchingpointdensity} The dimensionless branching point density defined across Eqs.~(\ref{eq:branchingpointmu}) and (\ref{eq:branchingpoint}). The relative increase is even greater than that experienced by the vortex density in Fig.~\ref{fig:vortexdensity}, and there again appears to be a stable trend for $N_f \gtrsim 5$.}
\end{figure}

With the semantics of the branching point density established, its $N_f$ dependence is displayed in Fig.~\ref{fig:branchingpointdensity}.
Its qualitative behavior is very similar to the vortex density, with the amount of branching points rapidly growing with $N_f$. Once again, it seems that $\rho_\mathrm{branch}$ is roughly linear for moderate $N_f \gtrsim 5$, though there does appear to be a gentle curve in Fig.~\ref{fig:branchingpointdensity} when compared against Fig.~\ref{fig:vortexdensity}. Still, this continues to raise the possibility of something interesting occurring around $N_f \approx 5$. This will be verified and quantified in greater detail by our next measure of vortex geometry.

\subsubsection{Vortex structure correlation}
So far, we have only considered describing the amount of vortex matter, though it is also of interest to quantify the shape of the two-dimensional vortex sheet that exists in four dimensions. This is achieved by defining a correlation of the vortex structure along each of the four dimensions, ascribing a number to how similar the structure is across the corresponding three-dimensional slices. For this we follow Refs.~\cite{Mickley:2024zyg, Mickley:2024vkm}, where the correlation was introduced to measure an alignment of the vortex sheet with the temporal dimension that exists at very high temperatures. There, large correlation values along the temporal dimension indicate a high probability that a vortex remains invariant between consecutive temporal slices, representing the alignment.

For our purposes, the intuitive picture of increased roughness in the vortex field as $N_f$ rises would imply a smaller correlation, as there would be lots of small ``wiggles" in the vortex sheet that cause the three-dimensional structure to differ between slices. Therefore, it is interesting to ascertain the extent to which this develops as the number of flavors is varied.

For simplicity, suppose we are taking temporal slices of the lattice (as done to produce the visualizations in Fig.~\ref{fig:nfvis}). To quantify the similarity of the vortex structure between the various temporal slices, an indicator function is defined,
\begin{equation} \label{eq:indicator}
	\chi_{ij}(\mathbf{x}, t; \Delta t) = \begin{cases}
		1, & m_{ij}(\mathbf{x}, t) \, m_{ij}(\mathbf{x}, t+\Delta t) > 0 \\
		0, & \text{otherwise} \\
	\end{cases} ,
\end{equation}
where $m_{ij}(\mathbf{x},t)$ is the center charge of the plaquette $P_{ij}(\mathbf{x},t)$ as defined in Eq.~(\ref{eq:centerprojplaq}), and $i,j = 1,2,3$ enumerate the three spatial dimensions (as we are holding the Euclidean time coordinate fixed). It takes the value $1$ if a nontrivial plaquette at spatial position $\mathbf{x}$ in temporal slice $t$ is also pierced by a vortex in the same direction in the later temporal slice $t+\Delta t$. This naturally leads to the correlation
\begin{equation} \label{eq:correlation}
	C(\Delta t) = \frac{1}{N_\mathrm{vor} \, N_t} \,\sum_{\substack{\mathbf{x},\,t,\\i,\,j}} \chi_{ij}(\mathbf{x}, t; \Delta t) \,,
\end{equation}
where $N_\mathrm{vor}$ is the average number of vortices per temporal slice. The normalization factor ensures $C(\Delta t) \in [0,1]$, such that it gives the probability that a pierced plaquette remains invariant between slices $t$ and $t+\Delta t$. This is what we sought to quantify.

\begin{figure}
	\centering
	\includegraphics[width=\linewidth]{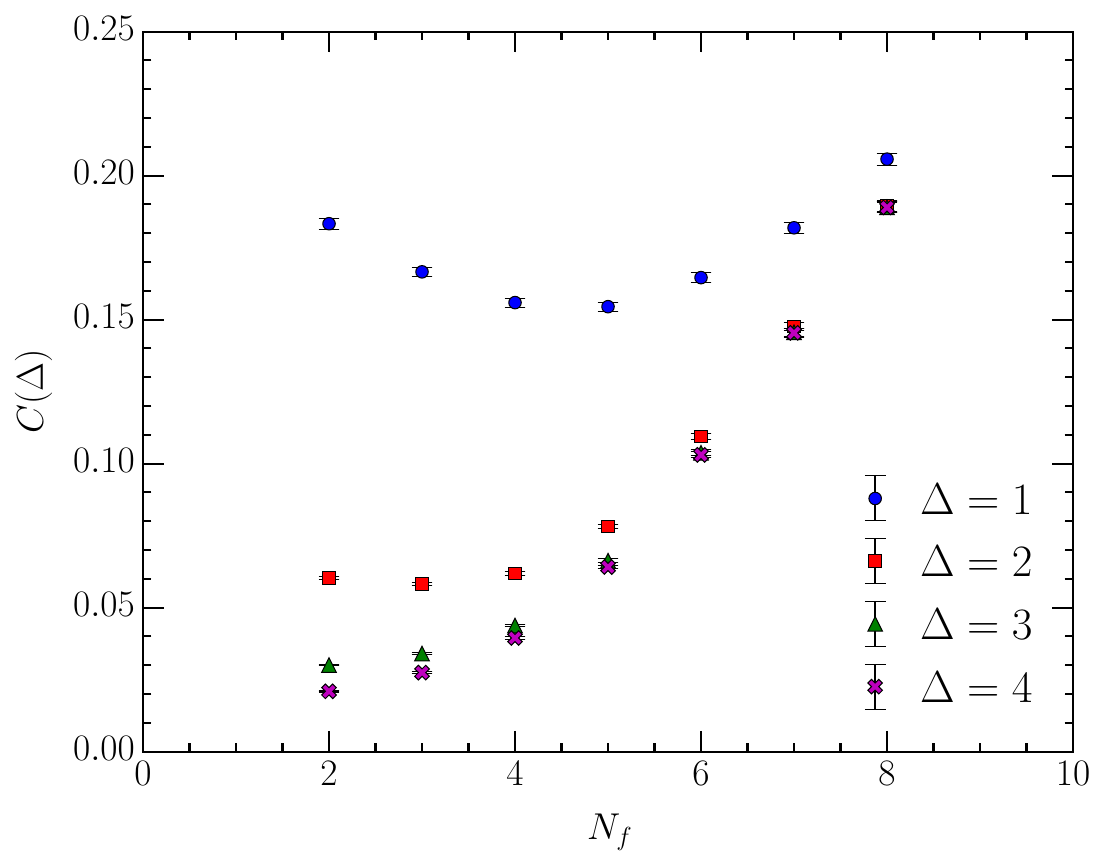}
	
	\vspace{-0.5em}
	
	\caption{\label{fig:correlation} The vortex structure correlation defined throughout Eqs.~(\ref{eq:indicator})--(\ref{eq:averaged_correlation}) for $\Delta \in \{1,2,3,4\}$ as a function of $N_f$. For consecutive slices ($\Delta = 1$), the correlation initially decreases in line with the ``roughness" argument, though features a turning point around $N_f \simeq 4$--$5$ past which it increases. In addition, $C(\Delta)$ decays rapidly as a function of $\Delta$, to the point that for $N_f = 8$ its values for all $\Delta \geq 2$ are indistinguishable from each other.}
\end{figure}

Of course, the above procedure could be repeated for any of the four dimensions to obtain a $C(\Delta x_\mu)$ for each dimension $\mu$. As with the vortex and branching point densities, this can finally be averaged over the four dimensions for common values of the separations $\Delta x_\mu$,
\begin{equation} \label{eq:averaged_correlation}
	C(\Delta) = \frac{1}{4} \sum_\mu C(\Delta x_\mu) \,.
\end{equation}
Due to periodic boundary conditions, the separation $\Delta$ is effectively restricted to $\Delta \leq \lfloor N_\mathrm{min}/2 \rfloor$, where $N_\mathrm{min}$ is the smallest number of lattice sites in any one dimension. In our case, this is $\Delta \leq 12$. Notwithstanding, for reasons that will become apparent, it is sufficient to compare only $\Delta = 1$--$4$ between each $N_f$. This evolution is presented in Fig.~\ref{fig:correlation}.

We start by observing the correlation on consecutive slices $C(\Delta = 1)$. Here, it initially experiences a decrease as hypothesized from the developing roughness of the vortex matter. With small but frequent protrusions in the vortex surface, there will be additional misses in the indicator function of Eq.~(\ref{eq:indicator}). However, it then encounters a turning point between $N_f \simeq 4$--$5$, beyond which the correlation instead increases.

There is a logical explanation for this shift in behavior. With the proportion of plaquettes pierced continuously growing, as seen from the vortex density, there is a greater chance that two plaquettes selected at random happen to possess the same center charge. For instance, at $N_f = 8$ upwards of $1/3$ of all plaquettes are pierced. If this is treated as a probability that a plaquette is pierced, there is already a considerable chance that two random plaquettes are pierced in the same direction (assuming independence). The turning point in $C(\Delta = 1)$ seen in Fig.~\ref{fig:correlation} identifies the point at which this becomes the dominant effect. That is, it marks the transition from increased roughness being the primary contribution, to \textit{randomness} in the vortex field underlying the $N_f$ trends.

It is curious to note that the point at which this transition occurs coincides with the onset of stable, approximately linear trends for the vortex and branching point densities. The correlation measure has now explicitly singled this out as an important point in the vortex evolution with $N_f$.

Next, we analyse $C(\Delta)$ for separations $\Delta \geq 2$. The decrease in value at low $N_f$ is significantly weaker than for $\Delta = 1$, and there is no decrease at all for $\Delta \geq 3$. This is unsurprising. Any ``wiggles" in the vortex sheet are just that---small-scale fluctuations in the surface that are only relevant at short distances. It is intuitive that these would have a considerable impact for three-dimensional slices only one step apart, but would become insignificant as the slices in question are moved farther apart. The crucial finding is the turning point in $C(\Delta = 1)$, signifying that these fluctuations are overwhelmed by the growing randomness in the vortex content.

$C(\Delta)$ can also be treated as a function of $\Delta$ for fixed $N_f$. Here, we find it rapidly decays to plateau at a nonzero value. This represents some random chance that two plaquettes will possess the same center charge, regardless of how far apart they are. The values for $\Delta = 3$ and $\Delta = 4$ are already very close to each other at all $N_f$ shown in Fig.~\ref{fig:correlation}, justifying why displaying any $\Delta > 4$ would be superfluous. The distance between the $C(\Delta)$ for the various $\Delta$ further decreases as $N_f$ increases. Indeed, by $N_f = 8$ the values of $C(\Delta)$ for all $\Delta \geq 2$ lie completely on top of each other. It seems reasonable to propose that $C(\Delta)$ for $\Delta = 1$ and $\Delta \geq 2$ will eventually coincide with each other at some higher value of $N_f$, given that the increase in the latter is steeper than the former. This would signal an important point in the evolution of center vortex geometry with the number of fermion flavors. The value of $C(\Delta)$ at this point, and the corresponding value of $N_f$, is the subject of the following subsections.

\subsection{Haar measure} \label{subsec:haar}
With the above findings on center vortex structure, it is natural to question whether the various geometrical aspects of center vortices possess some ``upper limit" that the vortex content is approaching, but cannot exceed. With our understanding of increased roughness and randomness in the vortex field, we propose that this limit is given by a uniform gauge field in which all link variables are sampled independently from the Haar measure on $\mathrm{SU}(3)$. This would in some sense represent the ``roughest" possible vortex field, with a completely random chance that any given plaquette is pierced by a vortex.

\begin{figure}
	\centering
	\includegraphics[width=\linewidth]{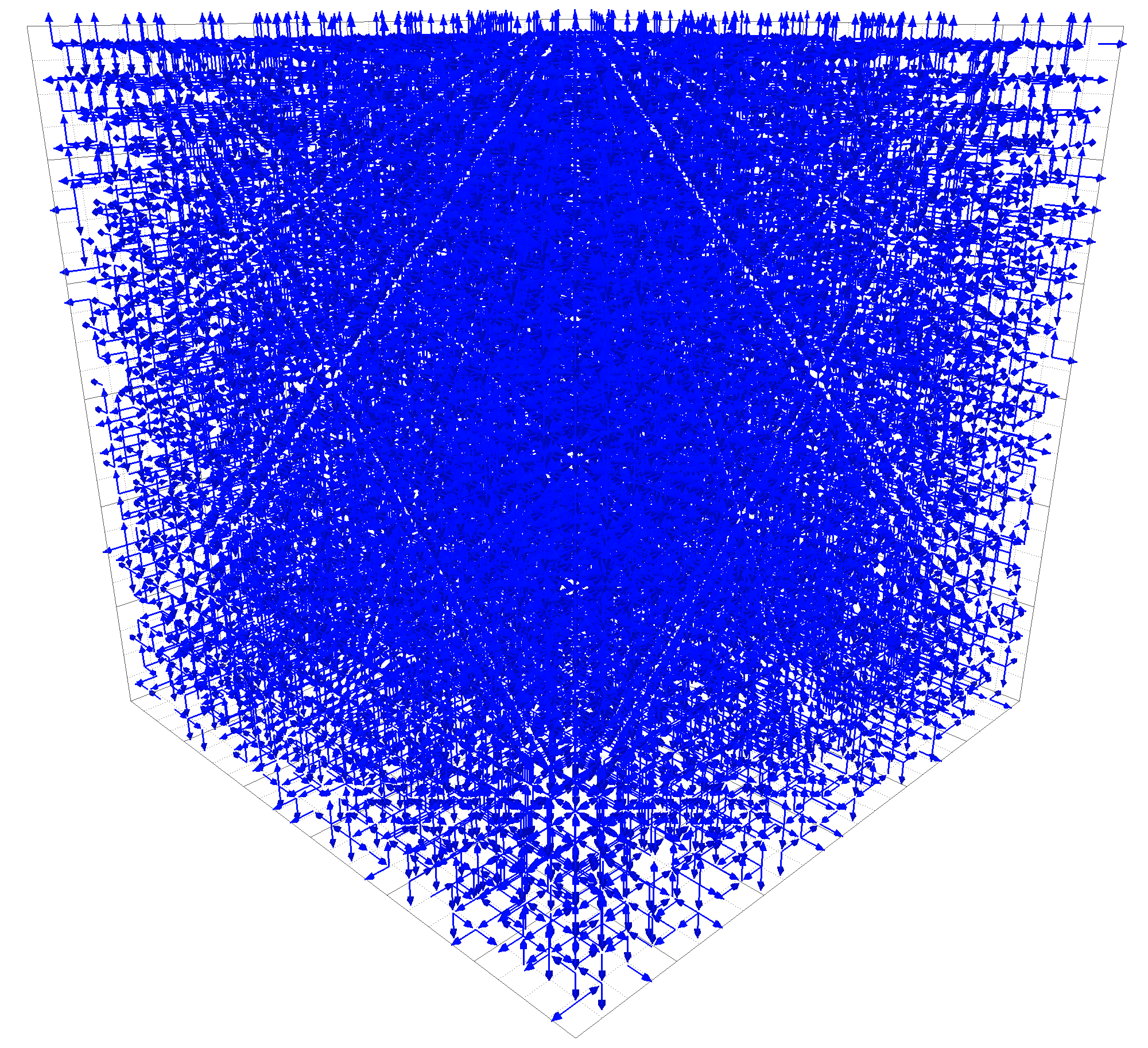}
	\caption{\label{fig:haarvis} An example visualization of the center vortex structure on a Haar-random gauge field, obtained with the standard vortex-identification procedure. One giant percolating cluster fills the entire volume, with approximately $2/3$ of all plaquettes pierced. There are no secondary clusters.}
\end{figure}

To test this hypothesis, we generate 100 independent uniformly random gauge fields, and proceed with the usual vortex identification procedure of fixing to maximal center gauge and center projection. A three-dimensional visualization of the resulting structure is presented in Fig.~\ref{fig:haarvis}, and the statistical quantities from Sec.~\ref{subsec:statistics} evaluated on this ensemble are provided in Table~\ref{tab:statistics}.

Turning first to the visualization, we uncover that the three-dimensional volume is essentially a wall of vortices, consisting of a single enormous percolating cluster that takes up nearly the entire available space. This is qualitatively in accordance with the visualizations at increasing $N_f$ in Fig.~\ref{fig:nfvis}, which saw the percolating cluster grow in density with a suppression of secondary clusters. In fact, fewer than $0.5\%$ of all slices from our Haar-random ensemble contain any secondary clusters at all, and no one slice contained more than a single secondary cluster. It is reassuring that the bulk Haar-random vortex content is a natural extension of that found at the upper end of our $N_f$ range.

Moving to the quantitative results in Table~\ref{tab:statistics}, we find that the vortex density $\rho_\mathrm{vortex}$ and correlation on consecutive slices $C(\Delta = 1)$ are consistent with $2/3$ and $1/3$ within statistical uncertainty, respectively. In fact, $C(\Delta)$ is consistent with $1/3$ for all $\Delta$, i.e. it is constant as a function of $\Delta$. It was clear from Fig.~\ref{fig:correlation} that the $C(\Delta)$ were converging to a common value as $N_f$ increases, and hence such a value is provided by a Haar-random gauge field.

Both of these values can be intuitively understood. Given that the links are drawn from the uniform distribution on $\mathrm{SU}(3)$, the three center phases will be equally populated. As such, one would naively assign a probability of $2/3$ for a plaquette to be pierced by a vortex (as there are two nontrivial center elements). This automatically translates to a vortex density of $\rho_\mathrm{vortex} = 2/3$. Similarly, there is a random $1/3$ chance that any two plaquettes share the same center charge, leading directly to a value of $C(\Delta) = 1/3$ irrespective of the separation $\Delta$. That said, it is not immediately clear that these probabilities should survive the vortex-identification procedure, wherein the gauge is rotated to MCG and the links subsequently projected onto the center. Thus, the fact that these naive probabilities do still produce the correct prediction is nonetheless a nontrivial result.

The case of the branching point density is more interesting. Its Haar-random value falls suspiciously close to $1/2$, though is certainly not consistent with $1/2$ within statistical error. This raises the question as to whether its exact value can be predicted in a similar fashion to the vortex density and correlation measure based on naive probabilities. For this, we need to count the proportion of vortex arrangements within an elementary cube that constitute a branching point. At first, one might assume there are $3^6 = 729$ \textit{total} possible arrangements, with three choices for each of the six faces of a cube. However, as the vortex flux is trivially conserved, the sixth face is uniquely determined from the other five to ensure the flux sums to zero modulo 3. Therefore, there are in fact only $3^5 = 243$ unique arrangements.

Now, we need to count the number of these that are three-, five- and six-way branching points.
\begin{itemize}
	\item Three-way: there are $\binom{6}{3} = 20$ ways to arrange the three vortices, which is then doubled to account for the choice in orientation (either converging to or emerging from the point). That is, there are $40$ distinct three-way branching points.
	\item Five-way: the starting point is $\binom{6}{3} \times \binom{3}{2} \times 4 = 240$, which in turn accounts for the possible arrangements of the branching point, the continuous vortex line, and their orientations. However, there is multiple counting at play. Referring to the five-way branching example in Fig.~\ref{fig:branching_examples}, any one of the four outgoing vortices could be attributed to the continuous vortex line. This implies we have over-counted by a factor of four, and the actual number of potential five-way branching points is $60$.
	\item Six-way: for the monopole plus anti-monopole case, there are again $\binom{6}{3} = 20$ arrangements. This should not be doubled, as the reverse orientation has already been accounted for by the presence of the anti-monopole. Finally, there are two extra cases in which all six vortices converge to or emerge from the point, totalling $22$ six-way branching points.
\end{itemize}
Accumulating these counts amounts to a probability of $122/243 \approx 0.50206$ for an elementary cube to contain a branching point. This is remarkably consistent with the numerical Haar-random value for $\rho_\mathrm{branch}$ in Table~\ref{tab:statistics} of $0.50208(4)$.

With these exact values established, we now return to the matter at hand. The idea is that the Haar-random properties provide an upper bound for vortex behavior in the confining broken chiral symmetry phase. It is clear that all quantities under consideration are quickly approaching these values as the number of fermion flavors increases, and we propose that they cannot be exceeded in this phase. An immediate consequence of this conjecture is that the point at with the Haar values are reached should be an important point in the evolution of vacuum field structure as a function of $N_f$, signalling a sudden change to vortex behavior. In the next subsection, we seek to provide an estimation of this critical point.

\subsection{Extrapolations} \label{subsec:fits}
We now look to perform fits to the various vortex statistics as a function of $N_f$, allowing an extrapolation to the critical point $N_f^*$ at which the Haar-random properties might be realized. As elucidated by the behavior of the correlation on consecutive slices $C(\Delta = 1)$, these fits should only be performed over the range $N_f \geq 5$. Recall that the turning point in $C(\Delta = 1)$ marks the point at which randomness in the vortex field becomes the dominant feature, effecting an increase in the correlation that represents a drive towards the Haar-random limit. As such, we have clear motivation for considering these ``restricted" fits.

As mentioned prior, the vortex and branching point densities also appear to have settled into stable trends for $N_f \gtrsim 5$, and we believe the consistency of this with the turning point in $C(\Delta = 1)$ is no coincidence. Thereby, it is sufficient to consider simple two-parameter ans{\"a}tze in each case. For the vortex density $\rho_\mathrm{vortex}$, the evolution over $N_f \geq 5$ is captured by a linear ansatz,
\begin{equation} \label{eq:linear}
	\rho_\mathrm{vortex}(N_f) = aN_f + b \,.
\end{equation}
The branching point density $\rho_\mathrm{branch}$ has a gentle curve that suggests a two-parameter quadratic,
\begin{equation} \label{eq:quadratic}
	\rho_\mathrm{branch}(N_f) = aN_f^2 + b \,.
\end{equation}
Finally, one might be tempted to also fit Eq.~(\ref{eq:quadratic}) to the correlation on consecutive slices $C(\Delta = 1)$. However, due to the horizontal offset of the turning point a three-parameter quadratic would have to be employed. We find that the $\chi^2/\mathrm{dof}$ value from utilizing such a three-parameter quadratic is $\sim\order{10^{-3}}$, suggesting that the data has been over-fitted. Therefore, in place of introducing an extra parameter to the fit, we instead consider a basic two-parameter cubic,
\begin{equation} \label{eq:cubic}
	C_{\Delta=1}(N_f) = aN_f^3 + b \,,
\end{equation}
which we will see is sufficient for describing the data and produces a more reasonable reduced $\chi^2$ statistic.

\begin{figure}
	\centering
	\includegraphics[width=\linewidth]{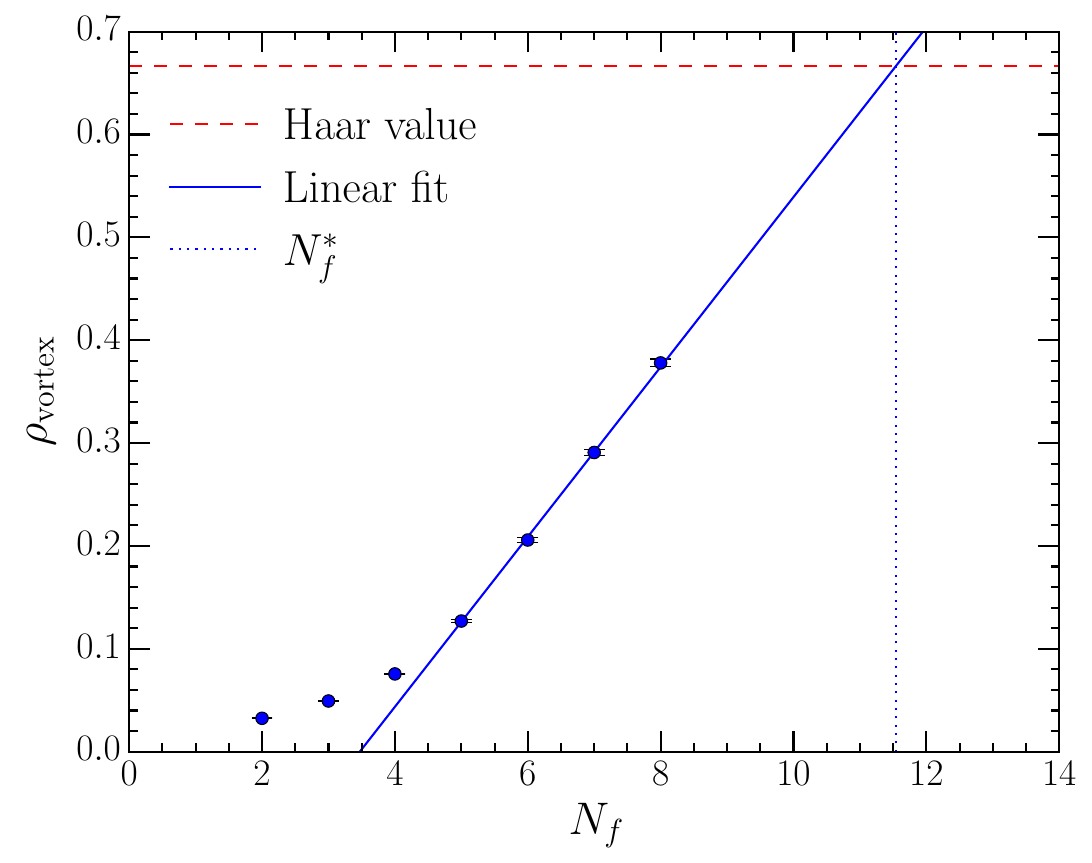}
	
	\vspace{-0.5em}
	
	\caption{\label{fig:vortexdensityextrapolation} The linear ansatz of Eq.~(\ref{eq:linear}) fit to the vortex density for $N_f \geq 5$. It is seen to accurately describe the data over the fitted range. The critical value $N_f^*$ at which the extrapolation reaches the Haar-random value of $2/3$ is also indicated, landing near $N_f \simeq 11.5$.}
\end{figure}

\begin{figure}
	\centering
	\includegraphics[width=\linewidth]{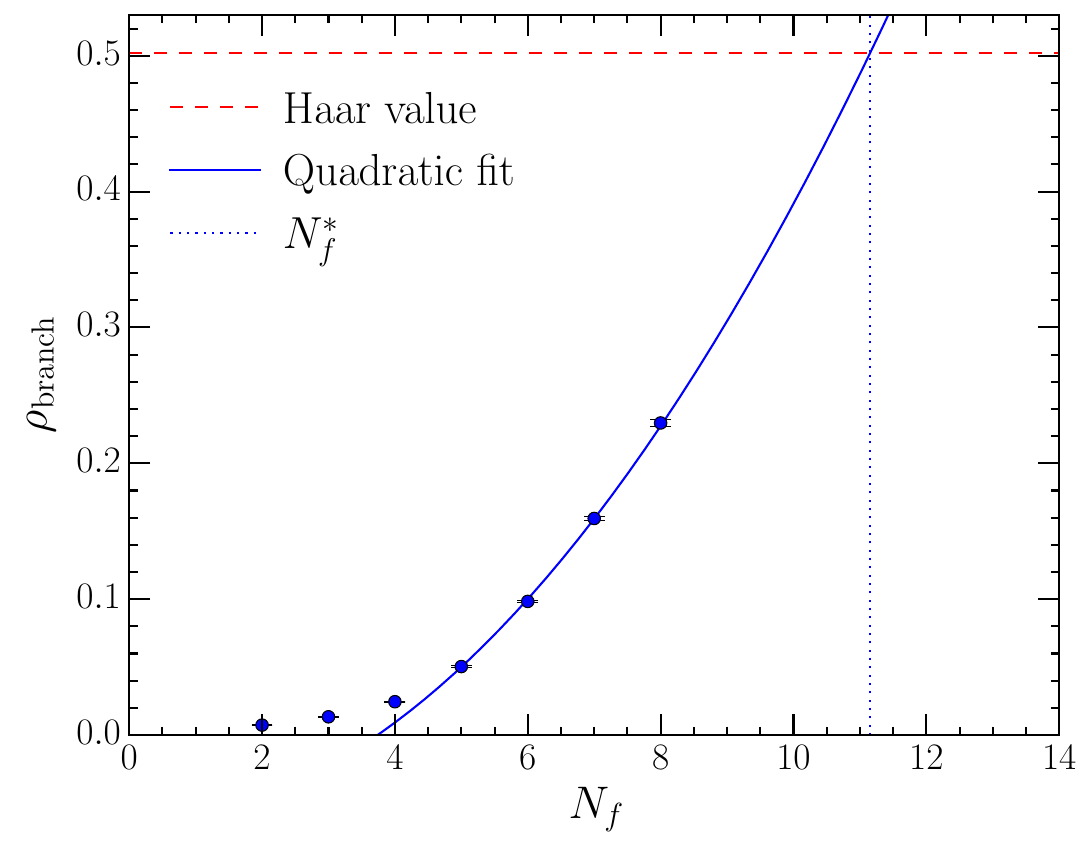}
	
	\vspace{-0.5em}
	
	\caption{\label{fig:branchingpointdensityextrapolation} The two-parameter quadratic ansatz of Eq.~(\ref{eq:quadratic}) fit to the branching point density for $N_f \geq 5$. It is seen to accurately describe the data over the fitted range. The critical value $N_f^*$ at which the extrapolation reaches the Haar-random value of $122/243$ is also indicated, landing just above $N_f \simeq 11$.}
\end{figure}

\begin{figure}
	\centering
	\includegraphics[width=\linewidth]{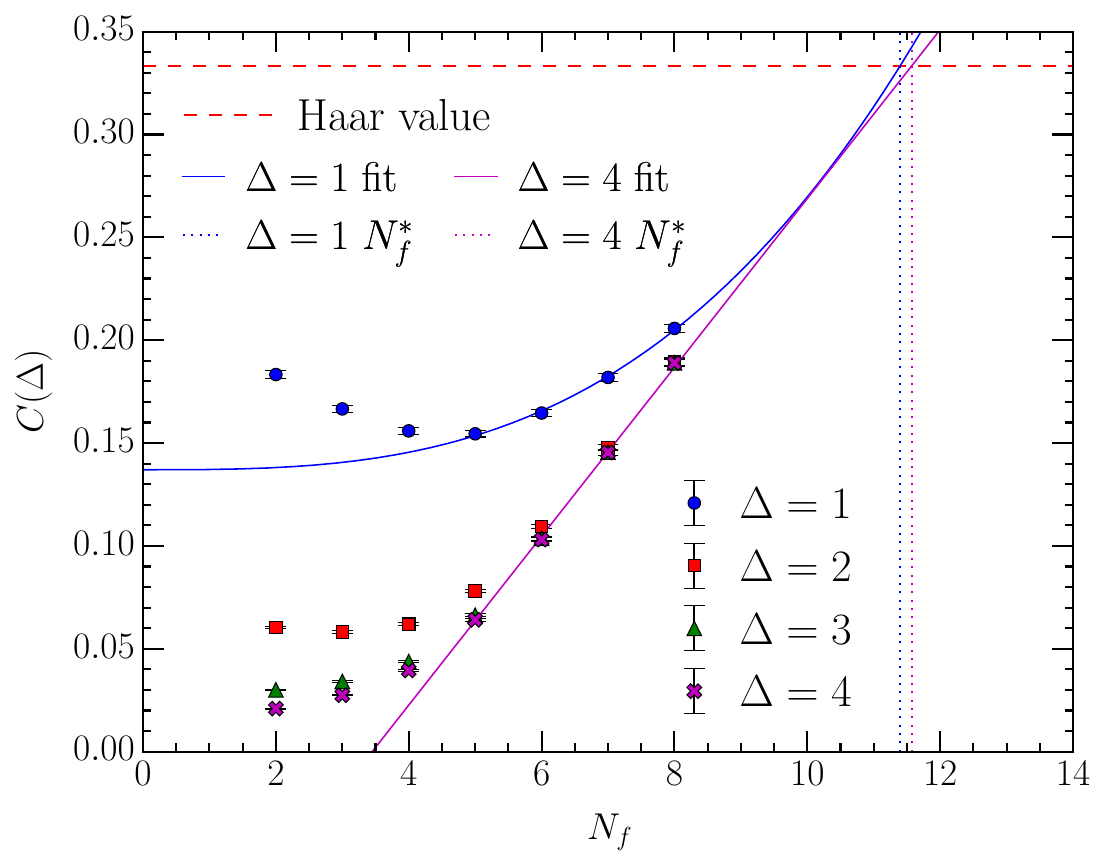}
	
	\vspace{-0.5em}
	
	\caption{\label{fig:correlationextrapolation} The two-parameter cubic ansatz of Eq.~(\ref{eq:cubic}) fit to the correlation $C(\Delta)$ on consecutive slices $C(\Delta = 1)$ for $N_f \geq 5$. The ansatz is seen to accurately describe the data over the fitted range. The critical value $N_f^*$ at which the extrapolation reaches the Haar-random value of $1/3$ is also indicated, landing near $N_f \simeq 11.5$. An example linear fit for $\Delta = 4$ is also given, and intersects the Haar limit within the vicinity of $\Delta = 1$ as required.}
\end{figure}

\begin{table}
	\centering
	\caption{\label{tab:fits} Details of the extrapolations performed for each vortex quantity, including the ansatz, the critical value $N_f^*$ at which the Haar-random value is attained, the fit parameters and the reduced $\chi^2$ statistic. The values associated with $C(\Delta\geq 2)$ are obtained by averaging the results for $C(\Delta)$ over all $\Delta \geq 2$; the errors are a combination of the statistical uncertainty and the spread of results for the various $\Delta$. The ans{\"a}tze are found to describe the data well over the fitted range $N_f\geq 5$, with $\chi^2/\mathrm{dof}$ values $\sim\order{1}$. Additionally, the various estimates of $N_f^*$ predominantly agree with each within statistical uncertainty and consistently fall between $N_f = 11$--$12$. The one exception is the branching-point-density estimate, which is slightly lower than the others though still within the same range.}
	\begin{ruledtabular}
		\begin{tabular}{ccD{.}{.}{2.6}cD{.}{.}{2.7}D{.}{.}{1.5}}
			Quantity & Ansatz &  \multicolumn{1}{c}{$N_f^*$} & \multicolumn{1}{c}{$a$} &  \multicolumn{1}{c}{$b$} & \multicolumn{1}{c}{$\!\!\!\chi^2/\mathrm{dof}$} \\
			\colrule \\[-0.9em]
			$\rho_\mathrm{vortex}$ & $ax + b$     & 11.54(17) & 0.0826(11)\, & -0.287(6)  & 1.846 \\
			$\rho_\mathrm{branch}$ & $ax^2 \!+ b$ & 11.15(6)  & 0.00455(5)\, & -0.064(1)  & 2.245 \\
			$C(\Delta \!=\! 1)$    & $ax^3 \!+ b$ & 11.40(18) & 0.00013(1)\, &  0.137(2)  & 0.519 \\
                        $C(\Delta \!\geq\! 2)$ & $ax + b$     & 11.61(24) & 0.0407(16)\, & -0.139(12) & 1.89(4) \\[-0.2em]
		\end{tabular}
	\end{ruledtabular}
\end{table}
Furthermore, it is prescient to consider extrapolations of $C(\Delta)$ for all $\Delta$, given that it is predicted they should reach the Haar-random value of $1/3$ at the same critical value $N_f^*$. As can be seen from Fig.~\ref{fig:correlation}, a linear ansatz is adequate for describing $C(\Delta \geq 2)$ over $N_f \geq 5$. The details of these fits are provided in Table~\ref{tab:fits}, and the fits themselves are displayed throughout Figs.~\ref{fig:vortexdensityextrapolation}--\ref{fig:correlationextrapolation}. As it is infeasible to show the fits for all $\Delta$, we choose to present $\Delta = 4$ (in addition to $\Delta = 1$) as an example; this was the largest $\Delta$ considered earlier in Fig.~\ref{fig:correlation}.

Looking at the fits, it can be seen that they accurately describe their respective data over the fitted region $N_f \geq 5$, passing through the majority of points within statistical uncertainty. Their $\chi^2/\mathrm{dof}$ values are provided in Table~\ref{tab:fits} and are all $\sim\order{1}$, indicating reasonable fits within the bounds of statistical errors. The fact that the fits drop below zero at low $N_f$ for the vortex and branching point densities is unproblematic given their restricted range.

The extrapolated estimates of the critical value $N_f^*$ at which the fits intersect their Haar-random lines are also indicated throughout Figs.~\ref{fig:vortexdensityextrapolation}--\ref{fig:correlationextrapolation}, accompanied by their precise values in Table~\ref{tab:fits}. We find that they consistently fall between $N_f = 11$--$12$, tending to be near $N_f \simeq 11.5$. The various estimates are compatible with each other within statistical uncertainty, barring the estimate from $\rho_\mathrm{branch}$ which lies slightly lower but still above $N_f = 11$. This consistency extends to the independent estimates from each individual choice of $\Delta$ in the correlation measure, which are averaged over in Table~\ref{tab:fits}. This is a requisite for the Haar-random interpretation to be valid, with each $C(\Delta)$ reaching $1/3$ at the same critical point. The example $\Delta = 4$ extrapolation in Fig.~\ref{fig:correlationextrapolation} is in very close proximity to the $\Delta = 1$ estimate, and they agree within uncertainty. Given that for $N_f \geq 5$, $C(\Delta)$ is already effectively constant as a function of $\Delta$ for $\Delta \geq 3$, it is clear that this holds true for all larger separations as well.

Taking an average of the individual estimates for $N_f^*$ in Table~\ref{tab:fits} and calculating a systematic error through the standard deviation of their distribution leads to our final estimate,
\begin{equation} \label{eq:Nfstar_estimate}
	N_f^* = 11.43(16)(17) \,,
\end{equation}
where the errors are statistical and systematic, respectively. It lies entirely between $N_f = 11$--$12$, even accounting for the worst-case scenario with the errors. The above estimate should of course be understood as the result along our line of constant physics, with fixed lattice spacing and pion mass, only. The approach to the continuum and chiral limits is of great importance and is left for future work.

Tying this back to the phase structure of gauge theories as a function of number of fermion flavors $N_f$, we interpret this result as an estimate for the lower bound of the conformal window. With the expectation that a Haar-random center vortex field cannot be exceeded, the critical point at which this is realized necessarily signals a sudden change in vacuum field structure. Although it is currently unclear exactly how this change will manifest, we can note that our final estimate for $N_f^*$ in Eq.~(\ref{eq:Nfstar_estimate}) is certainly consistent with various other methods for extracting the conformal window's lower end, which place it between $N_f^* = 8$--$13$ \cite{Appelquist:1988yc, Cohen:1988sq, Sannino:2004qp, Dietrich:2006cm, Armoni:2009jn, Braun:2009ns, Frandsen:2010ej, Ryttov:2016ner, Kim:2020yvr, Lee:2020ihn}. It is also worthwhile to mention that the proximity of our result to $N_f = 12$ could explain the inconclusive nature of this system in previous studies. The most recent of these investigations into the infrared conformality of the $N_f = 12$ theory places it within the conformal window \cite{Hasenfratz:2024fad}. This is in agreement with our result, which suggests the transition takes place at around twelve flavors.

\section{Conclusion} \label{sec:conclusion}
In this work, we have studied the center vortex structure of QCD ground-state fields as a function of the number of fermion flavors $N_f$ as a means to study the $\mathrm{SU}(3)$ conformal window. Through the consideration of an upper bound for vortex behavior, we have extracted an estimate of the conformal window's lower end as corresponding to when the vortex content of a uniform-random gauge field is attained.

We utilize ensembles covering $N_f = 2$--$8$, with a line of constant physics implemented by fixing $a m_\pi \simeq 0.248$ and $a f_\pi \simeq 0.0568$. Visualizations of the center vortex structure at increasing $N_f$ revealed a ``roughness" that permeated the structure at high $N_f$. This manifested specifically as an increase in density of the percolating cluster, with a substantial suppression of secondary clusters approaching the upper end of our $N_f$ range. The growing roughness was linked intuitively to the decrease in $\beta$ value necessary to maintain the line of constant physics.

To quantify the evolution of vortex structure as a function of $N_f$, a collection of vortex statistics was introduced, starting with the vortex and branching point densities. These revealed a rapid rise in value with $N_f$, matching the qualitative trends captured in the visualizations. A third quantity, a correlation of the vortex structure, was introduced to describe the shape of the two-dimensional vortex sheet that exists in four dimensions. The correlation measure exhibited interesting behavior, initially decreasing for $N_f \lesssim 5$ before experiencing a turning point beyond which it instead increased. This reflects a growing randomness in the vortex field, with the significant amount of vortex matter resulting in a greater random chance that any two plaquettes happen to possess the same center charge. The turning point marks the place at which this becomes the dominant effect. As a consequence, the value of the correlation for various separations $\Delta$ also became closer together, seemingly approaching a common value.

These findings naturally raised the question of whether there exists an upper limit to center vortex behavior. With the observation of growing randomness for $N_f \gtrsim 5$, the vortex content of a Haar-random gauge field forms such an appropriate bound and was found to feature many desirable properties. These include a single percolating cluster that fills the volume with an absence of secondary clusters, a constant value of the correlation measure $C(\Delta)$ as a function of $\Delta$, and statistics that can be justified by simple counting arguments assuming each center phase in the links is equally likely. These include a vortex density consistent with $2/3$ (the probability of a plaquette having a nontrivial center charge), a branching point density consistent with $122/243$ (the proportion of vortex arrangements in an elementary cube that constitute a branching point), and a correlation $C(\Delta)$ consistent with $1/3$ for all $\Delta$ (the probability that two random plaquettes have the same center charge).

With this understanding, fits were performed to the various quantities for $N_f \geq 5$, over which they are well described by various two-parameter ans{\"a}tze. By extrapolating outwards, estimates for the proposed critical point at which the Haar-random vortex content is attained were extracted. These wholly fell between $N_f = 11$--$12$,\linebreak suggesting a transition to the conformal window in moving from eleven to twelve fermion flavors. This is consistent with prior determinations of its lower bound \cite{Appelquist:1988yc, Cohen:1988sq, Sannino:2004qp, Dietrich:2006cm, Armoni:2009jn, Braun:2009ns, Frandsen:2010ej, Ryttov:2016ner, Kim:2020yvr, Lee:2020ihn}. Our final result combining the various estimates is provided in Eq.~(\ref{eq:Nfstar_estimate}) and lies close to $N_f^* \simeq 11.5$ for the lattice spacing and pion mass considered herein.

There are two main avenues for extending our analysis in future work. First, the approach to the continuum and chiral limits needs to be addressed. Second, it will be interesting to extend the simulations to larger values of $N_f$. The present study is inhibited by an unphysical phase in the $(\beta, m)$ phase space which must be circumvented \cite{Cheng:2011ic}. It has been demonstrated that this lattice artefact can be mitigated by the addition of Pauli-Villars fields \cite{Pauli:1949zm} with nHYP-smeared staggered fermions \cite{Hasenfratz:2021zsl, Hasenfratz:2001hp, Hasenfratz:2007rf}, paving the way for an improved estimation of $N_f^*$. Of course, examining center vortices in the unphysical phase would be an interesting study in its own right to see whether the vortex information captures its key features, such as the broken shift symmetry.

Since projected center vortices are infinitely thin, representing the vortex ``axis of rotation", there is no problem in principle with taking the continuum limit. How the various vortex properties and associated extrapolations depend on the lattice spacing, and whether this impacts the critical point, will be crucial in ascertaining the precise location of the transition to the conformal window.

\begin{acknowledgments}
This work was supported with supercomputing resources provided by the Phoenix HPC service at the University of Adelaide. This research was undertaken with the assistance of resources and services from the National Computational Infrastructure (NCI), which is supported by the Australian Government. This research was supported by the Australian Research Council through Grant No. DP210103706. DN was supported by the NKFIH excellence grant TKP2021-NKTA-64. DN also acknowledges support from the George Southgate Fellowship and thanks the CSSM and the Department of Physics at the University of Adelaide for their kind hospitality during his visit where parts of this work were completed. 
\end{acknowledgments}


\bibliography{main}

\providecommand{\noopsort}[1]{}\providecommand{\singleletter}[1]{#1}%
\begin{thebibliography}{86}%
\makeatletter
\providecommand \@ifxundefined [1]{%
 \@ifx{#1\undefined}
}%
\providecommand \@ifnum [1]{%
 \ifnum #1\expandafter \@firstoftwo
 \else \expandafter \@secondoftwo
 \fi
}%
\providecommand \@ifx [1]{%
 \ifx #1\expandafter \@firstoftwo
 \else \expandafter \@secondoftwo
 \fi
}%
\providecommand \natexlab [1]{#1}%
\providecommand \enquote  [1]{``#1''}%
\providecommand \bibnamefont  [1]{#1}%
\providecommand \bibfnamefont [1]{#1}%
\providecommand \citenamefont [1]{#1}%
\providecommand \href@noop [0]{\@secondoftwo}%
\providecommand \href [0]{\begingroup \@sanitize@url \@href}%
\providecommand \@href[1]{\@@startlink{#1}\@@href}%
\providecommand \@@href[1]{\endgroup#1\@@endlink}%
\providecommand \@sanitize@url [0]{\catcode `\\12\catcode `\$12\catcode
  `\&12\catcode `\#12\catcode `\^12\catcode `\_12\catcode `\%12\relax}%
\providecommand \@@startlink[1]{}%
\providecommand \@@endlink[0]{}%
\providecommand \url  [0]{\begingroup\@sanitize@url \@url }%
\providecommand \@url [1]{\endgroup\@href {#1}{\urlprefix }}%
\providecommand \urlprefix  [0]{URL }%
\providecommand \Eprint [0]{\href }%
\providecommand \doibase [0]{https://doi.org/}%
\providecommand \selectlanguage [0]{\@gobble}%
\providecommand \bibinfo  [0]{\@secondoftwo}%
\providecommand \bibfield  [0]{\@secondoftwo}%
\providecommand \translation [1]{[#1]}%
\providecommand \BibitemOpen [0]{}%
\providecommand \bibitemStop [0]{}%
\providecommand \bibitemNoStop [0]{.\EOS\space}%
\providecommand \EOS [0]{\spacefactor3000\relax}%
\providecommand \BibitemShut  [1]{\csname bibitem#1\endcsname}%
\let\auto@bib@innerbib\@empty
\bibitem [{\citenamefont {Banks}\ and\ \citenamefont
  {Zaks}(1982)}]{Banks:1981nn}%
  \BibitemOpen
  \bibfield  {author} {\bibinfo {author} {\bibfnamefont {T.}~\bibnamefont
  {Banks}}\ and\ \bibinfo {author} {\bibfnamefont {A.}~\bibnamefont {Zaks}},\
  }\bibfield  {title} {\bibinfo {title} {{On the phase structure of vector-like
  gauge theories with massless fermions}},\ }\href
  {https://doi.org/10.1016/0550-3213(82)90035-9} {\bibfield  {journal}
  {\bibinfo  {journal} {Nucl. Phys. B}\ }\textbf {\bibinfo {volume} {196}},\
  \bibinfo {pages} {189} (\bibinfo {year} {1982})}\BibitemShut {NoStop}%
\bibitem [{\citenamefont {Nogradi}\ and\ \citenamefont
  {Patella}(2016)}]{Nogradi:2016qek}%
  \BibitemOpen
  \bibfield  {author} {\bibinfo {author} {\bibfnamefont {D.}~\bibnamefont
  {Nogradi}}\ and\ \bibinfo {author} {\bibfnamefont {A.}~\bibnamefont
  {Patella}},\ }\bibfield  {title} {\bibinfo {title} {{Strong dynamics,
  composite Higgs and the conformal window}},\ }\href
  {https://doi.org/10.1142/S0217751X1643003X} {\bibfield  {journal} {\bibinfo
  {journal} {Int. J. Mod. Phys. A}\ }\textbf {\bibinfo {volume} {31}},\
  \bibinfo {pages} {1643003} (\bibinfo {year} {2016})},\ \Eprint
  {https://arxiv.org/abs/1607.07638} {arXiv:1607.07638 [hep-lat]} \BibitemShut
  {NoStop}%
\bibitem [{\citenamefont {Drach}(2020)}]{Drach:2020qpj}%
  \BibitemOpen
  \bibfield  {author} {\bibinfo {author} {\bibfnamefont {V.}~\bibnamefont
  {Drach}},\ }\bibfield  {title} {\bibinfo {title} {{Composite electroweak
  sectors on the lattice}},\ }\href {https://doi.org/10.22323/1.363.0242}
  {\bibfield  {journal} {\bibinfo  {journal} {PoS}\ }\textbf {\bibinfo {volume}
  {LATTICE2019}},\ \bibinfo {pages} {242} (\bibinfo {year} {2020})},\ \Eprint
  {https://arxiv.org/abs/2005.01002} {arXiv:2005.01002 [hep-lat]} \BibitemShut
  {NoStop}%
\bibitem [{\citenamefont {Appelquist}\ \emph {et~al.}(1988)\citenamefont
  {Appelquist}, \citenamefont {Lane},\ and\ \citenamefont
  {Mahanta}}]{Appelquist:1988yc}%
  \BibitemOpen
  \bibfield  {author} {\bibinfo {author} {\bibfnamefont {T.}~\bibnamefont
  {Appelquist}}, \bibinfo {author} {\bibfnamefont {K.~D.}\ \bibnamefont
  {Lane}},\ and\ \bibinfo {author} {\bibfnamefont {U.}~\bibnamefont
  {Mahanta}},\ }\bibfield  {title} {\bibinfo {title} {{Ladder Approximation for
  Spontaneous Chiral-Symmetry Breaking}},\ }\href
  {https://doi.org/10.1103/PhysRevLett.61.1553} {\bibfield  {journal} {\bibinfo
   {journal} {Phys. Rev. Lett.}\ }\textbf {\bibinfo {volume} {61}},\ \bibinfo
  {pages} {1553} (\bibinfo {year} {1988})}\BibitemShut {NoStop}%
\bibitem [{\citenamefont {Cohen}\ and\ \citenamefont
  {Georgi}(1989)}]{Cohen:1988sq}%
  \BibitemOpen
  \bibfield  {author} {\bibinfo {author} {\bibfnamefont {A.~G.}\ \bibnamefont
  {Cohen}}\ and\ \bibinfo {author} {\bibfnamefont {H.}~\bibnamefont {Georgi}},\
  }\bibfield  {title} {\bibinfo {title} {{Walking beyond the rainbow}},\ }\href
  {https://doi.org/10.1016/0550-3213(89)90109-0} {\bibfield  {journal}
  {\bibinfo  {journal} {Nucl. Phys. B}\ }\textbf {\bibinfo {volume} {314}},\
  \bibinfo {pages} {7} (\bibinfo {year} {1989})}\BibitemShut {NoStop}%
\bibitem [{\citenamefont {Sannino}\ and\ \citenamefont
  {Tuominen}(2005)}]{Sannino:2004qp}%
  \BibitemOpen
  \bibfield  {author} {\bibinfo {author} {\bibfnamefont {F.}~\bibnamefont
  {Sannino}}\ and\ \bibinfo {author} {\bibfnamefont {K.}~\bibnamefont
  {Tuominen}},\ }\bibfield  {title} {\bibinfo {title} {{Orientifold theory
  dynamics and symmetry breaking}},\ }\href
  {https://doi.org/10.1103/PhysRevD.71.051901} {\bibfield  {journal} {\bibinfo
  {journal} {Phys. Rev. D}\ }\textbf {\bibinfo {volume} {71}},\ \bibinfo
  {pages} {051901} (\bibinfo {year} {2005})},\ \Eprint
  {https://arxiv.org/abs/hep-ph/0405209} {arXiv:hep-ph/0405209} \BibitemShut
  {NoStop}%
\bibitem [{\citenamefont {Dietrich}\ and\ \citenamefont
  {Sannino}(2007)}]{Dietrich:2006cm}%
  \BibitemOpen
  \bibfield  {author} {\bibinfo {author} {\bibfnamefont {D.~D.}\ \bibnamefont
  {Dietrich}}\ and\ \bibinfo {author} {\bibfnamefont {F.}~\bibnamefont
  {Sannino}},\ }\bibfield  {title} {\bibinfo {title} {{Conformal window of
  $\mathrm{SU}(N)$ gauge theories with fermions in higher dimensional
  representations}},\ }\href {https://doi.org/10.1103/PhysRevD.75.085018}
  {\bibfield  {journal} {\bibinfo  {journal} {Phys. Rev. D}\ }\textbf {\bibinfo
  {volume} {75}},\ \bibinfo {pages} {085018} (\bibinfo {year} {2007})},\
  \Eprint {https://arxiv.org/abs/hep-ph/0611341} {arXiv:hep-ph/0611341}
  \BibitemShut {NoStop}%
\bibitem [{\citenamefont {Armoni}(2010)}]{Armoni:2009jn}%
  \BibitemOpen
  \bibfield  {author} {\bibinfo {author} {\bibfnamefont {A.}~\bibnamefont
  {Armoni}},\ }\bibfield  {title} {\bibinfo {title} {{The conformal window from
  the worldline formalism}},\ }\href
  {https://doi.org/10.1016/j.nuclphysb.2009.10.010} {\bibfield  {journal}
  {\bibinfo  {journal} {Nucl. Phys. B}\ }\textbf {\bibinfo {volume} {826}},\
  \bibinfo {pages} {328} (\bibinfo {year} {2010})},\ \Eprint
  {https://arxiv.org/abs/0907.4091} {arXiv:0907.4091 [hep-ph]} \BibitemShut
  {NoStop}%
\bibitem [{\citenamefont {Braun}\ and\ \citenamefont
  {Gies}(2010)}]{Braun:2009ns}%
  \BibitemOpen
  \bibfield  {author} {\bibinfo {author} {\bibfnamefont {J.}~\bibnamefont
  {Braun}}\ and\ \bibinfo {author} {\bibfnamefont {H.}~\bibnamefont {Gies}},\
  }\bibfield  {title} {\bibinfo {title} {{Scaling laws near the conformal
  window of many-flavor QCD}},\ }\href
  {https://doi.org/10.1007/JHEP05(2010)060} {\bibfield  {journal} {\bibinfo
  {journal} {J. High Energy Phys.}\ }\textbf {\bibinfo {volume} {05}},\
  \bibinfo {pages} {060}},\ \Eprint {https://arxiv.org/abs/0912.4168}
  {arXiv:0912.4168 [hep-ph]} \BibitemShut {NoStop}%
\bibitem [{\citenamefont {Frandsen}\ \emph {et~al.}(2011)\citenamefont
  {Frandsen}, \citenamefont {Pickup},\ and\ \citenamefont
  {Teper}}]{Frandsen:2010ej}%
  \BibitemOpen
  \bibfield  {author} {\bibinfo {author} {\bibfnamefont {M.~T.}\ \bibnamefont
  {Frandsen}}, \bibinfo {author} {\bibfnamefont {T.}~\bibnamefont {Pickup}},\
  and\ \bibinfo {author} {\bibfnamefont {M.}~\bibnamefont {Teper}},\ }\bibfield
   {title} {\bibinfo {title} {{Delineating the conformal window}},\ }\href
  {https://doi.org/10.1016/j.physletb.2010.10.064} {\bibfield  {journal}
  {\bibinfo  {journal} {Phys. Lett. B}\ }\textbf {\bibinfo {volume} {695}},\
  \bibinfo {pages} {231} (\bibinfo {year} {2011})},\ \Eprint
  {https://arxiv.org/abs/1007.1614} {arXiv:1007.1614 [hep-ph]} \BibitemShut
  {NoStop}%
\bibitem [{\citenamefont {Ryttov}\ and\ \citenamefont
  {Shrock}(2016)}]{Ryttov:2016ner}%
  \BibitemOpen
  \bibfield  {author} {\bibinfo {author} {\bibfnamefont {T.~A.}\ \bibnamefont
  {Ryttov}}\ and\ \bibinfo {author} {\bibfnamefont {R.}~\bibnamefont
  {Shrock}},\ }\bibfield  {title} {\bibinfo {title} {{Infrared zero of $\beta$
  and value of $\gamma_m$ for an SU(3) gauge theory at the five-loop level}},\
  }\href {https://doi.org/10.1103/PhysRevD.94.105015} {\bibfield  {journal}
  {\bibinfo  {journal} {Phys. Rev. D}\ }\textbf {\bibinfo {volume} {94}},\
  \bibinfo {pages} {105015} (\bibinfo {year} {2016})},\ \Eprint
  {https://arxiv.org/abs/1607.06866} {arXiv:1607.06866 [hep-th]} \BibitemShut
  {NoStop}%
\bibitem [{\citenamefont {Kim}\ \emph {et~al.}(2020)\citenamefont {Kim},
  \citenamefont {Hong},\ and\ \citenamefont {Lee}}]{Kim:2020yvr}%
  \BibitemOpen
  \bibfield  {author} {\bibinfo {author} {\bibfnamefont {B.~S.}\ \bibnamefont
  {Kim}}, \bibinfo {author} {\bibfnamefont {D.~K.}\ \bibnamefont {Hong}},\ and\
  \bibinfo {author} {\bibfnamefont {J.-W.}\ \bibnamefont {Lee}},\ }\bibfield
  {title} {\bibinfo {title} {{Into the conformal window: Multirepresentation
  gauge theories}},\ }\href {https://doi.org/10.1103/PhysRevD.101.056008}
  {\bibfield  {journal} {\bibinfo  {journal} {Phys. Rev. D}\ }\textbf {\bibinfo
  {volume} {101}},\ \bibinfo {pages} {056008} (\bibinfo {year} {2020})},\
  \Eprint {https://arxiv.org/abs/2001.02690} {arXiv:2001.02690 [hep-ph]}
  \BibitemShut {NoStop}%
\bibitem [{\citenamefont {Lee}(2021)}]{Lee:2020ihn}%
  \BibitemOpen
  \bibfield  {author} {\bibinfo {author} {\bibfnamefont {J.-W.}\ \bibnamefont
  {Lee}},\ }\bibfield  {title} {\bibinfo {title} {{Conformal window from
  conformal expansion}},\ }\href {https://doi.org/10.1103/PhysRevD.103.076006}
  {\bibfield  {journal} {\bibinfo  {journal} {Phys. Rev. D}\ }\textbf {\bibinfo
  {volume} {103}},\ \bibinfo {pages} {076006} (\bibinfo {year} {2021})},\
  \Eprint {https://arxiv.org/abs/2008.12223} {arXiv:2008.12223 [hep-ph]}
  \BibitemShut {NoStop}%
\bibitem [{\citenamefont {Appelquist}\ \emph {et~al.}(2008)\citenamefont
  {Appelquist}, \citenamefont {Fleming},\ and\ \citenamefont
  {Neil}}]{Appelquist:2007hu}%
  \BibitemOpen
  \bibfield  {author} {\bibinfo {author} {\bibfnamefont {T.}~\bibnamefont
  {Appelquist}}, \bibinfo {author} {\bibfnamefont {G.~T.}\ \bibnamefont
  {Fleming}},\ and\ \bibinfo {author} {\bibfnamefont {E.~T.}\ \bibnamefont
  {Neil}},\ }\bibfield  {title} {\bibinfo {title} {{Lattice Study of the
  Conformal Window in QCD-like Theories}},\ }\href
  {https://doi.org/10.1103/PhysRevLett.100.171607} {\bibfield  {journal}
  {\bibinfo  {journal} {Phys. Rev. Lett.}\ }\textbf {\bibinfo {volume} {100}},\
  \bibinfo {pages} {171607} (\bibinfo {year} {2008})},\ \bibinfo {note}
  {[Erratum: Phys. Rev. Lett. 102, 149902 (2009)]},\ \Eprint
  {https://arxiv.org/abs/0712.0609} {arXiv:0712.0609 [hep-ph]} \BibitemShut
  {NoStop}%
\bibitem [{\citenamefont {Appelquist}\ \emph {et~al.}(2009)\citenamefont
  {Appelquist}, \citenamefont {Fleming},\ and\ \citenamefont
  {Neil}}]{Appelquist:2009ty}%
  \BibitemOpen
  \bibfield  {author} {\bibinfo {author} {\bibfnamefont {T.}~\bibnamefont
  {Appelquist}}, \bibinfo {author} {\bibfnamefont {G.~T.}\ \bibnamefont
  {Fleming}},\ and\ \bibinfo {author} {\bibfnamefont {E.~T.}\ \bibnamefont
  {Neil}},\ }\bibfield  {title} {\bibinfo {title} {{Lattice study of conformal
  behavior in SU(3) Yang-Mills theories}},\ }\href
  {https://doi.org/10.1103/PhysRevD.79.076010} {\bibfield  {journal} {\bibinfo
  {journal} {Phys. Rev. D}\ }\textbf {\bibinfo {volume} {79}},\ \bibinfo
  {pages} {076010} (\bibinfo {year} {2009})},\ \Eprint
  {https://arxiv.org/abs/0901.3766} {arXiv:0901.3766 [hep-ph]} \BibitemShut
  {NoStop}%
\bibitem [{\citenamefont {Fodor}\ \emph {et~al.}(2009)\citenamefont {Fodor},
  \citenamefont {Holland}, \citenamefont {Kuti}, \citenamefont {Nogradi},\ and\
  \citenamefont {Schroeder}}]{Fodor:2009wk}%
  \BibitemOpen
  \bibfield  {author} {\bibinfo {author} {\bibfnamefont {Z.}~\bibnamefont
  {Fodor}}, \bibinfo {author} {\bibfnamefont {K.}~\bibnamefont {Holland}},
  \bibinfo {author} {\bibfnamefont {J.}~\bibnamefont {Kuti}}, \bibinfo {author}
  {\bibfnamefont {D.}~\bibnamefont {Nogradi}},\ and\ \bibinfo {author}
  {\bibfnamefont {C.}~\bibnamefont {Schroeder}},\ }\bibfield  {title} {\bibinfo
  {title} {{Nearly conformal gauge theories in finite volume}},\ }\href
  {https://doi.org/10.1016/j.physletb.2009.10.040} {\bibfield  {journal}
  {\bibinfo  {journal} {Phys. Lett. B}\ }\textbf {\bibinfo {volume} {681}},\
  \bibinfo {pages} {353} (\bibinfo {year} {2009})},\ \Eprint
  {https://arxiv.org/abs/0907.4562} {arXiv:0907.4562 [hep-lat]} \BibitemShut
  {NoStop}%
\bibitem [{\citenamefont {Fodor}\ \emph {et~al.}(2011)\citenamefont {Fodor},
  \citenamefont {Holland}, \citenamefont {Kuti}, \citenamefont {Nogradi},
  \citenamefont {Schroeder}, \citenamefont {Holland}, \citenamefont {Kuti},
  \citenamefont {Nogradi},\ and\ \citenamefont {Schroeder}}]{Fodor:2011tu}%
  \BibitemOpen
  \bibfield  {author} {\bibinfo {author} {\bibfnamefont {Z.}~\bibnamefont
  {Fodor}}, \bibinfo {author} {\bibfnamefont {K.}~\bibnamefont {Holland}},
  \bibinfo {author} {\bibfnamefont {J.}~\bibnamefont {Kuti}}, \bibinfo {author}
  {\bibfnamefont {D.}~\bibnamefont {Nogradi}}, \bibinfo {author} {\bibfnamefont
  {C.}~\bibnamefont {Schroeder}}, \bibinfo {author} {\bibfnamefont
  {K.}~\bibnamefont {Holland}}, \bibinfo {author} {\bibfnamefont
  {J.}~\bibnamefont {Kuti}}, \bibinfo {author} {\bibfnamefont {D.}~\bibnamefont
  {Nogradi}},\ and\ \bibinfo {author} {\bibfnamefont {C.}~\bibnamefont
  {Schroeder}},\ }\bibfield  {title} {\bibinfo {title} {{Twelve massless
  flavors and three colors below the conformal window}},\ }\href
  {https://doi.org/10.1016/j.physletb.2011.07.037} {\bibfield  {journal}
  {\bibinfo  {journal} {Phys. Lett. B}\ }\textbf {\bibinfo {volume} {703}},\
  \bibinfo {pages} {348} (\bibinfo {year} {2011})},\ \Eprint
  {https://arxiv.org/abs/1104.3124} {arXiv:1104.3124 [hep-lat]} \BibitemShut
  {NoStop}%
\bibitem [{\citenamefont {Hasenfratz}(2012)}]{Hasenfratz:2011xn}%
  \BibitemOpen
  \bibfield  {author} {\bibinfo {author} {\bibfnamefont {A.}~\bibnamefont
  {Hasenfratz}},\ }\bibfield  {title} {\bibinfo {title} {{Infrared Fixed Point
  of the 12-Fermion SU(3) Gauge Model Based on 2-Lattice Monte Carlo
  Renormalization-Group Matching}},\ }\href
  {https://doi.org/10.1103/PhysRevLett.108.061601} {\bibfield  {journal}
  {\bibinfo  {journal} {Phys. Rev. Lett.}\ }\textbf {\bibinfo {volume} {108}},\
  \bibinfo {pages} {061601} (\bibinfo {year} {2012})},\ \Eprint
  {https://arxiv.org/abs/1106.5293} {arXiv:1106.5293 [hep-lat]} \BibitemShut
  {NoStop}%
\bibitem [{\citenamefont {DeGrand}(2011)}]{DeGrand:2011cu}%
  \BibitemOpen
  \bibfield  {author} {\bibinfo {author} {\bibfnamefont {T.}~\bibnamefont
  {DeGrand}},\ }\bibfield  {title} {\bibinfo {title} {{Finite-size scaling
  tests for spectra in SU(3) lattice gauge theory coupled to 12 fundamental
  flavor fermions}},\ }\href {https://doi.org/10.1103/PhysRevD.84.116901}
  {\bibfield  {journal} {\bibinfo  {journal} {Phys. Rev. D}\ }\textbf {\bibinfo
  {volume} {84}},\ \bibinfo {pages} {116901} (\bibinfo {year} {2011})},\
  \Eprint {https://arxiv.org/abs/1109.1237} {arXiv:1109.1237 [hep-lat]}
  \BibitemShut {NoStop}%
\bibitem [{\citenamefont {Lin}\ \emph {et~al.}(2012)\citenamefont {Lin},
  \citenamefont {Ogawa}, \citenamefont {Ohki},\ and\ \citenamefont
  {Shintani}}]{Lin:2012iw}%
  \BibitemOpen
  \bibfield  {author} {\bibinfo {author} {\bibfnamefont {C.~J.~D.}\
  \bibnamefont {Lin}}, \bibinfo {author} {\bibfnamefont {K.}~\bibnamefont
  {Ogawa}}, \bibinfo {author} {\bibfnamefont {H.}~\bibnamefont {Ohki}},\ and\
  \bibinfo {author} {\bibfnamefont {E.}~\bibnamefont {Shintani}},\ }\bibfield
  {title} {\bibinfo {title} {{Lattice study of infrared behaviour in SU(3)
  gauge theory with twelve massless flavours}},\ }\href
  {https://doi.org/10.1007/JHEP08(2012)096} {\bibfield  {journal} {\bibinfo
  {journal} {J. High Energy Phys.}\ }\textbf {\bibinfo {volume} {08}},\
  \bibinfo {pages} {096}},\ \Eprint {https://arxiv.org/abs/1205.6076}
  {arXiv:1205.6076 [hep-lat]} \BibitemShut {NoStop}%
\bibitem [{\citenamefont {Aoki}\ \emph {et~al.}(2012)\citenamefont {Aoki},
  \citenamefont {Aoyama}, \citenamefont {Kurachi}, \citenamefont {Maskawa},
  \citenamefont {Nagai}, \citenamefont {Ohki}, \citenamefont {Shibata},
  \citenamefont {Yamawaki},\ and\ \citenamefont {Yamazaki}}]{Aoki:2012eq}%
  \BibitemOpen
  \bibfield  {author} {\bibinfo {author} {\bibfnamefont {Y.}~\bibnamefont
  {Aoki}}, \bibinfo {author} {\bibfnamefont {T.}~\bibnamefont {Aoyama}},
  \bibinfo {author} {\bibfnamefont {M.}~\bibnamefont {Kurachi}}, \bibinfo
  {author} {\bibfnamefont {T.}~\bibnamefont {Maskawa}}, \bibinfo {author}
  {\bibfnamefont {K.-i.}\ \bibnamefont {Nagai}}, \bibinfo {author}
  {\bibfnamefont {H.}~\bibnamefont {Ohki}}, \bibinfo {author} {\bibfnamefont
  {A.}~\bibnamefont {Shibata}}, \bibinfo {author} {\bibfnamefont
  {K.}~\bibnamefont {Yamawaki}},\ and\ \bibinfo {author} {\bibfnamefont
  {T.}~\bibnamefont {Yamazaki}},\ }\bibfield  {title} {\bibinfo {title}
  {{Lattice study of conformality in twelve-flavor QCD}},\ }\href
  {https://doi.org/10.1103/PhysRevD.86.054506} {\bibfield  {journal} {\bibinfo
  {journal} {Phys. Rev. D}\ }\textbf {\bibinfo {volume} {86}},\ \bibinfo
  {pages} {054506} (\bibinfo {year} {2012})},\ \Eprint
  {https://arxiv.org/abs/1207.3060} {arXiv:1207.3060 [hep-lat]} \BibitemShut
  {NoStop}%
\bibitem [{\citenamefont {Aoki}\ \emph {et~al.}(2013)\citenamefont {Aoki},
  \citenamefont {Aoyama}, \citenamefont {Kurachi}, \citenamefont {Maskawa},
  \citenamefont {Nagai}, \citenamefont {Ohki}, \citenamefont {Rinaldi},
  \citenamefont {Shibata}, \citenamefont {Yamawaki},\ and\ \citenamefont
  {Yamazaki}}]{LatKMI:2013bhp}%
  \BibitemOpen
  \bibfield  {author} {\bibinfo {author} {\bibfnamefont {Y.}~\bibnamefont
  {Aoki}}, \bibinfo {author} {\bibfnamefont {T.}~\bibnamefont {Aoyama}},
  \bibinfo {author} {\bibfnamefont {M.}~\bibnamefont {Kurachi}}, \bibinfo
  {author} {\bibfnamefont {T.}~\bibnamefont {Maskawa}}, \bibinfo {author}
  {\bibfnamefont {K.-i.}\ \bibnamefont {Nagai}}, \bibinfo {author}
  {\bibfnamefont {H.}~\bibnamefont {Ohki}}, \bibinfo {author} {\bibfnamefont
  {E.}~\bibnamefont {Rinaldi}}, \bibinfo {author} {\bibfnamefont
  {A.}~\bibnamefont {Shibata}}, \bibinfo {author} {\bibfnamefont
  {K.}~\bibnamefont {Yamawaki}},\ and\ \bibinfo {author} {\bibfnamefont
  {T.}~\bibnamefont {Yamazaki}} (\bibinfo {collaboration} {LatKMI}),\
  }\bibfield  {title} {\bibinfo {title} {{Light Composite Scalar in
  Twelve-Flavor QCD on the Lattice}},\ }\href
  {https://doi.org/10.1103/PhysRevLett.111.162001} {\bibfield  {journal}
  {\bibinfo  {journal} {Phys. Rev. Lett.}\ }\textbf {\bibinfo {volume} {111}},\
  \bibinfo {pages} {162001} (\bibinfo {year} {2013})},\ \Eprint
  {https://arxiv.org/abs/1305.6006} {arXiv:1305.6006 [hep-lat]} \BibitemShut
  {NoStop}%
\bibitem [{\citenamefont {Fodor}\ \emph {et~al.}(2016)\citenamefont {Fodor},
  \citenamefont {Holland}, \citenamefont {Kuti}, \citenamefont {Mondal},
  \citenamefont {Nogradi},\ and\ \citenamefont {Wong}}]{Fodor:2016zil}%
  \BibitemOpen
  \bibfield  {author} {\bibinfo {author} {\bibfnamefont {Z.}~\bibnamefont
  {Fodor}}, \bibinfo {author} {\bibfnamefont {K.}~\bibnamefont {Holland}},
  \bibinfo {author} {\bibfnamefont {J.}~\bibnamefont {Kuti}}, \bibinfo {author}
  {\bibfnamefont {S.}~\bibnamefont {Mondal}}, \bibinfo {author} {\bibfnamefont
  {D.}~\bibnamefont {Nogradi}},\ and\ \bibinfo {author} {\bibfnamefont {C.~H.}\
  \bibnamefont {Wong}},\ }\bibfield  {title} {\bibinfo {title} {{Fate of the
  conformal fixed point with twelve massless fermions and SU(3) gauge group}},\
  }\href {https://doi.org/10.1103/PhysRevD.94.091501} {\bibfield  {journal}
  {\bibinfo  {journal} {Phys. Rev. D}\ }\textbf {\bibinfo {volume} {94}},\
  \bibinfo {pages} {091501} (\bibinfo {year} {2016})},\ \Eprint
  {https://arxiv.org/abs/1607.06121} {arXiv:1607.06121 [hep-lat]} \BibitemShut
  {NoStop}%
\bibitem [{\citenamefont {Hasenfratz}\ and\ \citenamefont
  {Schaich}(2018)}]{Hasenfratz:2016dou}%
  \BibitemOpen
  \bibfield  {author} {\bibinfo {author} {\bibfnamefont {A.}~\bibnamefont
  {Hasenfratz}}\ and\ \bibinfo {author} {\bibfnamefont {D.}~\bibnamefont
  {Schaich}},\ }\bibfield  {title} {\bibinfo {title} {{Nonperturbative $\beta$
  function of twelve-flavor SU(3) gauge theory}},\ }\href
  {https://doi.org/10.1007/JHEP02(2018)132} {\bibfield  {journal} {\bibinfo
  {journal} {J. High Energy Phys.}\ }\textbf {\bibinfo {volume} {02}},\
  \bibinfo {pages} {132}},\ \Eprint {https://arxiv.org/abs/1610.10004}
  {arXiv:1610.10004 [hep-lat]} \BibitemShut {NoStop}%
\bibitem [{\citenamefont {Fodor}\ \emph
  {et~al.}(2018{\natexlab{a}})\citenamefont {Fodor}, \citenamefont {Holland},
  \citenamefont {Kuti}, \citenamefont {Nogradi},\ and\ \citenamefont
  {Wong}}]{Fodor:2017gtj}%
  \BibitemOpen
  \bibfield  {author} {\bibinfo {author} {\bibfnamefont {Z.}~\bibnamefont
  {Fodor}}, \bibinfo {author} {\bibfnamefont {K.}~\bibnamefont {Holland}},
  \bibinfo {author} {\bibfnamefont {J.}~\bibnamefont {Kuti}}, \bibinfo {author}
  {\bibfnamefont {D.}~\bibnamefont {Nogradi}},\ and\ \bibinfo {author}
  {\bibfnamefont {C.~H.}\ \bibnamefont {Wong}},\ }\bibfield  {title} {\bibinfo
  {title} {{Extended investigation of the twelve-flavor $\beta$-function}},\
  }\href {https://doi.org/10.1016/j.physletb.2018.02.008} {\bibfield  {journal}
  {\bibinfo  {journal} {Phys. Lett. B}\ }\textbf {\bibinfo {volume} {779}},\
  \bibinfo {pages} {230} (\bibinfo {year} {2018}{\natexlab{a}})},\ \Eprint
  {https://arxiv.org/abs/1710.09262} {arXiv:1710.09262 [hep-lat]} \BibitemShut
  {NoStop}%
\bibitem [{\citenamefont {Hasenfratz}\ \emph {et~al.}(2019)\citenamefont
  {Hasenfratz}, \citenamefont {Rebbi},\ and\ \citenamefont
  {Witzel}}]{Hasenfratz:2017qyr}%
  \BibitemOpen
  \bibfield  {author} {\bibinfo {author} {\bibfnamefont {A.}~\bibnamefont
  {Hasenfratz}}, \bibinfo {author} {\bibfnamefont {C.}~\bibnamefont {Rebbi}},\
  and\ \bibinfo {author} {\bibfnamefont {O.}~\bibnamefont {Witzel}},\
  }\bibfield  {title} {\bibinfo {title} {{Nonperturbative determination of
  $\beta$ functions for SU(3) gauge theories with 10 and 12 fundamental flavors
  using domain wall fermions}},\ }\href
  {https://doi.org/10.1016/j.physletb.2019.134937} {\bibfield  {journal}
  {\bibinfo  {journal} {Phys. Lett. B}\ }\textbf {\bibinfo {volume} {798}},\
  \bibinfo {pages} {134937} (\bibinfo {year} {2019})},\ \Eprint
  {https://arxiv.org/abs/1710.11578} {arXiv:1710.11578 [hep-lat]} \BibitemShut
  {NoStop}%
\bibitem [{\citenamefont {Fodor}\ \emph
  {et~al.}(2018{\natexlab{b}})\citenamefont {Fodor}, \citenamefont {Holland},
  \citenamefont {Kuti}, \citenamefont {Nogradi},\ and\ \citenamefont
  {Wong}}]{Fodor:2017nlp}%
  \BibitemOpen
  \bibfield  {author} {\bibinfo {author} {\bibfnamefont {Z.}~\bibnamefont
  {Fodor}}, \bibinfo {author} {\bibfnamefont {K.}~\bibnamefont {Holland}},
  \bibinfo {author} {\bibfnamefont {J.}~\bibnamefont {Kuti}}, \bibinfo {author}
  {\bibfnamefont {D.}~\bibnamefont {Nogradi}},\ and\ \bibinfo {author}
  {\bibfnamefont {C.~H.}\ \bibnamefont {Wong}},\ }\bibfield  {title} {\bibinfo
  {title} {{The twelve-flavor $\beta$-function and dilaton tests of the sextet
  scalar}},\ }\href {https://doi.org/10.1051/epjconf/201817508015} {\bibfield
  {journal} {\bibinfo  {journal} {EPJ Web Conf.}\ }\textbf {\bibinfo {volume}
  {175}},\ \bibinfo {pages} {08015} (\bibinfo {year} {2018}{\natexlab{b}})},\
  \Eprint {https://arxiv.org/abs/1712.08594} {arXiv:1712.08594 [hep-lat]}
  \BibitemShut {NoStop}%
\bibitem [{\citenamefont {Hasenfratz}\ and\ \citenamefont
  {Peterson}(2024)}]{Hasenfratz:2024fad}%
  \BibitemOpen
  \bibfield  {author} {\bibinfo {author} {\bibfnamefont {A.}~\bibnamefont
  {Hasenfratz}}\ and\ \bibinfo {author} {\bibfnamefont {C.~T.}\ \bibnamefont
  {Peterson}},\ }\bibfield  {title} {\bibinfo {title} {{Infrared fixed point in
  the massless twelve-flavor SU(3) gauge-fermion system}},\ }\href
  {https://doi.org/10.1103/PhysRevD.109.114507} {\bibfield  {journal} {\bibinfo
   {journal} {Phys. Rev. D}\ }\textbf {\bibinfo {volume} {109}},\ \bibinfo
  {pages} {114507} (\bibinfo {year} {2024})},\ \Eprint
  {https://arxiv.org/abs/2402.18038} {arXiv:2402.18038 [hep-lat]} \BibitemShut
  {NoStop}%
\bibitem [{\citenamefont {'t~Hooft}(1978)}]{tHooft:1977nqb}%
  \BibitemOpen
  \bibfield  {author} {\bibinfo {author} {\bibfnamefont {G.}~\bibnamefont
  {'t~Hooft}},\ }\bibfield  {title} {\bibinfo {title} {{On the phase transition
  towards permanent quark confinement}},\ }\href
  {https://doi.org/10.1016/0550-3213(78)90153-0} {\bibfield  {journal}
  {\bibinfo  {journal} {Nucl. Phys. B}\ }\textbf {\bibinfo {volume} {138}},\
  \bibinfo {pages} {1} (\bibinfo {year} {1978})}\BibitemShut {NoStop}%
\bibitem [{\citenamefont {'t~Hooft}(1979)}]{tHooft:1979rtg}%
  \BibitemOpen
  \bibfield  {author} {\bibinfo {author} {\bibfnamefont {G.}~\bibnamefont
  {'t~Hooft}},\ }\bibfield  {title} {\bibinfo {title} {{A property of electric
  and magnetic flux in non-Abelian gauge theories}},\ }\href
  {https://doi.org/10.1016/0550-3213(79)90595-9} {\bibfield  {journal}
  {\bibinfo  {journal} {Nucl. Phys. B}\ }\textbf {\bibinfo {volume} {153}},\
  \bibinfo {pages} {141} (\bibinfo {year} {1979})}\BibitemShut {NoStop}%
\bibitem [{\citenamefont {Nielsen}\ and\ \citenamefont
  {Olesen}(1979)}]{Nielsen:1979xu}%
  \BibitemOpen
  \bibfield  {author} {\bibinfo {author} {\bibfnamefont {H.~B.}\ \bibnamefont
  {Nielsen}}\ and\ \bibinfo {author} {\bibfnamefont {P.}~\bibnamefont
  {Olesen}},\ }\bibfield  {title} {\bibinfo {title} {{A quantum liquid model
  for the QCD vacuum: Gauge and rotational invariance of domained and quantized
  homogeneous color fields}},\ }\href
  {https://doi.org/10.1016/0550-3213(79)90065-8} {\bibfield  {journal}
  {\bibinfo  {journal} {Nucl. Phys. B}\ }\textbf {\bibinfo {volume} {160}},\
  \bibinfo {pages} {380} (\bibinfo {year} {1979})}\BibitemShut {NoStop}%
\bibitem [{\citenamefont {Greensite}(2003)}]{Greensite:2003bk}%
  \BibitemOpen
  \bibfield  {author} {\bibinfo {author} {\bibfnamefont {J.}~\bibnamefont
  {Greensite}},\ }\bibfield  {title} {\bibinfo {title} {{The confinement
  problem in lattice gauge theory}},\ }\href
  {https://doi.org/10.1016/S0146-6410(03)90012-3} {\bibfield  {journal}
  {\bibinfo  {journal} {Prog. Part. Nucl. Phys.}\ }\textbf {\bibinfo {volume}
  {51}},\ \bibinfo {pages} {1} (\bibinfo {year} {2003})},\ \Eprint
  {https://arxiv.org/abs/hep-lat/0301023} {arXiv:hep-lat/0301023} \BibitemShut
  {NoStop}%
\bibitem [{\citenamefont {Del~Debbio}\ \emph {et~al.}(1997)\citenamefont
  {Del~Debbio}, \citenamefont {Faber}, \citenamefont {Greensite},\ and\
  \citenamefont {Olejn{\'i}k}}]{DelDebbio:1996lih}%
  \BibitemOpen
  \bibfield  {author} {\bibinfo {author} {\bibfnamefont {L.}~\bibnamefont
  {Del~Debbio}}, \bibinfo {author} {\bibfnamefont {M.}~\bibnamefont {Faber}},
  \bibinfo {author} {\bibfnamefont {J.}~\bibnamefont {Greensite}},\ and\
  \bibinfo {author} {\bibfnamefont {{\v S}.}~\bibnamefont {Olejn{\'i}k}},\
  }\bibfield  {title} {\bibinfo {title} {{Center dominance and $Z_2$ vortices
  in SU(2) lattice gauge theory}},\ }\href
  {https://doi.org/10.1103/PhysRevD.55.2298} {\bibfield  {journal} {\bibinfo
  {journal} {Phys. Rev. D}\ }\textbf {\bibinfo {volume} {55}},\ \bibinfo
  {pages} {2298} (\bibinfo {year} {1997})},\ \Eprint
  {https://arxiv.org/abs/hep-lat/9610005} {arXiv:hep-lat/9610005} \BibitemShut
  {NoStop}%
\bibitem [{\citenamefont {Del~Debbio}\ \emph
  {et~al.}(1998{\natexlab{a}})\citenamefont {Del~Debbio}, \citenamefont
  {Faber}, \citenamefont {Greensite},\ and\ \citenamefont
  {Olejn{\'i}k}}]{DelDebbio:1997ep}%
  \BibitemOpen
  \bibfield  {author} {\bibinfo {author} {\bibfnamefont {L.}~\bibnamefont
  {Del~Debbio}}, \bibinfo {author} {\bibfnamefont {M.}~\bibnamefont {Faber}},
  \bibinfo {author} {\bibfnamefont {J.}~\bibnamefont {Greensite}},\ and\
  \bibinfo {author} {\bibfnamefont {{\v S}.}~\bibnamefont {Olejn{\'i}k}},\
  }\bibfield  {title} {\bibinfo {title} {{Center vortices and the asymptotic
  string tension}},\ }\href {https://doi.org/10.1016/S0920-5632(97)00831-1}
  {\bibfield  {journal} {\bibinfo  {journal} {Nucl. Phys. B Proc. Suppl.}\
  }\textbf {\bibinfo {volume} {63}},\ \bibinfo {pages} {552} (\bibinfo {year}
  {1998}{\natexlab{a}})},\ \Eprint {https://arxiv.org/abs/hep-lat/9709032}
  {arXiv:hep-lat/9709032} \BibitemShut {NoStop}%
\bibitem [{\citenamefont {Langfeld}\ \emph {et~al.}(1998)\citenamefont
  {Langfeld}, \citenamefont {Reinhardt},\ and\ \citenamefont
  {Tennert}}]{Langfeld:1997jx}%
  \BibitemOpen
  \bibfield  {author} {\bibinfo {author} {\bibfnamefont {K.}~\bibnamefont
  {Langfeld}}, \bibinfo {author} {\bibfnamefont {H.}~\bibnamefont
  {Reinhardt}},\ and\ \bibinfo {author} {\bibfnamefont {O.}~\bibnamefont
  {Tennert}},\ }\bibfield  {title} {\bibinfo {title} {{Confinement and scaling
  of the vortex vacuum of SU(2) lattice gauge theory}},\ }\href
  {https://doi.org/10.1016/S0370-2693(97)01435-4} {\bibfield  {journal}
  {\bibinfo  {journal} {Phys. Lett. B}\ }\textbf {\bibinfo {volume} {419}},\
  \bibinfo {pages} {317} (\bibinfo {year} {1998})},\ \Eprint
  {https://arxiv.org/abs/hep-lat/9710068} {arXiv:hep-lat/9710068} \BibitemShut
  {NoStop}%
\bibitem [{\citenamefont {Del~Debbio}\ \emph
  {et~al.}(1998{\natexlab{b}})\citenamefont {Del~Debbio}, \citenamefont
  {Faber}, \citenamefont {Giedt}, \citenamefont {Greensite},\ and\
  \citenamefont {Olejn{\'i}k}}]{DelDebbio:1998luz}%
  \BibitemOpen
  \bibfield  {author} {\bibinfo {author} {\bibfnamefont {L.}~\bibnamefont
  {Del~Debbio}}, \bibinfo {author} {\bibfnamefont {M.}~\bibnamefont {Faber}},
  \bibinfo {author} {\bibfnamefont {J.}~\bibnamefont {Giedt}}, \bibinfo
  {author} {\bibfnamefont {J.}~\bibnamefont {Greensite}},\ and\ \bibinfo
  {author} {\bibfnamefont {{\v S}.}~\bibnamefont {Olejn{\'i}k}},\ }\bibfield
  {title} {\bibinfo {title} {{Detection of center vortices in the lattice
  Yang-Mills vacuum}},\ }\href {https://doi.org/10.1103/PhysRevD.58.094501}
  {\bibfield  {journal} {\bibinfo  {journal} {Phys. Rev. D}\ }\textbf {\bibinfo
  {volume} {58}},\ \bibinfo {pages} {094501} (\bibinfo {year}
  {1998}{\natexlab{b}})},\ \Eprint {https://arxiv.org/abs/hep-lat/9801027}
  {arXiv:hep-lat/9801027} \BibitemShut {NoStop}%
\bibitem [{\citenamefont {Faber}\ \emph {et~al.}(1998)\citenamefont {Faber},
  \citenamefont {Greensite},\ and\ \citenamefont {Olejn{\'i}k}}]{Faber:1997rp}%
  \BibitemOpen
  \bibfield  {author} {\bibinfo {author} {\bibfnamefont {M.}~\bibnamefont
  {Faber}}, \bibinfo {author} {\bibfnamefont {J.}~\bibnamefont {Greensite}},\
  and\ \bibinfo {author} {\bibfnamefont {{\v S}.}~\bibnamefont {Olejn{\'i}k}},\
  }\bibfield  {title} {\bibinfo {title} {{Casimir scaling from center vortices:
  Towards an understanding of the adjoint string tension}},\ }\href
  {https://doi.org/10.1103/PhysRevD.57.2603} {\bibfield  {journal} {\bibinfo
  {journal} {Phys. Rev. D}\ }\textbf {\bibinfo {volume} {57}},\ \bibinfo
  {pages} {2603} (\bibinfo {year} {1998})},\ \Eprint
  {https://arxiv.org/abs/hep-lat/9710039} {arXiv:hep-lat/9710039} \BibitemShut
  {NoStop}%
\bibitem [{\citenamefont {Faber}\ \emph
  {et~al.}(1999{\natexlab{a}})\citenamefont {Faber}, \citenamefont
  {Greensite},\ and\ \citenamefont {Olejn{\'i}k}}]{Faber:1998qn}%
  \BibitemOpen
  \bibfield  {author} {\bibinfo {author} {\bibfnamefont {M.}~\bibnamefont
  {Faber}}, \bibinfo {author} {\bibfnamefont {J.}~\bibnamefont {Greensite}},\
  and\ \bibinfo {author} {\bibfnamefont {{\v S}.}~\bibnamefont {Olejn{\'i}k}},\
  }\bibfield  {title} {\bibinfo {title} {{Evidence for a center vortex origin
  of the adjoint string tension}},\ }\href@noop {} {\bibfield  {journal}
  {\bibinfo  {journal} {Acta Phys. Slov.}\ }\textbf {\bibinfo {volume} {49}},\
  \bibinfo {pages} {177} (\bibinfo {year} {1999}{\natexlab{a}})},\ \Eprint
  {https://arxiv.org/abs/hep-lat/9807008} {arXiv:hep-lat/9807008} \BibitemShut
  {NoStop}%
\bibitem [{\citenamefont {Kov{\'a}cs}\ and\ \citenamefont
  {Tomboulis}(1998)}]{Kovacs:1998xm}%
  \BibitemOpen
  \bibfield  {author} {\bibinfo {author} {\bibfnamefont {T.~G.}\ \bibnamefont
  {Kov{\'a}cs}}\ and\ \bibinfo {author} {\bibfnamefont {E.~T.}\ \bibnamefont
  {Tomboulis}},\ }\bibfield  {title} {\bibinfo {title} {{Vortices and
  confinement at weak coupling}},\ }\href
  {https://doi.org/10.1103/PhysRevD.57.4054} {\bibfield  {journal} {\bibinfo
  {journal} {Phys. Rev. D}\ }\textbf {\bibinfo {volume} {57}},\ \bibinfo
  {pages} {4054} (\bibinfo {year} {1998})},\ \Eprint
  {https://arxiv.org/abs/hep-lat/9711009} {arXiv:hep-lat/9711009} \BibitemShut
  {NoStop}%
\bibitem [{\citenamefont {Langfeld}\ \emph {et~al.}(1999)\citenamefont
  {Langfeld}, \citenamefont {Tennert}, \citenamefont {Engelhardt},\ and\
  \citenamefont {Reinhardt}}]{Langfeld:1998cz}%
  \BibitemOpen
  \bibfield  {author} {\bibinfo {author} {\bibfnamefont {K.}~\bibnamefont
  {Langfeld}}, \bibinfo {author} {\bibfnamefont {O.}~\bibnamefont {Tennert}},
  \bibinfo {author} {\bibfnamefont {M.}~\bibnamefont {Engelhardt}},\ and\
  \bibinfo {author} {\bibfnamefont {H.}~\bibnamefont {Reinhardt}},\ }\bibfield
  {title} {\bibinfo {title} {{Center vortices of Yang-Mills theory at finite
  temperatures}},\ }\href {https://doi.org/10.1016/S0370-2693(99)00252-X}
  {\bibfield  {journal} {\bibinfo  {journal} {Phys. Lett. B}\ }\textbf
  {\bibinfo {volume} {452}},\ \bibinfo {pages} {301} (\bibinfo {year}
  {1999})},\ \Eprint {https://arxiv.org/abs/hep-lat/9805002}
  {arXiv:hep-lat/9805002} \BibitemShut {NoStop}%
\bibitem [{\citenamefont {Engelhardt}\ \emph {et~al.}(1998)\citenamefont
  {Engelhardt}, \citenamefont {Langfeld}, \citenamefont {Reinhardt},\ and\
  \citenamefont {Tennert}}]{Engelhardt:1998wu}%
  \BibitemOpen
  \bibfield  {author} {\bibinfo {author} {\bibfnamefont {M.}~\bibnamefont
  {Engelhardt}}, \bibinfo {author} {\bibfnamefont {K.}~\bibnamefont
  {Langfeld}}, \bibinfo {author} {\bibfnamefont {H.}~\bibnamefont
  {Reinhardt}},\ and\ \bibinfo {author} {\bibfnamefont {O.}~\bibnamefont
  {Tennert}},\ }\bibfield  {title} {\bibinfo {title} {{Interaction of confining
  vortices in SU(2) lattice gauge theory}},\ }\href
  {https://doi.org/10.1016/S0370-2693(98)00583-8} {\bibfield  {journal}
  {\bibinfo  {journal} {Phys. Lett. B}\ }\textbf {\bibinfo {volume} {431}},\
  \bibinfo {pages} {141} (\bibinfo {year} {1998})},\ \Eprint
  {https://arxiv.org/abs/hep-lat/9801030} {arXiv:hep-lat/9801030} \BibitemShut
  {NoStop}%
\bibitem [{\citenamefont {Bertle}\ \emph {et~al.}(1999)\citenamefont {Bertle},
  \citenamefont {Faber}, \citenamefont {Greensite},\ and\ \citenamefont
  {Olejn{\'i}k}}]{Bertle:1999tw}%
  \BibitemOpen
  \bibfield  {author} {\bibinfo {author} {\bibfnamefont {R.}~\bibnamefont
  {Bertle}}, \bibinfo {author} {\bibfnamefont {M.}~\bibnamefont {Faber}},
  \bibinfo {author} {\bibfnamefont {J.}~\bibnamefont {Greensite}},\ and\
  \bibinfo {author} {\bibfnamefont {{\v S}.}~\bibnamefont {Olejn{\'i}k}},\
  }\bibfield  {title} {\bibinfo {title} {{The structure of projected center
  vortices in lattice gauge theory}},\ }\href
  {https://doi.org/10.1088/1126-6708/1999/03/019} {\bibfield  {journal}
  {\bibinfo  {journal} {J. High Energy Phys.}\ }\textbf {\bibinfo {volume}
  {03}},\ \bibinfo {pages} {019}},\ \Eprint
  {https://arxiv.org/abs/hep-lat/9903023} {arXiv:hep-lat/9903023} \BibitemShut
  {NoStop}%
\bibitem [{\citenamefont {Engelhardt}\ \emph {et~al.}(2000)\citenamefont
  {Engelhardt}, \citenamefont {Langfeld}, \citenamefont {Reinhardt},\ and\
  \citenamefont {Tennert}}]{Engelhardt:1999fd}%
  \BibitemOpen
  \bibfield  {author} {\bibinfo {author} {\bibfnamefont {M.}~\bibnamefont
  {Engelhardt}}, \bibinfo {author} {\bibfnamefont {K.}~\bibnamefont
  {Langfeld}}, \bibinfo {author} {\bibfnamefont {H.}~\bibnamefont
  {Reinhardt}},\ and\ \bibinfo {author} {\bibfnamefont {O.}~\bibnamefont
  {Tennert}},\ }\bibfield  {title} {\bibinfo {title} {{Deconfinement in SU(2)
  Yang-Mills theory as a center vortex percolation transition}},\ }\href
  {https://doi.org/10.1103/PhysRevD.61.054504} {\bibfield  {journal} {\bibinfo
  {journal} {Phys. Rev. D}\ }\textbf {\bibinfo {volume} {61}},\ \bibinfo
  {pages} {054504} (\bibinfo {year} {2000})},\ \Eprint
  {https://arxiv.org/abs/hep-lat/9904004} {arXiv:hep-lat/9904004} \BibitemShut
  {NoStop}%
\bibitem [{\citenamefont {Engelhardt}\ and\ \citenamefont
  {Reinhardt}(2000)}]{Engelhardt:1999wr}%
  \BibitemOpen
  \bibfield  {author} {\bibinfo {author} {\bibfnamefont {M.}~\bibnamefont
  {Engelhardt}}\ and\ \bibinfo {author} {\bibfnamefont {H.}~\bibnamefont
  {Reinhardt}},\ }\bibfield  {title} {\bibinfo {title} {{Center vortex model
  for the infrared sector of Yang-Mills theory --- confinement and
  deconfinement}},\ }\href {https://doi.org/10.1016/S0550-3213(00)00445-4}
  {\bibfield  {journal} {\bibinfo  {journal} {Nucl. Phys. B}\ }\textbf
  {\bibinfo {volume} {585}},\ \bibinfo {pages} {591} (\bibinfo {year}
  {2000})},\ \Eprint {https://arxiv.org/abs/hep-lat/9912003}
  {arXiv:hep-lat/9912003} \BibitemShut {NoStop}%
\bibitem [{\citenamefont {Faber}\ \emph {et~al.}(2000)\citenamefont {Faber},
  \citenamefont {Greensite},\ and\ \citenamefont {Olejn{\'i}k}}]{Faber:1999sq}%
  \BibitemOpen
  \bibfield  {author} {\bibinfo {author} {\bibfnamefont {M.}~\bibnamefont
  {Faber}}, \bibinfo {author} {\bibfnamefont {J.}~\bibnamefont {Greensite}},\
  and\ \bibinfo {author} {\bibfnamefont {{\v S}.}~\bibnamefont {Olejn{\'i}k}},\
  }\bibfield  {title} {\bibinfo {title} {{First evidence for center dominance
  in SU(3) lattice gauge theory}},\ }\href
  {https://doi.org/10.1016/S0370-2693(00)00013-7} {\bibfield  {journal}
  {\bibinfo  {journal} {Phys. Lett. B}\ }\textbf {\bibinfo {volume} {474}},\
  \bibinfo {pages} {177} (\bibinfo {year} {2000})},\ \Eprint
  {https://arxiv.org/abs/hep-lat/9911006} {arXiv:hep-lat/9911006} \BibitemShut
  {NoStop}%
\bibitem [{\citenamefont {de~Forcrand}\ and\ \citenamefont
  {D'Elia}(1999)}]{deForcrand:1999our}%
  \BibitemOpen
  \bibfield  {author} {\bibinfo {author} {\bibfnamefont {P.}~\bibnamefont
  {de~Forcrand}}\ and\ \bibinfo {author} {\bibfnamefont {M.}~\bibnamefont
  {D'Elia}},\ }\bibfield  {title} {\bibinfo {title} {{Relevance of Center
  Vortices to QCD}},\ }\href {https://doi.org/10.1103/PhysRevLett.82.4582}
  {\bibfield  {journal} {\bibinfo  {journal} {Phys. Rev. Lett.}\ }\textbf
  {\bibinfo {volume} {82}},\ \bibinfo {pages} {4582} (\bibinfo {year}
  {1999})},\ \Eprint {https://arxiv.org/abs/hep-lat/9901020}
  {arXiv:hep-lat/9901020} \BibitemShut {NoStop}%
\bibitem [{\citenamefont {de~Forcrand}\ and\ \citenamefont
  {Pepe}(2001)}]{deForcrand:2000pg}%
  \BibitemOpen
  \bibfield  {author} {\bibinfo {author} {\bibfnamefont {P.}~\bibnamefont
  {de~Forcrand}}\ and\ \bibinfo {author} {\bibfnamefont {M.}~\bibnamefont
  {Pepe}},\ }\bibfield  {title} {\bibinfo {title} {{Center vortices and
  monopoles without lattice Gribov copies}},\ }\href
  {https://doi.org/10.1016/S0550-3213(01)00009-8} {\bibfield  {journal}
  {\bibinfo  {journal} {Nucl. Phys. B}\ }\textbf {\bibinfo {volume} {598}},\
  \bibinfo {pages} {557} (\bibinfo {year} {2001})},\ \Eprint
  {https://arxiv.org/abs/hep-lat/0008016} {arXiv:hep-lat/0008016} \BibitemShut
  {NoStop}%
\bibitem [{\citenamefont {Kov{\'a}cs}\ and\ \citenamefont
  {Tomboulis}(2000)}]{Kovacs:2000sy}%
  \BibitemOpen
  \bibfield  {author} {\bibinfo {author} {\bibfnamefont {T.~G.}\ \bibnamefont
  {Kov{\'a}cs}}\ and\ \bibinfo {author} {\bibfnamefont {E.~T.}\ \bibnamefont
  {Tomboulis}},\ }\bibfield  {title} {\bibinfo {title} {{Computation of the
  Vortex Free Energy in SU(2) Gauge Theory}},\ }\href
  {https://doi.org/10.1103/PhysRevLett.85.704} {\bibfield  {journal} {\bibinfo
  {journal} {Phys. Rev. Lett.}\ }\textbf {\bibinfo {volume} {85}},\ \bibinfo
  {pages} {704} (\bibinfo {year} {2000})},\ \Eprint
  {https://arxiv.org/abs/hep-lat/0002004} {arXiv:hep-lat/0002004} \BibitemShut
  {NoStop}%
\bibitem [{\citenamefont {Langfeld}\ \emph {et~al.}(2002)\citenamefont
  {Langfeld}, \citenamefont {Reinhardt},\ and\ \citenamefont
  {Gattnar}}]{Langfeld:2001cz}%
  \BibitemOpen
  \bibfield  {author} {\bibinfo {author} {\bibfnamefont {K.}~\bibnamefont
  {Langfeld}}, \bibinfo {author} {\bibfnamefont {H.}~\bibnamefont
  {Reinhardt}},\ and\ \bibinfo {author} {\bibfnamefont {J.}~\bibnamefont
  {Gattnar}},\ }\bibfield  {title} {\bibinfo {title} {{Gluon propagator and
  quark confinement}},\ }\href {https://doi.org/10.1016/S0550-3213(01)00574-0}
  {\bibfield  {journal} {\bibinfo  {journal} {Nucl. Phys. B}\ }\textbf
  {\bibinfo {volume} {621}},\ \bibinfo {pages} {131} (\bibinfo {year}
  {2002})},\ \Eprint {https://arxiv.org/abs/hep-ph/0107141}
  {arXiv:hep-ph/0107141} \BibitemShut {NoStop}%
\bibitem [{\citenamefont {Langfeld}(2004)}]{Langfeld:2003ev}%
  \BibitemOpen
  \bibfield  {author} {\bibinfo {author} {\bibfnamefont {K.}~\bibnamefont
  {Langfeld}},\ }\bibfield  {title} {\bibinfo {title} {{Vortex structures in
  pure SU(3) lattice gauge theory}},\ }\href
  {https://doi.org/10.1103/PhysRevD.69.014503} {\bibfield  {journal} {\bibinfo
  {journal} {Phys. Rev. D}\ }\textbf {\bibinfo {volume} {69}},\ \bibinfo
  {pages} {014503} (\bibinfo {year} {2004})},\ \Eprint
  {https://arxiv.org/abs/hep-lat/0307030} {arXiv:hep-lat/0307030} \BibitemShut
  {NoStop}%
\bibitem [{\citenamefont {Engelhardt}\ \emph {et~al.}(2004)\citenamefont
  {Engelhardt}, \citenamefont {Quandt},\ and\ \citenamefont
  {Reinhardt}}]{Engelhardt:2003wm}%
  \BibitemOpen
  \bibfield  {author} {\bibinfo {author} {\bibfnamefont {M.}~\bibnamefont
  {Engelhardt}}, \bibinfo {author} {\bibfnamefont {M.}~\bibnamefont {Quandt}},\
  and\ \bibinfo {author} {\bibfnamefont {H.}~\bibnamefont {Reinhardt}},\
  }\bibfield  {title} {\bibinfo {title} {{Center vortex model for the infrared
  sector of $SU(3)$ Yang-Mills theory---confinement and deconfinement}},\
  }\href {https://doi.org/10.1016/j.nuclphysb.2004.02.036} {\bibfield
  {journal} {\bibinfo  {journal} {Nucl. Phys. B}\ }\textbf {\bibinfo {volume}
  {685}},\ \bibinfo {pages} {227} (\bibinfo {year} {2004})},\ \Eprint
  {https://arxiv.org/abs/hep-lat/0311029} {arXiv:hep-lat/0311029} \BibitemShut
  {NoStop}%
\bibitem [{\citenamefont {Gattnar}\ \emph {et~al.}(2005)\citenamefont
  {Gattnar}, \citenamefont {Gattringer}, \citenamefont {Langfeld},
  \citenamefont {Reinhardt}, \citenamefont {Sch{\"a}fer}, \citenamefont
  {Solbrig},\ and\ \citenamefont {Tok}}]{Gattnar:2004gx}%
  \BibitemOpen
  \bibfield  {author} {\bibinfo {author} {\bibfnamefont {J.}~\bibnamefont
  {Gattnar}}, \bibinfo {author} {\bibfnamefont {C.}~\bibnamefont {Gattringer}},
  \bibinfo {author} {\bibfnamefont {K.}~\bibnamefont {Langfeld}}, \bibinfo
  {author} {\bibfnamefont {H.}~\bibnamefont {Reinhardt}}, \bibinfo {author}
  {\bibfnamefont {A.}~\bibnamefont {Sch{\"a}fer}}, \bibinfo {author}
  {\bibfnamefont {S.}~\bibnamefont {Solbrig}},\ and\ \bibinfo {author}
  {\bibfnamefont {T.}~\bibnamefont {Tok}},\ }\bibfield  {title} {\bibinfo
  {title} {{Center vortices and Dirac eigenmodes in SU(2) lattice gauge
  theory}},\ }\href {https://doi.org/10.1016/j.nuclphysb.2005.03.027}
  {\bibfield  {journal} {\bibinfo  {journal} {Nucl. Phys. B}\ }\textbf
  {\bibinfo {volume} {716}},\ \bibinfo {pages} {105} (\bibinfo {year}
  {2005})},\ \Eprint {https://arxiv.org/abs/hep-lat/0412032}
  {arXiv:hep-lat/0412032} \BibitemShut {NoStop}%
\bibitem [{\citenamefont {Bornyakov}\ \emph {et~al.}(2008)\citenamefont
  {Bornyakov}, \citenamefont {Ilgenfritz}, \citenamefont {Martemyanov},
  \citenamefont {Morozov}, \citenamefont {Muller-Preussker},\ and\
  \citenamefont {Veselov}}]{Bornyakov:2007fz}%
  \BibitemOpen
  \bibfield  {author} {\bibinfo {author} {\bibfnamefont {V.~G.}\ \bibnamefont
  {Bornyakov}}, \bibinfo {author} {\bibfnamefont {E.~M.}\ \bibnamefont
  {Ilgenfritz}}, \bibinfo {author} {\bibfnamefont {B.~V.}\ \bibnamefont
  {Martemyanov}}, \bibinfo {author} {\bibfnamefont {S.~M.}\ \bibnamefont
  {Morozov}}, \bibinfo {author} {\bibfnamefont {M.}~\bibnamefont
  {Muller-Preussker}},\ and\ \bibinfo {author} {\bibfnamefont {A.~I.}\
  \bibnamefont {Veselov}},\ }\bibfield  {title} {\bibinfo {title}
  {{Interrelation between monopoles, vortices, topological charge and chiral
  symmetry breaking: Analysis using overlap fermions for $SU(2)$}},\ }\href
  {https://doi.org/10.1103/PhysRevD.77.074507} {\bibfield  {journal} {\bibinfo
  {journal} {Phys. Rev. D}\ }\textbf {\bibinfo {volume} {77}},\ \bibinfo
  {pages} {074507} (\bibinfo {year} {2008})},\ \Eprint
  {https://arxiv.org/abs/0708.3335} {arXiv:0708.3335 [hep-lat]} \BibitemShut
  {NoStop}%
\bibitem [{\citenamefont {Bowman}\ \emph {et~al.}(2008)\citenamefont {Bowman},
  \citenamefont {Langfeld}, \citenamefont {Leinweber}, \citenamefont {O'~Cais},
  \citenamefont {Sternbeck}, \citenamefont {von Smekal},\ and\ \citenamefont
  {Williams}}]{Bowman:2008qd}%
  \BibitemOpen
  \bibfield  {author} {\bibinfo {author} {\bibfnamefont {P.~O.}\ \bibnamefont
  {Bowman}}, \bibinfo {author} {\bibfnamefont {K.}~\bibnamefont {Langfeld}},
  \bibinfo {author} {\bibfnamefont {D.~B.}\ \bibnamefont {Leinweber}}, \bibinfo
  {author} {\bibfnamefont {A.}~\bibnamefont {O'~Cais}}, \bibinfo {author}
  {\bibfnamefont {A.}~\bibnamefont {Sternbeck}}, \bibinfo {author}
  {\bibfnamefont {L.}~\bibnamefont {von Smekal}},\ and\ \bibinfo {author}
  {\bibfnamefont {A.~G.}\ \bibnamefont {Williams}},\ }\bibfield  {title}
  {\bibinfo {title} {{Center vortices and the quark propagator in SU(2) gauge
  theory}},\ }\href {https://doi.org/10.1103/PhysRevD.78.054509} {\bibfield
  {journal} {\bibinfo  {journal} {Phys. Rev. D}\ }\textbf {\bibinfo {volume}
  {78}},\ \bibinfo {pages} {054509} (\bibinfo {year} {2008})},\ \Eprint
  {https://arxiv.org/abs/0806.4219} {arXiv:0806.4219 [hep-lat]} \BibitemShut
  {NoStop}%
\bibitem [{\citenamefont {Bowman}\ \emph {et~al.}(2011)\citenamefont {Bowman},
  \citenamefont {Langfeld}, \citenamefont {Leinweber}, \citenamefont
  {Sternbeck}, \citenamefont {von Smekal},\ and\ \citenamefont
  {Williams}}]{Bowman:2010zr}%
  \BibitemOpen
  \bibfield  {author} {\bibinfo {author} {\bibfnamefont {P.~O.}\ \bibnamefont
  {Bowman}}, \bibinfo {author} {\bibfnamefont {K.}~\bibnamefont {Langfeld}},
  \bibinfo {author} {\bibfnamefont {D.~B.}\ \bibnamefont {Leinweber}}, \bibinfo
  {author} {\bibfnamefont {A.}~\bibnamefont {Sternbeck}}, \bibinfo {author}
  {\bibfnamefont {L.}~\bibnamefont {von Smekal}},\ and\ \bibinfo {author}
  {\bibfnamefont {A.~G.}\ \bibnamefont {Williams}},\ }\bibfield  {title}
  {\bibinfo {title} {{Role of center vortices in chiral symmetry breaking in
  SU(3) gauge theory}},\ }\href {https://doi.org/10.1103/PhysRevD.84.034501}
  {\bibfield  {journal} {\bibinfo  {journal} {Phys. Rev. D}\ }\textbf {\bibinfo
  {volume} {84}},\ \bibinfo {pages} {034501} (\bibinfo {year} {2011})},\
  \Eprint {https://arxiv.org/abs/1010.4624} {arXiv:1010.4624 [hep-lat]}
  \BibitemShut {NoStop}%
\bibitem [{\citenamefont {O'Malley}\ \emph {et~al.}(2012)\citenamefont
  {O'Malley}, \citenamefont {Kamleh}, \citenamefont {Leinweber},\ and\
  \citenamefont {Moran}}]{OMalley:2011aa}%
  \BibitemOpen
  \bibfield  {author} {\bibinfo {author} {\bibfnamefont {E.-A.}\ \bibnamefont
  {O'Malley}}, \bibinfo {author} {\bibfnamefont {W.}~\bibnamefont {Kamleh}},
  \bibinfo {author} {\bibfnamefont {D.}~\bibnamefont {Leinweber}},\ and\
  \bibinfo {author} {\bibfnamefont {P.}~\bibnamefont {Moran}},\ }\bibfield
  {title} {\bibinfo {title} {{$SU(3)$ centre vortices underpin confinement and
  dynamical chiral symmetry breaking}},\ }\href
  {https://doi.org/10.1103/PhysRevD.86.054503} {\bibfield  {journal} {\bibinfo
  {journal} {Phys. Rev. D}\ }\textbf {\bibinfo {volume} {86}},\ \bibinfo
  {pages} {054503} (\bibinfo {year} {2012})},\ \Eprint
  {https://arxiv.org/abs/1112.2490} {arXiv:1112.2490 [hep-lat]} \BibitemShut
  {NoStop}%
\bibitem [{\citenamefont {H\"ollwieser}\ \emph {et~al.}(2013)\citenamefont
  {H\"ollwieser}, \citenamefont {Schweigler}, \citenamefont {Faber},\ and\
  \citenamefont {Heller}}]{Hollwieser:2013xja}%
  \BibitemOpen
  \bibfield  {author} {\bibinfo {author} {\bibfnamefont {R.}~\bibnamefont
  {H\"ollwieser}}, \bibinfo {author} {\bibfnamefont {T.}~\bibnamefont
  {Schweigler}}, \bibinfo {author} {\bibfnamefont {M.}~\bibnamefont {Faber}},\
  and\ \bibinfo {author} {\bibfnamefont {U.~M.}\ \bibnamefont {Heller}},\
  }\bibfield  {title} {\bibinfo {title} {{Center vortices and chiral symmetry
  breaking in $SU(2)$ lattice gauge theory}},\ }\href
  {https://doi.org/10.1103/PhysRevD.88.114505} {\bibfield  {journal} {\bibinfo
  {journal} {Phys. Rev. D}\ }\textbf {\bibinfo {volume} {88}},\ \bibinfo
  {pages} {114505} (\bibinfo {year} {2013})},\ \Eprint
  {https://arxiv.org/abs/1304.1277} {arXiv:1304.1277 [hep-lat]} \BibitemShut
  {NoStop}%
\bibitem [{\citenamefont {H\"ollwieser}\ \emph {et~al.}(2014)\citenamefont
  {H\"ollwieser}, \citenamefont {Faber}, \citenamefont {Schweigler},\ and\
  \citenamefont {Heller}}]{Hollwieser:2014soz}%
  \BibitemOpen
  \bibfield  {author} {\bibinfo {author} {\bibfnamefont {R.}~\bibnamefont
  {H\"ollwieser}}, \bibinfo {author} {\bibfnamefont {M.}~\bibnamefont {Faber}},
  \bibinfo {author} {\bibfnamefont {T.}~\bibnamefont {Schweigler}},\ and\
  \bibinfo {author} {\bibfnamefont {U.~M.}\ \bibnamefont {Heller}},\ }\bibfield
   {title} {\bibinfo {title} {{Chiral Symmetry Breaking from Center
  Vortices}},\ }\href {https://doi.org/10.22323/1.187.0505} {\bibfield
  {journal} {\bibinfo  {journal} {PoS}\ }\textbf {\bibinfo {volume}
  {LATTICE2013}},\ \bibinfo {pages} {505} (\bibinfo {year} {2014})},\ \Eprint
  {https://arxiv.org/abs/1410.2333} {arXiv:1410.2333 [hep-lat]} \BibitemShut
  {NoStop}%
\bibitem [{\citenamefont {Trewartha}\ \emph {et~al.}(2015)\citenamefont
  {Trewartha}, \citenamefont {Kamleh},\ and\ \citenamefont
  {Leinweber}}]{Trewartha:2015nna}%
  \BibitemOpen
  \bibfield  {author} {\bibinfo {author} {\bibfnamefont {A.}~\bibnamefont
  {Trewartha}}, \bibinfo {author} {\bibfnamefont {W.}~\bibnamefont {Kamleh}},\
  and\ \bibinfo {author} {\bibfnamefont {D.}~\bibnamefont {Leinweber}},\
  }\bibfield  {title} {\bibinfo {title} {{Evidence that centre vortices
  underpin dynamical chiral symmetry breaking in SU(3) gauge theory}},\ }\href
  {https://doi.org/10.1016/j.physletb.2015.06.025} {\bibfield  {journal}
  {\bibinfo  {journal} {Phys. Lett. B}\ }\textbf {\bibinfo {volume} {747}},\
  \bibinfo {pages} {373} (\bibinfo {year} {2015})},\ \Eprint
  {https://arxiv.org/abs/1502.06753} {arXiv:1502.06753 [hep-lat]} \BibitemShut
  {NoStop}%
\bibitem [{\citenamefont {Greensite}(2017)}]{Greensite:2016pfc}%
  \BibitemOpen
  \bibfield  {author} {\bibinfo {author} {\bibfnamefont {J.}~\bibnamefont
  {Greensite}},\ }\bibfield  {title} {\bibinfo {title} {{Confinement from
  Center Vortices: A review of old and new results}},\ }\href
  {https://doi.org/10.1051/epjconf/201713701009} {\bibfield  {journal}
  {\bibinfo  {journal} {EPJ Web Conf.}\ }\textbf {\bibinfo {volume} {137}},\
  \bibinfo {pages} {01009} (\bibinfo {year} {2017})},\ \Eprint
  {https://arxiv.org/abs/1610.06221} {arXiv:1610.06221 [hep-lat]} \BibitemShut
  {NoStop}%
\bibitem [{\citenamefont {Trewartha}\ \emph {et~al.}(2017)\citenamefont
  {Trewartha}, \citenamefont {Kamleh},\ and\ \citenamefont
  {Leinweber}}]{Trewartha:2017ive}%
  \BibitemOpen
  \bibfield  {author} {\bibinfo {author} {\bibfnamefont {A.}~\bibnamefont
  {Trewartha}}, \bibinfo {author} {\bibfnamefont {W.}~\bibnamefont {Kamleh}},\
  and\ \bibinfo {author} {\bibfnamefont {D.}~\bibnamefont {Leinweber}},\
  }\bibfield  {title} {\bibinfo {title} {{Centre vortex removal restores chiral
  symmetry}},\ }\href {https://doi.org/10.1088/1361-6471/aa9443} {\bibfield
  {journal} {\bibinfo  {journal} {J. Phys. G}\ }\textbf {\bibinfo {volume}
  {44}},\ \bibinfo {pages} {125002} (\bibinfo {year} {2017})},\ \Eprint
  {https://arxiv.org/abs/1708.06789} {arXiv:1708.06789 [hep-lat]} \BibitemShut
  {NoStop}%
\bibitem [{\citenamefont {Biddle}\ \emph
  {et~al.}(2022{\natexlab{a}})\citenamefont {Biddle}, \citenamefont {Kamleh},\
  and\ \citenamefont {Leinweber}}]{Biddle:2022zgw}%
  \BibitemOpen
  \bibfield  {author} {\bibinfo {author} {\bibfnamefont {J.~C.}\ \bibnamefont
  {Biddle}}, \bibinfo {author} {\bibfnamefont {W.}~\bibnamefont {Kamleh}},\
  and\ \bibinfo {author} {\bibfnamefont {D.~B.}\ \bibnamefont {Leinweber}},\
  }\bibfield  {title} {\bibinfo {title} {{Static quark potential from center
  vortices in the presence of dynamical fermions}},\ }\href
  {https://doi.org/10.1103/PhysRevD.106.054505} {\bibfield  {journal} {\bibinfo
   {journal} {Phys. Rev. D}\ }\textbf {\bibinfo {volume} {106}},\ \bibinfo
  {pages} {054505} (\bibinfo {year} {2022}{\natexlab{a}})},\ \Eprint
  {https://arxiv.org/abs/2206.00844} {arXiv:2206.00844 [hep-lat]} \BibitemShut
  {NoStop}%
\bibitem [{\citenamefont {Biddle}\ \emph
  {et~al.}(2022{\natexlab{b}})\citenamefont {Biddle}, \citenamefont {Kamleh},\
  and\ \citenamefont {Leinweber}}]{Biddle:2022acd}%
  \BibitemOpen
  \bibfield  {author} {\bibinfo {author} {\bibfnamefont {J.~C.}\ \bibnamefont
  {Biddle}}, \bibinfo {author} {\bibfnamefont {W.}~\bibnamefont {Kamleh}},\
  and\ \bibinfo {author} {\bibfnamefont {D.~B.}\ \bibnamefont {Leinweber}},\
  }\bibfield  {title} {\bibinfo {title} {{Impact of dynamical fermions on the
  center vortex gluon propagator}},\ }\href
  {https://doi.org/10.1103/PhysRevD.106.014506} {\bibfield  {journal} {\bibinfo
   {journal} {Phys. Rev. D}\ }\textbf {\bibinfo {volume} {106}},\ \bibinfo
  {pages} {014506} (\bibinfo {year} {2022}{\natexlab{b}})},\ \Eprint
  {https://arxiv.org/abs/2206.02320} {arXiv:2206.02320 [hep-lat]} \BibitemShut
  {NoStop}%
\bibitem [{\citenamefont {Mickley}\ \emph
  {et~al.}(2024{\natexlab{a}})\citenamefont {Mickley}, \citenamefont {Kamleh},\
  and\ \citenamefont {Leinweber}}]{Mickley:2024zyg}%
  \BibitemOpen
  \bibfield  {author} {\bibinfo {author} {\bibfnamefont {J.~A.}\ \bibnamefont
  {Mickley}}, \bibinfo {author} {\bibfnamefont {W.}~\bibnamefont {Kamleh}},\
  and\ \bibinfo {author} {\bibfnamefont {D.~B.}\ \bibnamefont {Leinweber}},\
  }\bibfield  {title} {\bibinfo {title} {{Center vortex geometry at finite
  temperature}},\ }\href {https://doi.org/10.1103/PhysRevD.110.034516}
  {\bibfield  {journal} {\bibinfo  {journal} {Phys. Rev. D}\ }\textbf {\bibinfo
  {volume} {110}},\ \bibinfo {pages} {034516} (\bibinfo {year}
  {2024}{\natexlab{a}})},\ \Eprint {https://arxiv.org/abs/2405.10670}
  {arXiv:2405.10670 [hep-lat]} \BibitemShut {NoStop}%
\bibitem [{\citenamefont {Morningstar}\ and\ \citenamefont
  {Peardon}(2004)}]{Morningstar:2003gk}%
  \BibitemOpen
  \bibfield  {author} {\bibinfo {author} {\bibfnamefont {C.}~\bibnamefont
  {Morningstar}}\ and\ \bibinfo {author} {\bibfnamefont {M.~J.}\ \bibnamefont
  {Peardon}},\ }\bibfield  {title} {\bibinfo {title} {{Analytic smearing of
  SU(3) link variables in lattice QCD}},\ }\href
  {https://doi.org/10.1103/PhysRevD.69.054501} {\bibfield  {journal} {\bibinfo
  {journal} {Phys. Rev. D}\ }\textbf {\bibinfo {volume} {69}},\ \bibinfo
  {pages} {054501} (\bibinfo {year} {2004})},\ \Eprint
  {https://arxiv.org/abs/hep-lat/0311018} {arXiv:hep-lat/0311018} \BibitemShut
  {NoStop}%
\bibitem [{\citenamefont {Durr}\ \emph {et~al.}(2011)\citenamefont {Durr},
  \citenamefont {Fodor}, \citenamefont {Hoelbling}, \citenamefont {Katz},
  \citenamefont {Krieg}, \citenamefont {Kurth}, \citenamefont {Lellouch},
  \citenamefont {Lippert}, \citenamefont {Szabo},\ and\ \citenamefont
  {Vulvert}}]{BMW:2010skj}%
  \BibitemOpen
  \bibfield  {author} {\bibinfo {author} {\bibfnamefont {S.}~\bibnamefont
  {Durr}}, \bibinfo {author} {\bibfnamefont {Z.}~\bibnamefont {Fodor}},
  \bibinfo {author} {\bibfnamefont {C.}~\bibnamefont {Hoelbling}}, \bibinfo
  {author} {\bibfnamefont {S.~D.}\ \bibnamefont {Katz}}, \bibinfo {author}
  {\bibfnamefont {S.}~\bibnamefont {Krieg}}, \bibinfo {author} {\bibfnamefont
  {T.}~\bibnamefont {Kurth}}, \bibinfo {author} {\bibfnamefont
  {L.}~\bibnamefont {Lellouch}}, \bibinfo {author} {\bibfnamefont
  {T.}~\bibnamefont {Lippert}}, \bibinfo {author} {\bibfnamefont {K.~K.}\
  \bibnamefont {Szabo}},\ and\ \bibinfo {author} {\bibfnamefont
  {G.}~\bibnamefont {Vulvert}} (\bibinfo {collaboration} {BMW}),\ }\bibfield
  {title} {\bibinfo {title} {{Lattice QCD at the physical point: Simulation and
  analysis details}},\ }\href {https://doi.org/10.1007/JHEP08(2011)148}
  {\bibfield  {journal} {\bibinfo  {journal} {J. High Energy Phys.}\ }\textbf
  {\bibinfo {volume} {08}},\ \bibinfo {pages} {148}},\ \Eprint
  {https://arxiv.org/abs/1011.2711} {arXiv:1011.2711 [hep-lat]} \BibitemShut
  {NoStop}%
\bibitem [{\citenamefont {Nogradi}\ \emph {et~al.}(2019)\citenamefont
  {Nogradi}, \citenamefont {Nogradi}, \citenamefont {Szikszai},\ and\
  \citenamefont {Szikszai}}]{Nogradi:2019iek}%
  \BibitemOpen
  \bibfield  {author} {\bibinfo {author} {\bibfnamefont {D.}~\bibnamefont
  {Nogradi}}, \bibinfo {author} {\bibfnamefont {D.}~\bibnamefont {Nogradi}},
  \bibinfo {author} {\bibfnamefont {L.}~\bibnamefont {Szikszai}},\ and\
  \bibinfo {author} {\bibfnamefont {L.}~\bibnamefont {Szikszai}},\ }\bibfield
  {title} {\bibinfo {title} {{The flavor dependence of $m_\varrho / f_\pi$}},\
  }\href {https://doi.org/10.1007/JHEP05(2019)197} {\bibfield  {journal}
  {\bibinfo  {journal} {J. High Energy Phys.}\ }\textbf {\bibinfo {volume}
  {05}},\ \bibinfo {pages} {197}},\ \bibinfo {note} {[Erratum: J. High Energy
  Phys. 06, 031 (2022)]},\ \Eprint {https://arxiv.org/abs/1905.01909}
  {arXiv:1905.01909 [hep-lat]} \BibitemShut {NoStop}%
\bibitem [{\citenamefont {Nogradi}\ and\ \citenamefont
  {Szikszai}(2019)}]{Nogradi:2019auv}%
  \BibitemOpen
  \bibfield  {author} {\bibinfo {author} {\bibfnamefont {D.}~\bibnamefont
  {Nogradi}}\ and\ \bibinfo {author} {\bibfnamefont {L.}~\bibnamefont
  {Szikszai}},\ }\bibfield  {title} {\bibinfo {title} {{The model dependence of
  $m_\varrho / f_\pi$}},\ }\href {https://doi.org/10.22323/1.363.0237}
  {\bibfield  {journal} {\bibinfo  {journal} {PoS}\ }\textbf {\bibinfo {volume}
  {LATTICE2019}},\ \bibinfo {pages} {237} (\bibinfo {year} {2019})},\ \Eprint
  {https://arxiv.org/abs/1912.04114} {arXiv:1912.04114 [hep-lat]} \BibitemShut
  {NoStop}%
\bibitem [{\citenamefont {Kotov}\ \emph {et~al.}(2021)\citenamefont {Kotov},
  \citenamefont {Nogradi}, \citenamefont {Szabo},\ and\ \citenamefont
  {Szikszai}}]{Kotov:2021mgp}%
  \BibitemOpen
  \bibfield  {author} {\bibinfo {author} {\bibfnamefont {A.~Y.}\ \bibnamefont
  {Kotov}}, \bibinfo {author} {\bibfnamefont {D.}~\bibnamefont {Nogradi}},
  \bibinfo {author} {\bibfnamefont {K.~K.}\ \bibnamefont {Szabo}},\ and\
  \bibinfo {author} {\bibfnamefont {L.}~\bibnamefont {Szikszai}},\ }\bibfield
  {title} {\bibinfo {title} {{More on the flavor dependence of $m_\varrho /
  f_\pi$}},\ }\href {https://doi.org/10.1007/JHEP07(2021)202} {\bibfield
  {journal} {\bibinfo  {journal} {J. High Energy Phys.}\ }\textbf {\bibinfo
  {volume} {07}},\ \bibinfo {pages} {202}},\ \bibinfo {note} {[Erratum: J. High
  Energy Phys. 06, 032 (2022)]},\ \Eprint {https://arxiv.org/abs/2107.05996}
  {arXiv:2107.05996 [hep-lat]} \BibitemShut {NoStop}%
\bibitem [{\citenamefont {Cheng}\ \emph {et~al.}(2012)\citenamefont {Cheng},
  \citenamefont {Hasenfratz},\ and\ \citenamefont {Schaich}}]{Cheng:2011ic}%
  \BibitemOpen
  \bibfield  {author} {\bibinfo {author} {\bibfnamefont {A.}~\bibnamefont
  {Cheng}}, \bibinfo {author} {\bibfnamefont {A.}~\bibnamefont {Hasenfratz}},\
  and\ \bibinfo {author} {\bibfnamefont {D.}~\bibnamefont {Schaich}},\
  }\bibfield  {title} {\bibinfo {title} {{Novel phase in SU(3) lattice gauge
  theory with 12 light fermions}},\ }\href
  {https://doi.org/10.1103/PhysRevD.85.094509} {\bibfield  {journal} {\bibinfo
  {journal} {Phys. Rev. D}\ }\textbf {\bibinfo {volume} {85}},\ \bibinfo
  {pages} {094509} (\bibinfo {year} {2012})},\ \Eprint
  {https://arxiv.org/abs/1111.2317} {arXiv:1111.2317 [hep-lat]} \BibitemShut
  {NoStop}%
\bibitem [{\citenamefont {Duane}\ \emph {et~al.}(1987)\citenamefont {Duane},
  \citenamefont {Kennedy}, \citenamefont {Pendleton},\ and\ \citenamefont
  {Roweth}}]{Duane:1987de}%
  \BibitemOpen
  \bibfield  {author} {\bibinfo {author} {\bibfnamefont {S.}~\bibnamefont
  {Duane}}, \bibinfo {author} {\bibfnamefont {A.~D.}\ \bibnamefont {Kennedy}},
  \bibinfo {author} {\bibfnamefont {B.~J.}\ \bibnamefont {Pendleton}},\ and\
  \bibinfo {author} {\bibfnamefont {D.}~\bibnamefont {Roweth}},\ }\bibfield
  {title} {\bibinfo {title} {{Hybrid Monte Carlo}},\ }\href
  {https://doi.org/10.1016/0370-2693(87)91197-X} {\bibfield  {journal}
  {\bibinfo  {journal} {Phys. Lett. B}\ }\textbf {\bibinfo {volume} {195}},\
  \bibinfo {pages} {216} (\bibinfo {year} {1987})}\BibitemShut {NoStop}%
\bibitem [{\citenamefont {Clark}\ and\ \citenamefont
  {Kennedy}(2007)}]{Clark:2006fx}%
  \BibitemOpen
  \bibfield  {author} {\bibinfo {author} {\bibfnamefont {M.~A.}\ \bibnamefont
  {Clark}}\ and\ \bibinfo {author} {\bibfnamefont {A.~D.}\ \bibnamefont
  {Kennedy}},\ }\bibfield  {title} {\bibinfo {title} {{Accelerating
  Dynamical-Fermion Computations Using the Rational Hybrid Monte Carlo
  Algorithm with Multiple Pseudofermion Fields}},\ }\href
  {https://doi.org/10.1103/PhysRevLett.98.051601} {\bibfield  {journal}
  {\bibinfo  {journal} {Phys. Rev. Lett.}\ }\textbf {\bibinfo {volume} {98}},\
  \bibinfo {pages} {051601} (\bibinfo {year} {2007})},\ \Eprint
  {https://arxiv.org/abs/hep-lat/0608015} {arXiv:hep-lat/0608015} \BibitemShut
  {NoStop}%
\bibitem [{\citenamefont {Sexton}\ and\ \citenamefont
  {Weingarten}(1992)}]{Sexton:1992nu}%
  \BibitemOpen
  \bibfield  {author} {\bibinfo {author} {\bibfnamefont {J.~C.}\ \bibnamefont
  {Sexton}}\ and\ \bibinfo {author} {\bibfnamefont {D.~H.}\ \bibnamefont
  {Weingarten}},\ }\bibfield  {title} {\bibinfo {title} {{Hamiltonian evolution
  for the hybrid Monte Carlo algorithm}},\ }\href
  {https://doi.org/10.1016/0550-3213(92)90263-B} {\bibfield  {journal}
  {\bibinfo  {journal} {Nucl. Phys. B}\ }\textbf {\bibinfo {volume} {380}},\
  \bibinfo {pages} {665} (\bibinfo {year} {1992})}\BibitemShut {NoStop}%
\bibitem [{\citenamefont {Omelyan}\ \emph {et~al.}(2003)\citenamefont
  {Omelyan}, \citenamefont {Mryglod},\ and\ \citenamefont
  {Folk}}]{Omelyan:2002qkh}%
  \BibitemOpen
  \bibfield  {author} {\bibinfo {author} {\bibfnamefont {I.~P.}\ \bibnamefont
  {Omelyan}}, \bibinfo {author} {\bibfnamefont {I.~M.}\ \bibnamefont
  {Mryglod}},\ and\ \bibinfo {author} {\bibfnamefont {R.}~\bibnamefont
  {Folk}},\ }\bibfield  {title} {\bibinfo {title} {{Symplectic analytically
  integrable decomposition algorithms: classification, derivation, and
  application to molecular dynamics, quantum and celestial mechanics
  simulations}},\ }\href {https://doi.org/10.1016/S0010-4655(02)00754-3}
  {\bibfield  {journal} {\bibinfo  {journal} {Comput. Phys. Commun.}\ }\textbf
  {\bibinfo {volume} {151}},\ \bibinfo {pages} {272} (\bibinfo {year}
  {2003})}\BibitemShut {NoStop}%
\bibitem [{\citenamefont {Takaishi}\ and\ \citenamefont
  {de~Forcrand}(2006)}]{Takaishi:2005tz}%
  \BibitemOpen
  \bibfield  {author} {\bibinfo {author} {\bibfnamefont {T.}~\bibnamefont
  {Takaishi}}\ and\ \bibinfo {author} {\bibfnamefont {P.}~\bibnamefont
  {de~Forcrand}},\ }\bibfield  {title} {\bibinfo {title} {{Testing and tuning
  symplectic integrators for hybrid Monte Carlo algorithm in lattice QCD}},\
  }\href {https://doi.org/10.1103/PhysRevE.73.036706} {\bibfield  {journal}
  {\bibinfo  {journal} {Phys. Rev. E}\ }\textbf {\bibinfo {volume} {73}},\
  \bibinfo {pages} {036706} (\bibinfo {year} {2006})},\ \Eprint
  {https://arxiv.org/abs/hep-lat/0505020} {arXiv:hep-lat/0505020} \BibitemShut
  {NoStop}%
\bibitem [{\citenamefont {Montero}(1999)}]{Montero:1999by}%
  \BibitemOpen
  \bibfield  {author} {\bibinfo {author} {\bibfnamefont {A.}~\bibnamefont
  {Montero}},\ }\bibfield  {title} {\bibinfo {title} {{Study of SU(3)
  vortex-like configurations with a new maximal center gauge fixing method}},\
  }\href {https://doi.org/10.1016/S0370-2693(99)01113-2} {\bibfield  {journal}
  {\bibinfo  {journal} {Phys. Lett. B}\ }\textbf {\bibinfo {volume} {467}},\
  \bibinfo {pages} {106} (\bibinfo {year} {1999})},\ \Eprint
  {https://arxiv.org/abs/hep-lat/9906010} {arXiv:hep-lat/9906010} \BibitemShut
  {NoStop}%
\bibitem [{\citenamefont {Spengler}\ \emph {et~al.}(2018)\citenamefont
  {Spengler}, \citenamefont {Quandt},\ and\ \citenamefont
  {Reinhardt}}]{Spengler:2018dxt}%
  \BibitemOpen
  \bibfield  {author} {\bibinfo {author} {\bibfnamefont {F.}~\bibnamefont
  {Spengler}}, \bibinfo {author} {\bibfnamefont {M.}~\bibnamefont {Quandt}},\
  and\ \bibinfo {author} {\bibfnamefont {H.}~\bibnamefont {Reinhardt}},\
  }\bibfield  {title} {\bibinfo {title} {{Branching of center vortices in SU(3)
  lattice gauge theory}},\ }\href {https://doi.org/10.1103/PhysRevD.98.094508}
  {\bibfield  {journal} {\bibinfo  {journal} {Phys. Rev. D}\ }\textbf {\bibinfo
  {volume} {98}},\ \bibinfo {pages} {094508} (\bibinfo {year} {2018})},\
  \Eprint {https://arxiv.org/abs/1810.04072} {arXiv:1810.04072 [hep-th]}
  \BibitemShut {NoStop}%
\bibitem [{\citenamefont {Faber}\ \emph
  {et~al.}(1999{\natexlab{b}})\citenamefont {Faber}, \citenamefont {Greensite},
  \citenamefont {Olejn{\'i}k},\ and\ \citenamefont {Yamada}}]{Faber:1999gu}%
  \BibitemOpen
  \bibfield  {author} {\bibinfo {author} {\bibfnamefont {M.}~\bibnamefont
  {Faber}}, \bibinfo {author} {\bibfnamefont {J.}~\bibnamefont {Greensite}},
  \bibinfo {author} {\bibfnamefont {{\v S}.}~\bibnamefont {Olejn{\'i}k}},\ and\
  \bibinfo {author} {\bibfnamefont {D.}~\bibnamefont {Yamada}},\ }\bibfield
  {title} {\bibinfo {title} {{The vortex-finding property of maximal center
  (and other) gauges}},\ }\href {https://doi.org/10.1088/1126-6708/1999/12/012}
  {\bibfield  {journal} {\bibinfo  {journal} {J. High Energy Phys.}\ }\textbf
  {\bibinfo {volume} {12}},\ \bibinfo {pages} {012}},\ \Eprint
  {https://arxiv.org/abs/hep-lat/9910033} {arXiv:hep-lat/9910033} \BibitemShut
  {NoStop}%
\bibitem [{\citenamefont {Wilson}(1974)}]{Wilson:1974sk}%
  \BibitemOpen
  \bibfield  {author} {\bibinfo {author} {\bibfnamefont {K.~G.}\ \bibnamefont
  {Wilson}},\ }\bibfield  {title} {\bibinfo {title} {{Confinement of quarks}},\
  }\href {https://doi.org/10.1103/PhysRevD.10.2445} {\bibfield  {journal}
  {\bibinfo  {journal} {Phys. Rev. D}\ }\textbf {\bibinfo {volume} {10}},\
  \bibinfo {pages} {2445} (\bibinfo {year} {1974})}\BibitemShut {NoStop}%
\bibitem [{\citenamefont {Biddle}\ \emph {et~al.}(2020)\citenamefont {Biddle},
  \citenamefont {Kamleh},\ and\ \citenamefont {Leinweber}}]{Biddle:2019gke}%
  \BibitemOpen
  \bibfield  {author} {\bibinfo {author} {\bibfnamefont {J.~C.}\ \bibnamefont
  {Biddle}}, \bibinfo {author} {\bibfnamefont {W.}~\bibnamefont {Kamleh}},\
  and\ \bibinfo {author} {\bibfnamefont {D.~B.}\ \bibnamefont {Leinweber}},\
  }\bibfield  {title} {\bibinfo {title} {{Visualization of center vortex
  structure}},\ }\href {https://doi.org/10.1103/PhysRevD.102.034504} {\bibfield
   {journal} {\bibinfo  {journal} {Phys. Rev. D}\ }\textbf {\bibinfo {volume}
  {102}},\ \bibinfo {pages} {034504} (\bibinfo {year} {2020})},\ \Eprint
  {https://arxiv.org/abs/1912.09531} {arXiv:1912.09531 [hep-lat]} \BibitemShut
  {NoStop}%
\bibitem [{\citenamefont {Biddle}\ \emph {et~al.}(2023)\citenamefont {Biddle},
  \citenamefont {Kamleh},\ and\ \citenamefont {Leinweber}}]{Biddle:2023lod}%
  \BibitemOpen
  \bibfield  {author} {\bibinfo {author} {\bibfnamefont {J.~C.}\ \bibnamefont
  {Biddle}}, \bibinfo {author} {\bibfnamefont {W.}~\bibnamefont {Kamleh}},\
  and\ \bibinfo {author} {\bibfnamefont {D.~B.}\ \bibnamefont {Leinweber}},\
  }\bibfield  {title} {\bibinfo {title} {{Center vortex structure in the
  presence of dynamical fermions}},\ }\href
  {https://doi.org/10.1103/PhysRevD.107.094507} {\bibfield  {journal} {\bibinfo
   {journal} {Phys. Rev. D}\ }\textbf {\bibinfo {volume} {107}},\ \bibinfo
  {pages} {094507} (\bibinfo {year} {2023})},\ \Eprint
  {https://arxiv.org/abs/2302.05897} {arXiv:2302.05897 [hep-lat]} \BibitemShut
  {NoStop}%
\bibitem [{\citenamefont {Mickley}\ \emph
  {et~al.}(2024{\natexlab{b}})\citenamefont {Mickley}, \citenamefont {Allton},
  \citenamefont {Bignell},\ and\ \citenamefont {Leinweber}}]{Mickley:2024vkm}%
  \BibitemOpen
  \bibfield  {author} {\bibinfo {author} {\bibfnamefont {J.~A.}\ \bibnamefont
  {Mickley}}, \bibinfo {author} {\bibfnamefont {C.}~\bibnamefont {Allton}},
  \bibinfo {author} {\bibfnamefont {R.}~\bibnamefont {Bignell}},\ and\ \bibinfo
  {author} {\bibfnamefont {D.~B.}\ \bibnamefont {Leinweber}},\ }\bibfield
  {title} {\bibinfo {title} {{Centre vortex evidence for a second
  finite-temperature QCD transition}},\ }\href@noop {} {\  (\bibinfo {year}
  {2024}{\natexlab{b}})},\ \Eprint {https://arxiv.org/abs/2411.19446}
  {arXiv:2411.19446 [hep-lat]} \BibitemShut {NoStop}%
\bibitem [{\citenamefont {Pauli}\ and\ \citenamefont
  {Villars}(1949)}]{Pauli:1949zm}%
  \BibitemOpen
  \bibfield  {author} {\bibinfo {author} {\bibfnamefont {W.}~\bibnamefont
  {Pauli}}\ and\ \bibinfo {author} {\bibfnamefont {F.}~\bibnamefont
  {Villars}},\ }\bibfield  {title} {\bibinfo {title} {{On the Invariant
  Regularization in Relativistic Quantum Theory}},\ }\href
  {https://doi.org/10.1103/RevModPhys.21.434} {\bibfield  {journal} {\bibinfo
  {journal} {Rev. Mod. Phys.}\ }\textbf {\bibinfo {volume} {21}},\ \bibinfo
  {pages} {434} (\bibinfo {year} {1949})}\BibitemShut {NoStop}%
\bibitem [{\citenamefont {Hasenfratz}\ \emph {et~al.}(2021)\citenamefont
  {Hasenfratz}, \citenamefont {Shamir},\ and\ \citenamefont
  {Svetitsky}}]{Hasenfratz:2021zsl}%
  \BibitemOpen
  \bibfield  {author} {\bibinfo {author} {\bibfnamefont {A.}~\bibnamefont
  {Hasenfratz}}, \bibinfo {author} {\bibfnamefont {Y.}~\bibnamefont {Shamir}},\
  and\ \bibinfo {author} {\bibfnamefont {B.}~\bibnamefont {Svetitsky}},\
  }\bibfield  {title} {\bibinfo {title} {{Taming lattice artifacts with
  Pauli-Villars fields}},\ }\href {https://doi.org/10.1103/PhysRevD.104.074509}
  {\bibfield  {journal} {\bibinfo  {journal} {Phys. Rev. D}\ }\textbf {\bibinfo
  {volume} {104}},\ \bibinfo {pages} {074509} (\bibinfo {year} {2021})},\
  \Eprint {https://arxiv.org/abs/2109.02790} {arXiv:2109.02790 [hep-lat]}
  \BibitemShut {NoStop}%
\bibitem [{\citenamefont {Hasenfratz}\ and\ \citenamefont
  {Knechtli}(2001)}]{Hasenfratz:2001hp}%
  \BibitemOpen
  \bibfield  {author} {\bibinfo {author} {\bibfnamefont {A.}~\bibnamefont
  {Hasenfratz}}\ and\ \bibinfo {author} {\bibfnamefont {F.}~\bibnamefont
  {Knechtli}},\ }\bibfield  {title} {\bibinfo {title} {{Flavor symmetry and the
  static potential with hypercubic blocking}},\ }\href
  {https://doi.org/10.1103/PhysRevD.64.034504} {\bibfield  {journal} {\bibinfo
  {journal} {Phys. Rev. D}\ }\textbf {\bibinfo {volume} {64}},\ \bibinfo
  {pages} {034504} (\bibinfo {year} {2001})},\ \Eprint
  {https://arxiv.org/abs/hep-lat/0103029} {arXiv:hep-lat/0103029} \BibitemShut
  {NoStop}%
\bibitem [{\citenamefont {Hasenfratz}\ \emph {et~al.}(2007)\citenamefont
  {Hasenfratz}, \citenamefont {Hoffmann},\ and\ \citenamefont
  {Schaefer}}]{Hasenfratz:2007rf}%
  \BibitemOpen
  \bibfield  {author} {\bibinfo {author} {\bibfnamefont {A.}~\bibnamefont
  {Hasenfratz}}, \bibinfo {author} {\bibfnamefont {R.}~\bibnamefont
  {Hoffmann}},\ and\ \bibinfo {author} {\bibfnamefont {S.}~\bibnamefont
  {Schaefer}},\ }\bibfield  {title} {\bibinfo {title} {{Hypercubic smeared
  links for dynamical fermions}},\ }\href
  {https://doi.org/10.1088/1126-6708/2007/05/029} {\bibfield  {journal}
  {\bibinfo  {journal} {J. High Energy Phys.}\ }\textbf {\bibinfo {volume}
  {05}},\ \bibinfo {pages} {029}},\ \Eprint
  {https://arxiv.org/abs/hep-lat/0702028} {arXiv:hep-lat/0702028} \BibitemShut
  {NoStop}%
\end{thebibliography}%

\end{document}